\documentclass[aps,prb,twocolumn,superscriptaddress,bibliography]{revtex4-2}

\usepackage{amsmath}
\usepackage{amssymb}
\usepackage{graphicx}
\usepackage{epstopdf} 
\AppendGraphicsExtensions{.tif}

\usepackage{xcolor}
\usepackage{comment}
\usepackage{xr}
\usepackage{soul}

\makeatletter
\newcommand*{\addFileDependency}[1]{
  \typeout{(#1)}
  \@addtofilelist{#1}
  \IfFileExists{#1}{}{\typeout{No file #1.}}
}
\makeatother

\newcommand*{\myexternaldocument}[1]{%
    \externaldocument{#1}%
    \addFileDependency{#1.tex}%
    \addFileDependency{#1.aux}%
}

\myexternaldocument{Na2Co2TeO6-Supp}

\newcommand{\VZ}[1] {{\color{blue}#1}}

\newcommand{\NCTO} {Na$_2$Co$_2$TeO$_6$}
\newcommand{\RuCl} {$\alpha$-RuCl$_{3}$}

\begin{document}


\title{Electronic and magnetic phase diagrams of Kitaev quantum spin liquid candidate Na$_2$Co$_2$TeO$_6$}



\author{Shengzhi Zhang}
\email{shengzhi@lanl.gov}
\affiliation{National High Magnetic Field Laboratory, Los Alamos National Laboratory, Los Alamos, New Mexico 87545, USA.}

\author{Sangyun Lee}
\affiliation{MPA-Q, Los Alamos National Laboratory, Los Alamos, New Mexico 87545, USA.}
\author{Andrew J. Woods}
\affiliation{MPA-Q, Los Alamos National Laboratory, Los Alamos, New Mexico 87545, USA.}
\author{William Peria}
\affiliation{National High Magnetic Field Laboratory, Los Alamos National Laboratory, Los Alamos, New Mexico 87545, USA.}
\author{Sean M. Thomas}
\affiliation{MPA-Q, Los Alamos National Laboratory, Los Alamos, New Mexico 87545, USA.}
\author{Roman Movshovich}
\affiliation{MPA-Q, Los Alamos National Laboratory, Los Alamos, New Mexico 87545, USA.}
\author{Eric Brosha}
\affiliation{Los Alamos National Laboratory, Los Alamos, New Mexico 87545, USA.}
\author{Qing Huang}
\affiliation{Department of Physics, University of Tennessee, Knoxville, Tennessee 37996, USA.}
\author{Haidong Zhou}
\affiliation{Department of Physics, University of Tennessee, Knoxville, Tennessee 37996, USA.}
\author{Vivien S. Zapf}
\email{vzapf@lanl.gov}
\affiliation{National High Magnetic Field Laboratory, Los Alamos National Laboratory, Los Alamos, New Mexico 87545, USA.}
\author{Minseong Lee}
\email{ml10k@lanl.gov}
\affiliation{National High Magnetic Field Laboratory, Los Alamos National Laboratory, Los Alamos, New Mexico 87545, USA.}

\date{\today}


\date{\today}

\begin{abstract}
The 3$d^7$ Co$^{2+}$-based insulating magnet \NCTO{} has recently been reported to have strong Kitaev interactions
on a honeycomb lattice, and is thus being considered as a Kitaev quantum spin liquid candidate. However, due to the existence of other types of interactions, a spontaneous long-range magnetic order occurs. This order is suppressed by applied magnetic fields leading to a succession of phases and ultimately saturation of the magnetic moments. The precise phase diagram, the nature of the phases, and the possibility that one of the field-induced phases is a Kitaev quantum spin liquid phase are still a matter of debate. Here we measured an extensive set of physical properties to build the complete temperature-field phase diagrams to magnetic saturation at 10 T for magnetic fields along the $a$- and $a^*$-axes, and a partial phase diagram up to 60 T along $c$. We probe the phases using magnetization, specific heat, magnetocaloric effect, magnetostriction, dielectric constant, and electric polarization, which is a symmetry-sensitive probe. With these measurements we identify all the previously incomplete phase boundaries and find new high-field phase boundaries. We find strong magnetoelectric coupling in the dielectric constant and moderate magnetostrictive coupling at several phase boundaries. Furthermore, we detect the symmetry of the magnetic order using electrical polarization measurements under magnetic fields. Based on our analysis, the absence of electric polarization under zero or finite magnetic field in any of the phases or after any combination of magnetic/electric field cooling suggests that a zigzag spin structure is more likely than a triple-Q spin structure at zero field. Finally we investigate the hysteresis and 1$^{st}$ or 2$^{nd}$ order nature of each phase transition and its entropy changes. With this information we establish a map of the magnetic phases of this compound and its magnetic, thermodynamic and magnetoelectric properties, and discuss
where spin liquid or other phases may be sought in future studies.
\end{abstract}


\maketitle

\section{Introduction}
Magnetism on a honeycomb lattice with antiferromagnetic nearest neighbor \textit{isotropic} exchange interactions is not frustrated. However, when the exchange interactions are dominated by \textit{bond-dependent} Kitaev interactions, strong magnetic frustration results and the intriguing Kitaev quantum spin liquid (KQSL) is predicted to form as the ground state \cite{kitaev2006anyons}. The Kitaev exchange interaction is of the form $K_{\gamma}S^{\gamma}_i S^{\gamma}_j$ where $\gamma \in x,y,z$ indicates the three types of bonds in a honeycomb lattice.


KQSLs are of particular interest to the quantum computing community because they host non-Abelian anyonic excitations \cite{kitaev2006anyons}. Non-Abelian anyons change the observable state of the system if they are braided (moved around each other) and these braiding operations have been shown to be capable of supporting fault-tolerant quantum computations \cite{kitaev2006anyons,nayak2008non}. However, the discovery of KQSLs in real magnets is still a significant challenge. KQSL candidates show both Kitaev and Heisenberg interactions, as well as other terms like off-diagonal exchange interactions and single-ion anisotropies. Thus most known candidates order at zero magnetic field (see below). Luckily, it has been predicted that if the non-Kitaev interactions are small enough, the long-range order can be suppressed by magnetic field in favor of a KQSL state \cite{Kasahara2018half-quantized,kasahara2022quantized}.

The Kitaev interaction can be realized at a certain balance of crystal field and spin-orbit coupling with 90$^{\circ}$ exchange paths. With octahedral crystal fields, orbitals split into $e_{g}$ and $t_{2g}$ levels. Five electrons residing in $t_{2g}$ lead to total $S = 1/2$ and $L = 1$. The strong spin-orbit coupling mixes $S$ and $L$ to form a spin-orbit entangled $j_{\text{eff}} = \frac{1}{2}$ Kramers\VZ{'} doublet, whose narrow band opens a Mott gap. The conventional Heisenberg type exchange interactions are suppressed due to quantum interference between multiple paths across ligand ions on edge-sharing octahedra  \cite{jackeli2009mott}.

Due to the need for strong spin-orbit coupling, most research on potential KQSLs has focused on the 4{\it d} and 5{\it d} ions with the low-spin $d^{5}$ electron configuration such as Ru$^{3+}$ and Ir$^{4+}$. The first prominent candidates were honeycomb iridium oxides $A_{2}$IrO$_{3}$ with $A$ = Na, Li \cite{chaloupka2010kitaev}, which have recently been extended to include $A_{3}$LiIr$_{2}$O$_{6}$ with $A = $ Ag and H \cite{kitagawa2018spin, bahrami2019thermodynamic, chakraborty2021unusual}. All of these candidates except for H$_{3}$LiIr$_{2}$O$_{6}$ show magnetic ordering at zero magnetic field while Ag$_{3}$LiIr$_{2}$O$_{6}$ was shown to form magnetic ordering in less disordered crystals \cite{Bahrami2021, LiValenti2022} and may show a KQSL in applied magnetic fields. Another highly promising candidate is \RuCl{} \cite{plumb2014alpha}.  It exhibits continuum spin excitations in neutron scattering experiments \cite{banerjee2016proximate,do2017RuClINS} around the Brillouin zone center, which signifies flux excitations in addition to the itinerant Majorana Fermions. Moreover, albeit seemingly dependent on growth techniques and the precise stacking structure, a potential half-quantized thermal Hall conductivity suggests a chiral quantum spin liquid phase stabilized by magnetic fields \cite{Kasahara2018half-quantized, czajka2021oscillations, bruin2022origin, bruin2022robustness, kasahara2022quantized}.

More recently, it has been proposed that certain 3{\it d} transition metals, which were previously dismissed by the conventional wisdom that they have small spin-orbit coupling, can also host the Kitaev exchange interactions playing a dominant role for their magnetism \cite{liu2018pseudospin,Sano18,liu2021towards}. It is noted that as long as the spin-orbit coupling is comparable to or larger than the exchange and orbital-lattice interactions, the Kitaev interaction can still dominate. Indeed, $d^7$ Co$^{2+}$ with the high spin configuration $t_{2g}^5e_g ^2$ has been shown to provide a strongly spin-orbit entangled $j_{\text{eff}}$ = 1/2 degree of freedom with Kitaev interactions \cite{liu2018pseudospin,Sano18,liu2021towards}  \textit{given that the crystal field is weak enough that it does not affect the spin-orbit coupling}. In $3d^7$ systems, the Kitaev interaction comes almost entirely from the $t_{2g}$-$e_g$ hopping process. It is calculated to dominate over the Heisenberg interactions because the $e_g$-$e_g$ Heisenberg and off-diagonal exchange interactions cancel those from the $t_{2g}$-$e_g$ hopping process \cite{liu2021towards}. In addition, it is helpful that the more localized nature of 3{\it d} orbitals compared to 4{\it d} or 5{\it d} helps suppress second- and third-nearest neighbor exchange interactions \cite{liu2018pseudospin,Sano18,liu2021towards}.

\NCTO{} has been proposed as a candidate 3{\it d} KQSL compound due to its honeycomb lattice and the observation in inelastic neutron scattering studies of Kitaev interactions \cite{kim2021antiferromagnetic, lin2021field, sanders2022dominant, yao2022excitations}.  Co$^{2+}$ has a $3d^7$ electronic configurations under octahedral crystal field and Co$^{2+}$-O$^{2-}$-Co$^{2+}$ form close to 90$^{\circ}$ bonds with spin-orbit coupling comparable with other energy scales \cite{viciu2007structure}. Such nearly ideal oxygen octahedron geometry \cite{viciu2007structure} helps suppress Heisenberg and symmetric off-diagonal terms \cite{liu2021towards}.  Some calculations predict a ferromagnetic Kitaev interaction \cite{Sano18, liu2018pseudospin} while the inelastic neutron scattering measurements \cite{kim2021antiferromagnetic, lin2021field, sanders2022dominant, yao2022excitations} support dominant antiferromagnetic Kitaev exchange. A theory by S. Winter considered both possibilities, and finds Kitaev interactions to be small \cite{winter2022magnetic}.

The space group of \NCTO{} is {\it P}6$_{3}$22 (No. 182), which is piezoelectric, providing another avenue for tracking phase transitions in high magnetic fields via their effect on electrical properties. The magnetic honeycomb layers of Co$^{2+}$ are separated by nonmagnetic Na$^{+}$ layers, which makes \NCTO{} a magnetically quasi-two-dimensional system \cite{viciu2007structure}. In comparison to $\alpha$-RuCl$_3$, \NCTO{} is structurally more robust and no other stacking sequence of layers has been detected \cite{bera2017zigzag}.

To explore the ground state and its evolution under magnetic field, several phase diagrams were constructed in the literature \cite{lin2021field, hong2021strongly, xiao2021magnetic, lin2022evidence}. However, these studies do not all extend to magnetic saturation and certain phase boundaries still need a closer investigation as they may surround a KQSL phase. Furthermore, although the ground state at low temperature in zero field is well established to be antiferromagnetic \cite{lefranccois2016magnetic}, its spin structure is still under debate. A zigzag structure \cite{lefranccois2016magnetic} similar to $\alpha$-RuCl$_3$ \cite{johnson2015monoclinic} was initially proposed due to commonality observed in both systems \cite{lin2021field}. The detected magnon dispersion from neutron diffraction can be well fitted using models based on the zigzag spin structure despite the discrepancies in fitting parameters among different studies \cite{songvilay2020kitaev, samarakoon2021static, kim2021antiferromagnetic, sanders2022dominant}. However, more recent inelastic neutron scattering \cite{chen2021spin} and nuclear magnetic resonance studies \cite{lee2021multistage} proposed a triple-Q order as the ground state spin structure. We also note that due to slightly different environments of the two Co, a ferrimagnetic magnetization was observed. \cite{yao2020ferrimagnetism}.

Multiple phases are observed in \NCTO{} with applied magnetic field. A phase emerging above 9.5 T is established as a mostly spin polarized phase from magnetization and specific heat measurements \cite{yao2020ferrimagnetism, hong2021strongly}. However, the nature of other phases are still not determined. For instance, a phase is observed in single-crystalline \NCTO{} above a metamagnetic phase transition at about 6 T. This phase mimics a putative KQSL phase in $\alpha$-RuCl$_3$ in many aspects such as observations of a plateau in field-dependent magnetic entropy and of an additional electron spin resonance mode \cite{lin2021field}. Therefore, it was also considered as a spin disordered phase. However, because of the lack of enough data points, the phase boundaries are not well defined and the nature of this phase is still not clear. Additionally, another phase was observed at between 8 T and saturation field in thermal conductivity measurement \cite{hong2021strongly} and in a combined study using torque magnetometry and inelastic neutron scattering \cite{lin2022evidence} measurements with magnetic field applied along the $a^*$-axis. This phase, instead of the above mentioned, is recently proposed to be a KQSL phase \cite{lin2022evidence}, but this also needs further confirmation from other measurements. Thus, it is necessary to establish a more complete phase diagram from many different experimental techniques that reaches the full saturation of the magnetization to clarify the behaviors of this material.

In this work, we construct comprehensive temperature-magnetic field ($\bf{T}$-$\bf{H}$) phase diagrams along both $a$- and $a^{*}$-axes based on the magnetic, thermodynamic, electric, and elastic properties of \NCTO{}. We also investigate a partial phase diagram for ${\bf H} \parallel c$. We observed a series of three phase transitions in magnetic fields for $\bf{H} \parallel $ $a$ and $a^{*}$. Most previous papers see only various subsets of these three phase transitions due to limited number of measurement techniques, though recent torque magnetometry and inelastic neutron scattering data \cite{lin2022evidence} show evidence of all three. For ${\bf H} \parallel a$, we also observe apparent phase transitions $T_{\text{1}}$ and $T_{\text{2}}$ as a function of temperature in the thermal expansion and specific heat at high fields that were not previous reported.

We do not observe temperature or field-dependent electric polarization onsetting at any of the magnetic phase boundaries in both single-crystals along $a^*$-axis with $\bf{H} \parallel$ $a$-axis and in a large polycrystal despite magnetic \textit{and} electric poling. This disagrees with the triple-Q spin structure, which should produce an electric polarization under magnetic field. Rather it favors the zigzag spin structure, whose symmetry does not support electric polarization with and without external magnetic fields. We note that in over 14 years of studying electric polarization at the National High Magnetic Field Laboratory in complex magnets, it was found that when the necessary symmetry conditions are fulfilled, we always observe the expected electrical polarization in insulators. We also note that in \NCTO{} we see strong coupling between magnetic and electric order parameters evidenced by magnetic-field- and temperature-dependent dielectric constants and peaks in the dielectric constant at some field-induced phase transitions (without concomitant peaks in the magnetostriction). Thus, the triple-Q ordering is not supported by our data. Therefore, either the zigzag or another spin structure that rules out linear magnetoelectric coupling is likely. At high fields for $\bf{H} \parallel$ $a$ and $a^*$ we repeatedly observed additional phase transitions as a function of temperature in dilatometry and heat capacity. These may indicate subtle structural changes. Finally, when $\bf{H} \parallel$ $c$-axis, five phases are observed in the ${\bf T-H}$ phase diagram below 16 T as shown in Fig. S7 of the S.I. \cite{supp} whereas the magnetization data became noisy at higher field (Fig. S6 in S.I \cite{supp}) and it was hard to identify the critical fields. The temperature-dependent dielectric constant is shown to be independent of magnetic field and three broad humps are observed that do not match with any of the magnetic transitions observed in magnetization measurements. As discussed in a later section, one possibility is that the dielectric humps are due to dynamics of different Na$^+$ configurations.

\section{Experiments}
\subsection{Crystal growth}
The single crystals were grown by the flux method. A polycrystalline sample of \NCTO{} was mixed with a flux of Na$_{2}$O and TeO$_{2}$ in a molar ratio of 1:0.5:2 and gradually heated to 900 $^{\circ}$C at 3 $^{\circ}$C/min in the air after grinding. The sample was kept at 900 $^{\circ}$C for 30 hours and then was cooled to 500 $^{\circ}$C at the rate of 3 $^{\circ}$C/hour. The furnace was then shut down to cool to room temperature. Crystal structure and purity were verified by X-ray diffraction and carefully oriented using Laue X-Ray diffractometer. Consistent magnetic susceptibilities of different single crystalline samples used in this work confirm that all samples maintain the same crystal qualities and retain the same magnetic properties.

\subsection{Magnetization and specific heat}
Vibrating sample magnetometry (VSM) for dc magnetization, ac magnetic susceptibility and specific heat measurements were performed in a 14 T Quantum Design Physical Property Measurement System (PPMS) using the built-in options with the magnetic field aligned along $a$- and $a^*$-axes.

Specific heat was obtained down to 1.9 K using the standard semi-adiabatic heat pulse method in the PPMS. To align the field orientation, one edge of a single crystal sample was carefully adjusted and mounted on the stage of the PPMS vertical puck.

\subsection{Electrical polarization, dielectric constant and electric field-induced magnetization}
Electrical polarization measurements were performed by the standard technique of integrating the current as a function of time between ground and the electrical contacts (two silver epoxy EPO-TEK H20E capacitor plates deposited on opposite sides of the sample) as the temperature or magnetic field changes. The areas of the electrodes for the {\it a} ({\it a$^{*}$}) direction were 0.28 mm$^{2}$ (1.11 mm$^{2}$) and the distance between the two electrodes was 0.46 mm (0.49 mm). Pt wires were used to electrically connect the electrodes on the samples to adjacent coaxial cables that in turn lead to the room temperature electronics. These measurements were performed on both single- and poly-crystals in millisecond 65 T pulsed fields using a Stanford Research 570 current-to-voltage converter \cite{Zapf2010, Zapf2021}, and in a PPMS using a custom coaxial cable probe and a Keithley 6517A electrometer. The data shown in the main text were taken after electric poling, \textit{i.e.}, applying electric fields as described while cooling from high temperature through $T_{\text{N}}$ to form an electric monodomain. We then measured the electric polarization both with and without applied electric and/or magnetic fields while sweeping the magnetic field as described in the S.I. \cite{supp}.

On the same samples we also measured the electric capacitance as a function of magnetic field using an Andeen-Hagerling AH2700A capacitance bridge at 12 kHz and 15 V excitation with a custom-built co-axial cable probe in the PPMS. Temperature-dependent capacitance was measured at different frequencies using the same probe and samples but with an LCR Meter (Keysight E4980A).

Finally we measured electric field-induced magnetization on the same polycrystals using the VSM in the PPMS, with a custom rod to apply electric field to the capacitor plates of the sample during measurement. The sample was poled by applying electric and magnetic fields of 2 kV/cm and 4 T, respectively, from 150 K. Electric field was then swept from -2 kV/cm to 2 kV/cm while measuring the magnetization.

\subsection{Magnetocaloric effect}
The magnetocaloric effect (sample temperature change vs. magnetic field) measurement was performed in pulsed magnetic fields. In this measurement, a nearly adiabatic condition was realized due to the ultra-fast field sweeping rate of $\sim$ 10,000 T/s. To obtain a strong thermal link between the sample and the thermometer on millisecond timescales in pulsed fields, a semiconducting 10 nm-thick AuGe thin film was directly deposited on the surface of the sample as a thermometer. The film was deposited by RF magnetron sputtering at 40 mTorr pressure of ultra-high purity Ar gas for 60 minutes with 100 W power. Au contact pads were then deposited on top of the AuGe film with a shadow mask, leaving a stripe of AuGe uncovered. A custom digital lock-in method with 100 kHz source current was used to measure the thermometer resistance in pulsed fields, with four point contacts, as is usually employed at the NHMFL-PFF. A detailed cartoon picture of the set-up can be found in the S.I.~\cite{supp}. The thermometer was calibrated in thermalized conditions with exchange gas to obtain resistance vs. temperature and an identical reference thermometer was used to obtain the magnetoresistance calibration.

\subsection{Thermal expansion/Magnetostriction}
Length changes of the sample were measured as a function of temperature (thermal expansion) and magnetic field (magnetostriction). The Fiber Bragg Grating (FBG) dilatometry measurement was adopted in the PPMS using a custom-built probe and optical fibers with 2 mm Bragg gratings \cite{jaime2017fiber}. A straight edge of the as-grown plate-like single crystal of \NCTO{} was carefully attached to the optical fiber using Henkel Ultra-gel superglue. A Pt wire connected the sample to the Cernox temperature sensor, providing a proper thermal link between the two and to the bath. The FBG spectra were recorded using an optical sensing interrogator (Micron Optics, si155). The $a$- and $a^*$-axis data were recorded {\it in situ} with a third empty Bragg Grating as a reference to be subtracted from the sample signals \cite{jaime2017fiber}. The obtained thermal expansion as a function of magnetic field and temperature are normalized respectively following $\Delta L (H, T_0) = \Delta L (H, T_0)/\Delta L(0, T_0)$, and $\Delta L (H_0,T) = \Delta L (H_0, T)/\Delta L(H_0, 3.3 \text{K})$. An illustration of the configuration of the sample attachment can be found in S.I. \cite{supp}. To ensure reproducibility, two different pieces of samples were measured for each crystallographic orientation. To investigate possible effects of the superglue, one sample was measured twice for each orientation with re-gluing in between and found to be consistent.

\section{Symmetry analysis to distinguish zigzag and triple-Q}

\begin{figure}
\includegraphics[width=\linewidth]{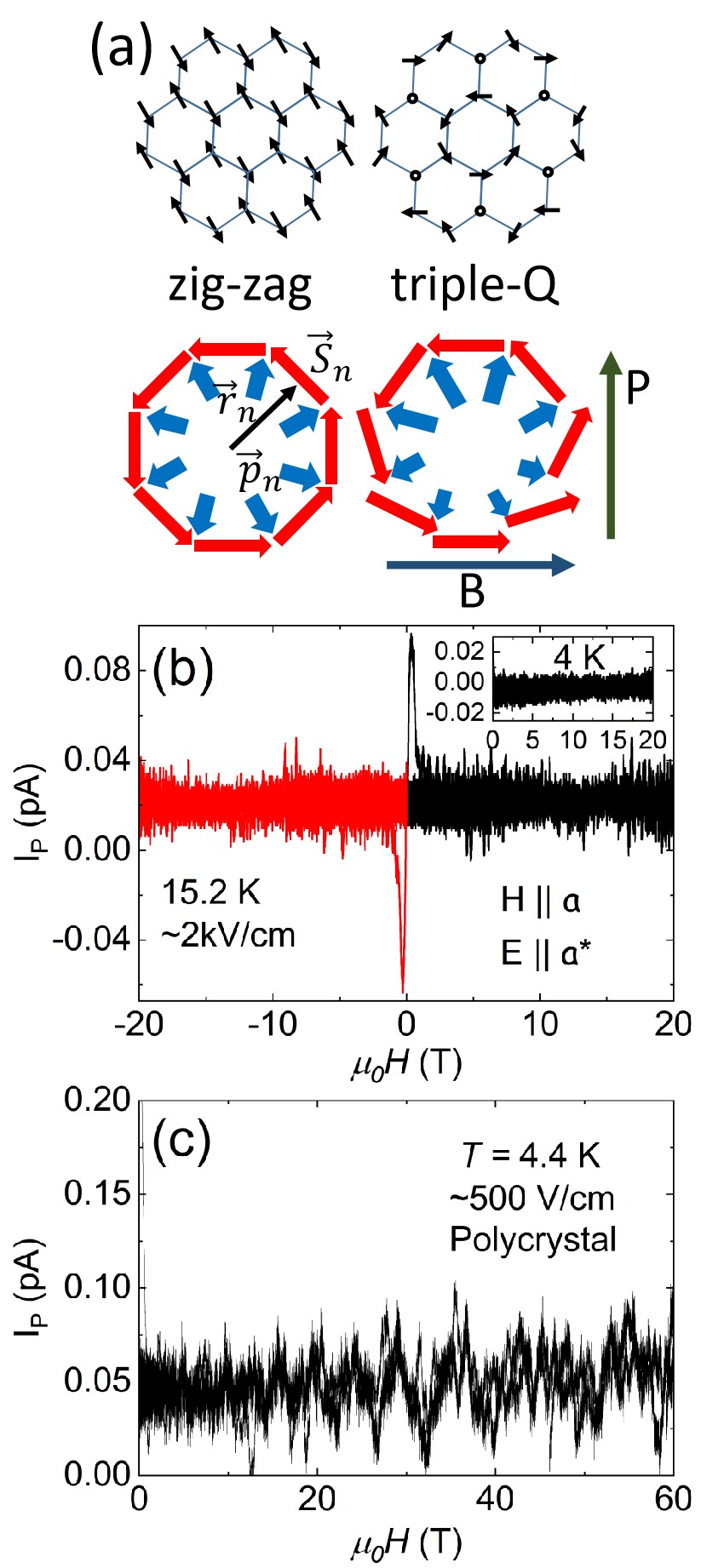}%
\caption {(a) Top two panels illustrate the zigzag and triple-Q spin structure for phase I. Bottom two panels demonstrate the toroidal moment in triple-Q structure (left) and how magnetic field induces a net electrical polarization (right). $\vec{S}_{n}$ and $\vec{r}_{n}$ are $n$th spin and the vector from the center of a toroidal moment to the $n$th spin, respectively. $\vec{p}_n$ is the electron dipoles. (b) Magnetoelectric current along $a^*$-axis measured at 15.2 K with ${\bf H} \parallel a$ for both positive and negative field sweeps. The polling voltage is about 2 kV/cm. Inset depicts the positive sweep of the same measurement configuration at 4 K. (b) Magnetoelectric current of a polycrystal measured at 4.4 K with ${\bf H} \perp {\bf E}$. The poling voltage is about 500V/cm. \label{Polarization}}
\end{figure}

One of the major debates for \NCTO{} is whether the ground state magnetic ordering at zero magnetic field forms a zigzag \cite{bera2017zigzag} or a triple-Q \cite{chen2021spin} spin structure, as illustrated in  Fig.~\ref{Polarization} (a). Here we show these two magnetic orderings can be distinguished using electric polarization measurements. Our results are consistent not with the triple-Q scenario but with the zigzag ordering, or another magnetic ordering with similar symmetry properties.

Electric polarization is a symmetry-sensitive measurement that detects the presence of a unique polar axis in the structure of the material. Linear magnetoelectric coupling is another symmetry-sensitive property, which detects the ability of a magnetic field to induce electric polarization, or an electric field to induce magnetization with an odd coupling between them \cite{Fiebig2005, agyei1990linear, rivera1994linear}. Linear magnetoelectric coupling is allowed in magnetic point groups that break time reversal and spatial inversion symmetry simultaneously at zero magnetic field. The field of multiferroics and magnetoelectrics has established over the past century that magnetic ordering influences the lattice and the orbital configurations and so the magnetic symmetry can imprint itself on the lattice and create magnetoelectric coupling \cite{Spaldin19,Fiebig2005}. This magnetoelectric coupling occurs because every term in a magnetic Hamiltonian depends in some way on the underlying lattice symmetry. Thus there is a back-coupling whereby the lattice deforms slightly to change the magnetic terms in the Hamiltonian and thereby lower the magnetic energy at the expense of the lattice deformation energy.  Magnetostriction (with or without electric polarization) usually creates lattice constant changes on the order of 1 part in 10$^3$ to 10$^5$ in inorganic crystals \cite{jaime2017fiber}. Electric polarization can also result from re-arrangement of electronic orbitals relative to their positively charged ions.

Firstly we note that \NCTO{} is electrically insulating below 150 K, with a measured loss of 0.01 (0.03) nS at 3.3 (80) K, and thus no conduction electrons can screen an electric polarization. We show our measured magnetoelectric current ($I_{p}(H)$) as a function of pulsed magnetic field  along the $a^*$-axis with magnetic field along $a$-axis at 15.2 K and 4 K in Fig.~\ref{Polarization} (b) and its inset, as well as on a polycrystal in Fig.~\ref{Polarization} for $\bf{H} \perp \bf{E}$ (c). These data were taken after cooling the sample in an electric field of 2 kV/cm (single crystal) and 500 V/cm (polycrystal) to align any polar domains. The magnetoelectric current is the derivative of the electric polarization with respect to time, and this current flows from ground onto and off the capacitor plates (not through the sample) in order to compensate changes in the electric field within the sample. These data show no resolvable change in the electric polarization up to 60 T, or at any of the field-induced phase transitions related to electric polarization in this compound. We note that due to fast magnetic field sweep rates up to $\sim$ 10 kT/s, this measurement in pulsed fields is particularly sensitive - the signal to noise scales as the square root of sweep rate. The observed peak at 0.3 T corresponds to a characteristic background noise at the beginning of the pulse and not to any phase transition observed by any other measurement in \NCTO{}. In the S.I. we also show the same measurement as a function of temperature instead of magnetic field (pyroelectric current) after poling in an electric field. The data in Fig. S3 shows only drift and no electric polarization below 150 K \cite{supp}. Above 150 K the data is affected by the onset of conductivity in the sample. These data were taken during the warming process after cooling the sample in an electric field of 2 kV/cm from 200K, 150K, 120K, and 70K as labeled in the figure.

We note that there is a report in the literature of ferroelectricity in \NCTO{} below 60 K by Mukherjee {\it et al.,} \cite{mukherjee2022ferroelectric}. We do not find any ferroelectricity in our sample at this temperature. Also, the observation from Mukherjee {\it et al.,} is inconsistent with the space group \NCTO{}, which has been probed at low and high temperatures by various groups \cite{bera2017zigzag, viciu2007structure,xiao2019crystal}.


Now we discuss the expected electric polarization in the zigzag versus triple-Q spin structures for the low field phase denoted as phase I in Figure.~\ref{NCTO-PD}. Combined with the crystal symmetry, the zigzag spin structure has a \textit{magnetic point group} of 2221$^{\prime}$(No. 6.2.18) with two-fold rotational symmetry along all three directions, regardless of whether the ground state is purely antiferromagnetic or ferrimagnetic as reported in Ref.~\cite{yao2020ferrimagnetism}. It forbids the spontaneous electric polarization. This point group has the magnetoelectric tensor $\alpha_{ij} = 0$, \textit{i.e.}, it also does not allow magnetic field-induced electric polarization. On the other hand, the triple-Q spin structure delineated in Ref. \cite{chen2021spin} breaks both inversion and time reversal symmetry. In particular, the triple-Q spin structure has non-zero off-diagonal components in \textbf{$\alpha$} within the plane ($\alpha_{12}=-\alpha_{21} \neq 0$) \cite{Spaldin2008toroidal}.
Hence, a linear magnetoelectric coupling is expected. That is, the electric polarization along $a^{*}$ should emerge for magnetic fields along $a$ and flip sign as the magnetic field sign is flipped. Below we demonstrate a detailed symmetry analysis. It is difficult to pinpoint the point group that the triple-Q spin structure possesses because no interlayer structure has been determined but our argument is valid as far as the net {\it toroidicity} defined below is nonzero.

In the triple-Q scenario \cite{chen2021spin}, there exists a spontaneous toroidal moment as shown in Fig.~\ref{Polarization} (a). The order parameter $\vec{t}$ is defined as
\begin{equation}
    \vec{t} = \sum_{n}\left(\vec{r}_{n}\times\vec{S}_{n}\right),
\end{equation}
where $\vec{S}_{n}$ and $\vec{r}_{n}$ are $n$th spin and the vector from the center of a toroidal moment to the $n$th spin, respectively. $\vec{t}$ is odd both under spatial inversion and time-reversal operations, which allows the following form of the free energy \cite{Spaldin2008toroidal},
\begin{equation}
F({\bf E},{\bf H}) = F_{0} - \frac{\varepsilon_{ij}E_{i}E_{j}}{8\pi}- \frac{\mu_{ij}H_{i}H_{j}}{8\pi} - \alpha_{ij}E_{i}H_{j} + \cdots,
\end{equation}
where $\varepsilon_{ij}, \mu_{ij},$ and $\alpha_{ij}$ are the dielectric permittivity, the magnetic permeability and the magnetoelectric tensor, respectively. Therefore, we expect the linearly increasing electric polarization as a function of the external magnetic field as follows:
\begin{equation}
P_{i} =  \left(\frac{\varepsilon_{ij} - \delta_{ij}}{4\pi}\right)E_{j} + \alpha_{ij}H_{j}.
\end{equation}

In our experiments, regardless of whether or not an electric field ($E_{i}$) is applied while sweeping the field ($H_{j}$) or while cooling from high temperatures, we observed no noticeable feature in the electric polarization vs. magnetic field or temperature under all conditions as shown in Fig.~\ref{Polarization} and Fig. S3 in S.I. \cite{supp}. These measurements were repeated for single and polycrystals.  In addition to that, we have also measured the electric field-induced magnetization, e.g. the converse magnetoelectric effect as shown in Fig. S3 in S.I. \cite{supp} and no such effect is observed either. One possibility is that electric polarization is too small to be measured, or that the lattice is too stiff to deform. We noted previously that both Zapf and Lee have never experienced such a scenario before. This scenario seems unlikely in the particular case of \NCTO{} because as shown in the next sections, we do observe both strong magnetodielectric effects (magnetocapacitance) and magnetostriction effects. Thus we know that \NCTO{}'s lattice deforms in response to magnetic order and does form electric dipoles - just not a net electric polarization. The last possibility is that each plane has the opposite sign of the net toroidal moment canceling each other to make the net toroidicity zero. This possibility could also be excluded with the electric field poling that may align all toroidal moments along the out-of-plane direction if the interlayer coupling is small \cite{chen2021spin}. However, if the poling energy is not large enough to overcome the interlayer coupling strength, i.e. the interlayer coupling is very strong, we then cannot rule out this possibility.

Therefore, our data are not consistent with the magnetic structure of the triple-Q phase. They are consistent with the zigzag spin structure or with another spin structure that does not allow electric polarization under magnetic field. We notice that in a recently uploaded elastic neutron scattering study \cite{yao2022magnetic}, a magnetic Bragg peak only recovered 2/3 of its intensity in the following field sweeps compared to the initial zero-field-cooled field-sweep. This seems like inconsistent with either triple-Q or zigzag structure. The inconsistency between our data and the neutron scattering study maybe due to some differences in sample which seems unlikely because all critical temperatures and magnetic fields are consistent among the literature. Therefore, this may imply another exotic spin structure that could be consistent with both experiments.

\section{The \textbf{T-H} phase diagram}
\subsection{Results}
\label{results}
\begin{figure*}
\includegraphics[width=\textwidth]{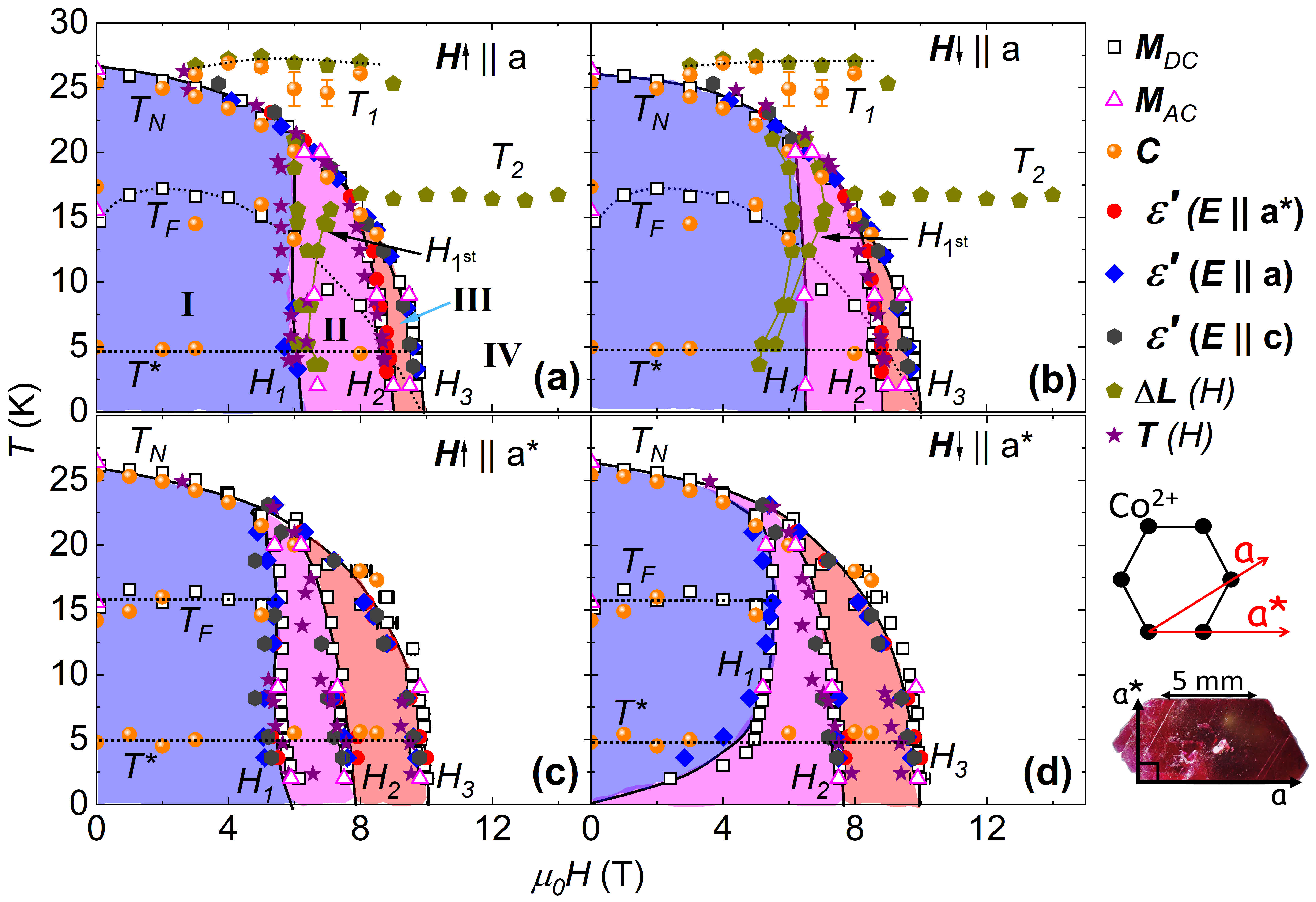}%
\caption{Phase diagrams for ${\bf H} \parallel a$ (panel (a) and (b)) and  ${\bf H} \parallel a^{*}$ (panel (c) and (d)) constructed from magnetization $M$, specific heat $C$, dielectric constant $\varepsilon^{\prime}$ for three directions of the applied electric field ${\bf E}$ as indicated, magnetostriction $\Delta L(H)$, and magnetocaloric effect $T(H)$. ${\bf H}\uparrow$ and ${\bf H}\downarrow$ are the up and down sweeps of the magnetic field, respectively. The solid or dotted lines are guides to the eyes. The hexagon in the legend defines the $a$- and $a^*$-axes with respect to the honeycomb lattice of Co$^{2+}$. At the bottom right is a photo of a single crystal with $a$- and $a^*$-axes indicated. \label{NCTO-PD}}
\end{figure*}

From the aforementioned various measurements, we constructed comprehensive ${\bf T-H}$ phase diagrams of \NCTO{} as illustrated in Fig.~\ref{NCTO-PD} with field along 
$a$- and $a^*$-axis. A phase diagram with ${\bf H} \parallel$ $c$ constructed from magnetization measurements can be found in the S.I.~\cite{supp}. For clarity, we discuss an overview of the phase diagram before describing the details of the individual measurements below. From field-dependent measurements, there are four successive phases (I-IV) including the polarized phase, separated by the critical fields $H_{\text{1}}$, $H_{\text{2}}$, and $H_{\text{3}}$. As a function of temperature, three major phase boundaries are observed as $T_{\text{N}}$, $T_{\text{F}}$, and $T^*$, consistent with the literature \cite{yao2020ferrimagnetism, hong2021strongly, xiao2021magnetic}. We tracked $T_{\text{F}}$ to higher fields than previously reported. It is noteworthy that the $T_{\text{F}}$ boundary is qualitatively different when magnetic field is applied along $a$- or $a^*$-axes. It is field-dependent and persists into phase II when magnetic field is applied along $a$-axis whereas it becomes field independent and stops within phase I when magnetic field is applied along $a^*$-axis. Several additional critical fields/temperatures ($T_1,T_2$ and $H_{1^{st}}$) were also observed in thermal expansion and specific heat measurements that were not previously described.

\begin{figure*}
\includegraphics[width=0.7\linewidth]{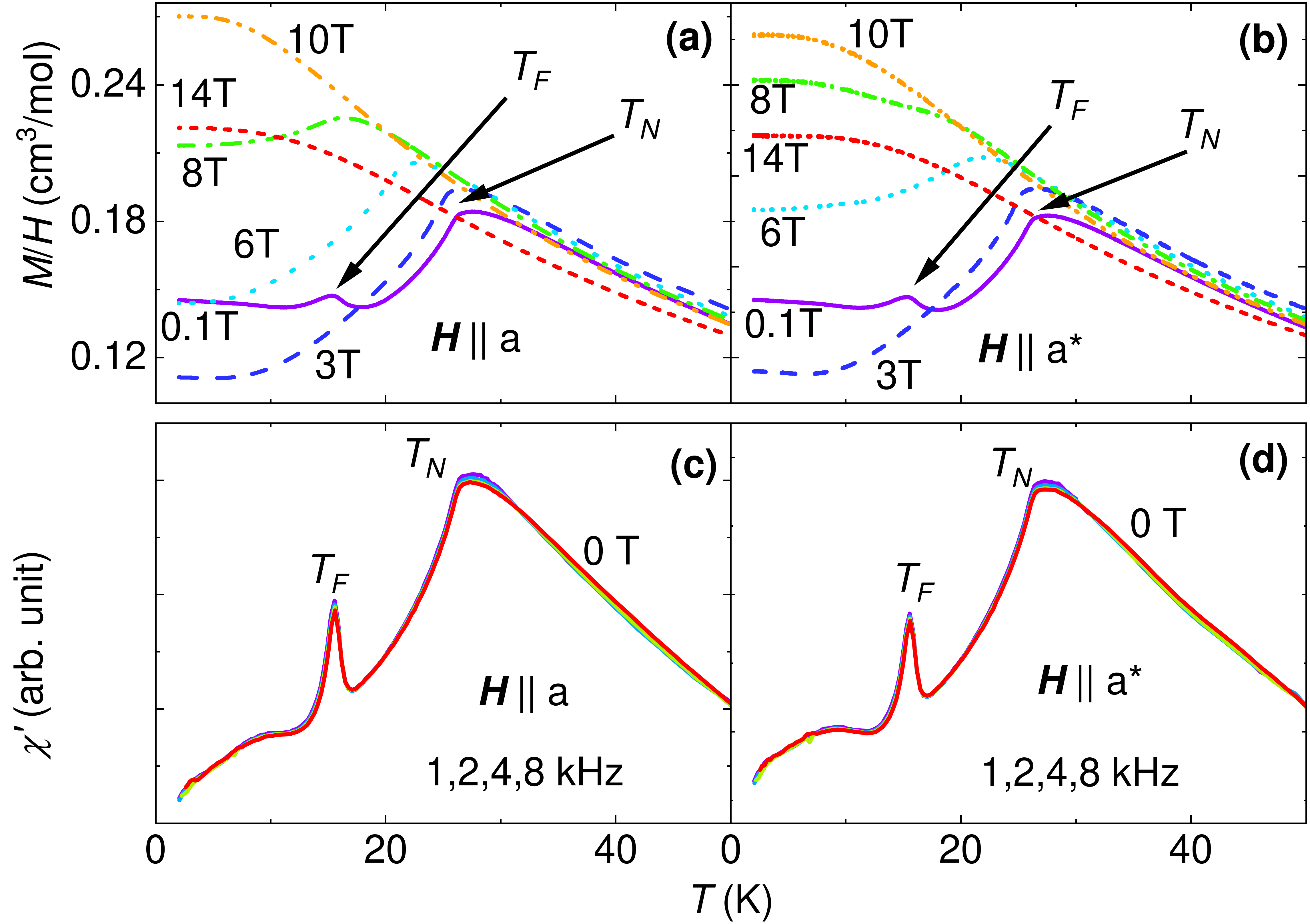}%
\caption{Longitudinal dc magnetic susceptibility $M/H$ vs. $T$ for various magnetic fields between 0.1 and 14 T along (a) $a$ and (b) $a^*$. Ac magnetic susceptibility $\chi^{\prime}$ vs $T$ for fields along the (c) $a$ axis and (d) $a^*$ axis. The ac field with amplitude 10 Oe is applied at 1,2,4,and 8 kHz with a dc field $H = 0$. $T_{\text{F}}$ and $T_{\text{N}}$ are the temperatures recorded in the phase diagram. The complete data sets are available in the S.I.~\cite{supp}. \label{MvsT}}
\end{figure*}

Fig.~\ref{MvsT} (a) and (b) show the dc magnetic susceptibility ($M/H$) as a function of $T$ taken at various magnetic fields $H$ between 0.1 T and 14 T applied along $a$ and $a^*$, respectively. $T_{\text{N}}$ indicates the antiferromagnetic phase transition temperature $\sim$27 K for both directions, consistent with other results \cite{lefranccois2016magnetic, bera2017zigzag, yao2020ferrimagnetism, lee2021multistage, lin2021field, mukherjee2022ferroelectric}. Another peak at around 16 K and at 0.1 T, denoted with $T_{\text{F}}$, is also observed for both directions. This feature has been interpreted as a signature of spin canting \cite{lin2021field} and it is the temperature at which a low-energy broad excitation spectra turn into a clear magnon band \cite{chen2021spin}. The peak at $T_{\text{F}}$ is quickly suppressed as the applied magnetic field increases from 0.1 T to 1 T so that its feature is only clearly visible in the derivatives at higher fields. On the other hand, $T_{\text{N}}$ shifts towards lower temperatures and the feature becomes broadened with increasing field and eventually disappears above 8 T.

{Fig.~\ref{MvsT}} (c) and (d) display the ac magnetic susceptibility ($\chi^{\prime}$) respectively along $a$- and $a^*$-axes as a function of $T$ taken under four different frequencies. No frequency dependence is observed below 8 kHz up to 100 K, in contrast with the frequency-dependent dielectric constant (see Fig. S8 of the S.I. \cite{supp}). We also observed $T_{\text{N}}$ and $T_{\text{F}}$, whose temperatures are consistent with the dc measurement while the feature of $T_{\text{F}}$ is much more pronounced in ac susceptibility.

\begin{figure}
\includegraphics[width=\linewidth]{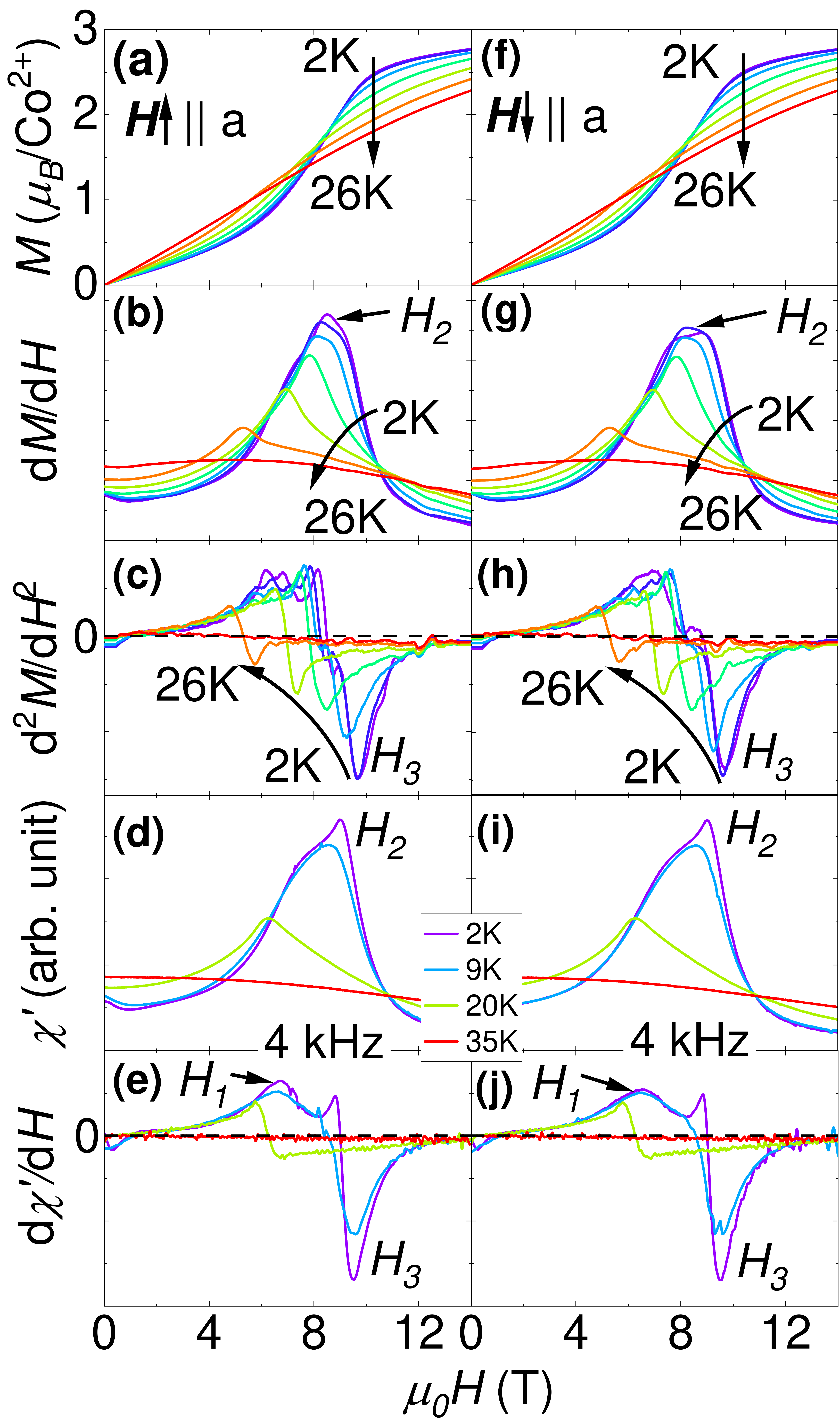}%
\caption{Dc Magnetization $M$ and ac magnetic susceptibility $\chi'$ and their first, and second derivatives with respect to magnetic field ${\bf H}$ as a function of ${\bf H} \parallel a$ taken in superconducting magnets. (a)-(e) display the up field-sweep (${\bf H}\uparrow$) data in which (a)-(c) and (d)-(e) are from dc and ac measurements, respectively. (f)-(j) display the down field-sweep (${\bf H}\downarrow$) data in which (f)-(h) and (i)-(j) are from dc and ac measurements, respectively. All ac magnetic susceptibility data shown here are measured at 4 kHz. $ H_{\text{1,2,3}}$ are the critical fields recorded in the phase diagram. The complete data sets are available in the S.I.~\cite{supp}. \label{MvsH-a}}
\end{figure}

Fig.~\ref{MvsH-a} (a)-(c) and (f)-(h) illustrate the dc magnetization ($M(H)$) at various {\bf T} when {\bf H} is applied along $a$-axis. Magnetization and its first and second derivatives of up (down) field-sweeps are presented in Fig.~\ref{MvsH-a} (a), (b), and (c) ((f), (g), and (h)), respectively. Data curves from up and down field-sweeps overlap with each other showing no hysteresis, consistent with previous reports \cite{yao2020ferrimagnetism, lin2021field, lee2021multistage}. Here we define $H_{\text{2}}$ as the inflection point of the magnetization curves found from the peak in the first derivative magnetization (Fig.~\ref{MvsH-a} (b) and (e)) and $H_{\text{3}}$ is the maximum curvature point defined as the peak in the second derivative in Fig.~\ref{MvsH-a} (c) and (h). While the $H_{\text{3}}$ phase boundary is consistent with those found in previous thermal conductivity and magnetization works {\cite{hong2021strongly, yao2020ferrimagnetism}}, $H_{\text{2}}$ has not been called out in all previous works, despite subtle features consistently observed in previous reports \cite{yao2020ferrimagnetism, hong2021strongly, lin2021field, lin2022evidence}. Above $H_{\text{3}}$, the magnetization increases with a downward curvature consistent with saturation. However a small linear component in $M(H)$ persists up to the highest measured fields of 60 T (Fig. S4), likely due to Van Vleck paramagnetism \cite{Lee2014VV}. 

Ac magnetic susceptibility ($\chi^{\prime}$) and its first derivative along $a$-axis are shown in {Fig.~\ref{MvsH-a}} (d)-(e) and (i)-(j) for the up and down field-sweeps, respectively. All of $H_{\text{1}}$ through $H_{\text{3}}$ are observed to have similar temperature evolution compared to dc measurements. $H_{\text{1}}$ and $H_{\text{3}}$ are defined as peak and dip in the first derivative and $H_{\text{2}}$ is defined as the peak in $\chi^{\prime}$. Note that $H_{\text{1}}$ is only observed in ac measurements, likely because it is too subtle to observe in dc measurements.

\begin{figure}
\includegraphics[width=\linewidth]{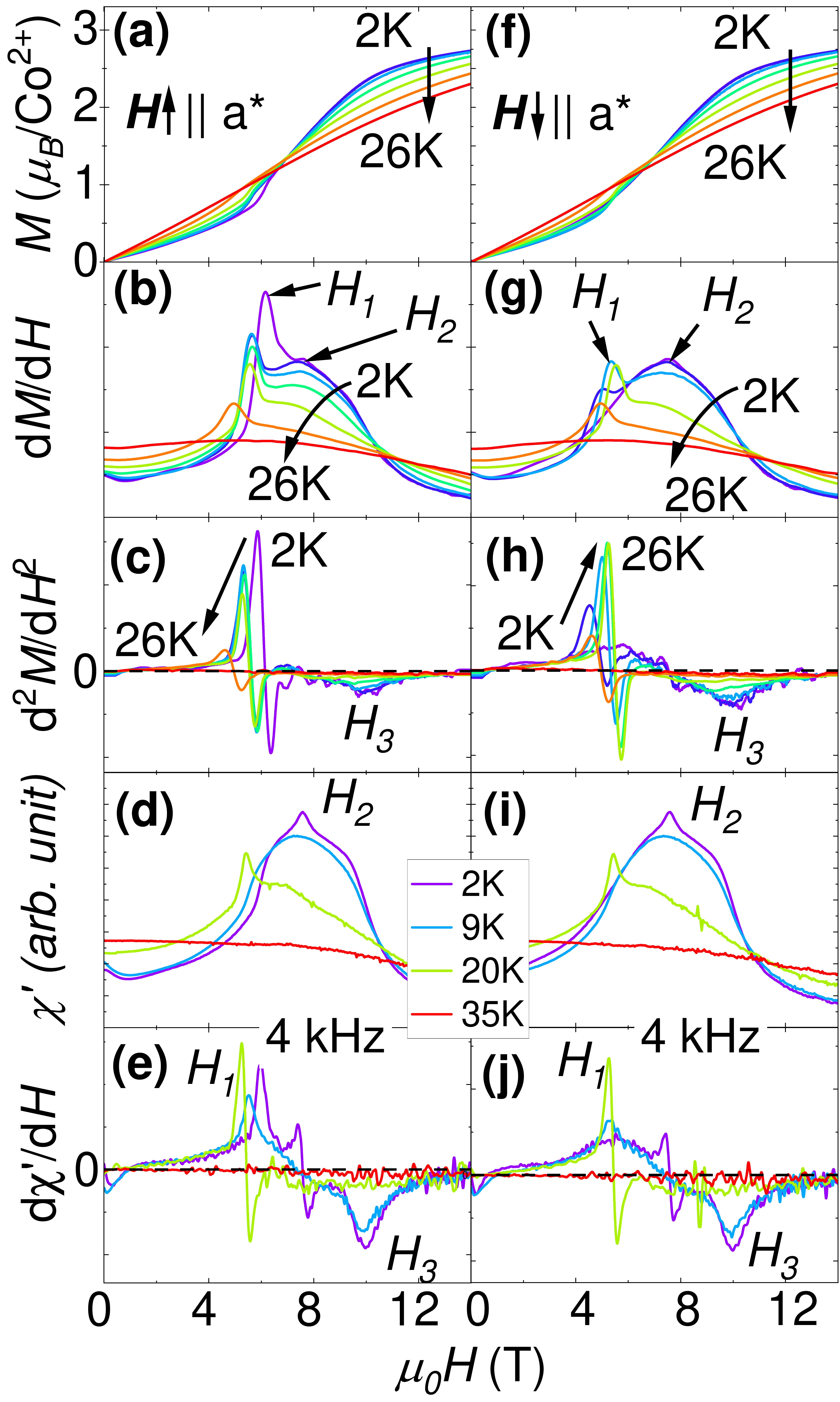}%
\caption{Dc Magnetization $M$ and ac magnetic susceptibility $\chi'$ and their first, and second derivatives with respect to magnetic field ${\bf H}$ as a function of ${\bf H} \parallel a$ taken in superconducting magnets. (a)-(e) display the up field-sweep (${\bf H} \uparrow$) data in which (a)-(c) and (d)-(e) are from dc and ac measurements, respectively. (f)-(j) display the down field-sweep (${\bf H} \downarrow$) data in which (f)-(h) and (i)-(j) are from dc and ac measurements, respectively. All ac magnetic susceptibility data shown here are measured at 4 kHz. $H_{\text{1,2,3}}$ are the critical fields recorded in the phase diagram. The complete data sets are available in the S.I.~\cite{supp}. \label{MvsH-astar}}
\end{figure}

Results of dc magnetization ($M(H)$) and its first and second derivatives at various{\bf T} when {\bf H} is applied along $a^*$-axis are shown in Fig.~\ref{MvsH-astar} (a)-(c) and (f)-(h) for field up- and down-sweeps, respectively. Three phase boundaries, $H_{\text{1}}$ through $H_{\text{3}}$, are observed in this direction. $H_{\text{1}}$ and $H_{\text{2}}$ are defined as the peak positions in the first derivative and $H_{\text{3}}$ is defined as the peak position in the second derivative of magnetization. In contrast to the $a$-axis data, there is noticeable hysteresis beginning at $H_{\text{1}}$ below 26 K ($\approx T_N$), as shown more clearly in Fig. S4 in the S.I. \cite{supp}, consistent with previous reports \cite{yao2020ferrimagnetism, lin2021field, lee2021multistage}. With increasing temperature, all critical fields shift towards lower fields with peak height decreasing for both up and down field-sweeps except for the $H_{\text{1}}$ peak in down-sweeps whose amplitude increases with increasing temperature. Similar to the magnetization with $H\parallel a$, we observed the linearly increasing magnetization above the saturation magnetization because of the Van Vleck paramagnetism.

{Fig.~\ref{MvsH-astar}} (d)-(e) and (i)-(j) illustrates the ac susceptibility ($\chi^{\prime}$) and its first derivative along $a^*$-axis for the up and down field-sweeps, respectively. All of $H_{\text{1}}$ through $H_{\text{3}}$ are observed with similar temperature evolution compared to dc measurements. $H_{\text{1}}$ and $H_{\text{3}}$ are defined as peak and dip in the first derivative and $H_{\text{2}}$ is defined as the peak in $\chi^{\prime}$.

In order to check if there are no additional magnetic phase transitions, we measured the magnetization with ${\bf H} \parallel a^*$ up to 60 T using pulsed-field magnet at 4.3 K as shown Fig.~S4. The low-field section of the data agrees with dc measurements, as illustrated more clearly by $dM/dH$ in the inset. We did not observe any additional magnetic phase transition above $H_{\text{3}}$, confirming that magnetization saturates above $H_{\text{3}}$.

\begin{figure*}
\includegraphics[width=\linewidth]{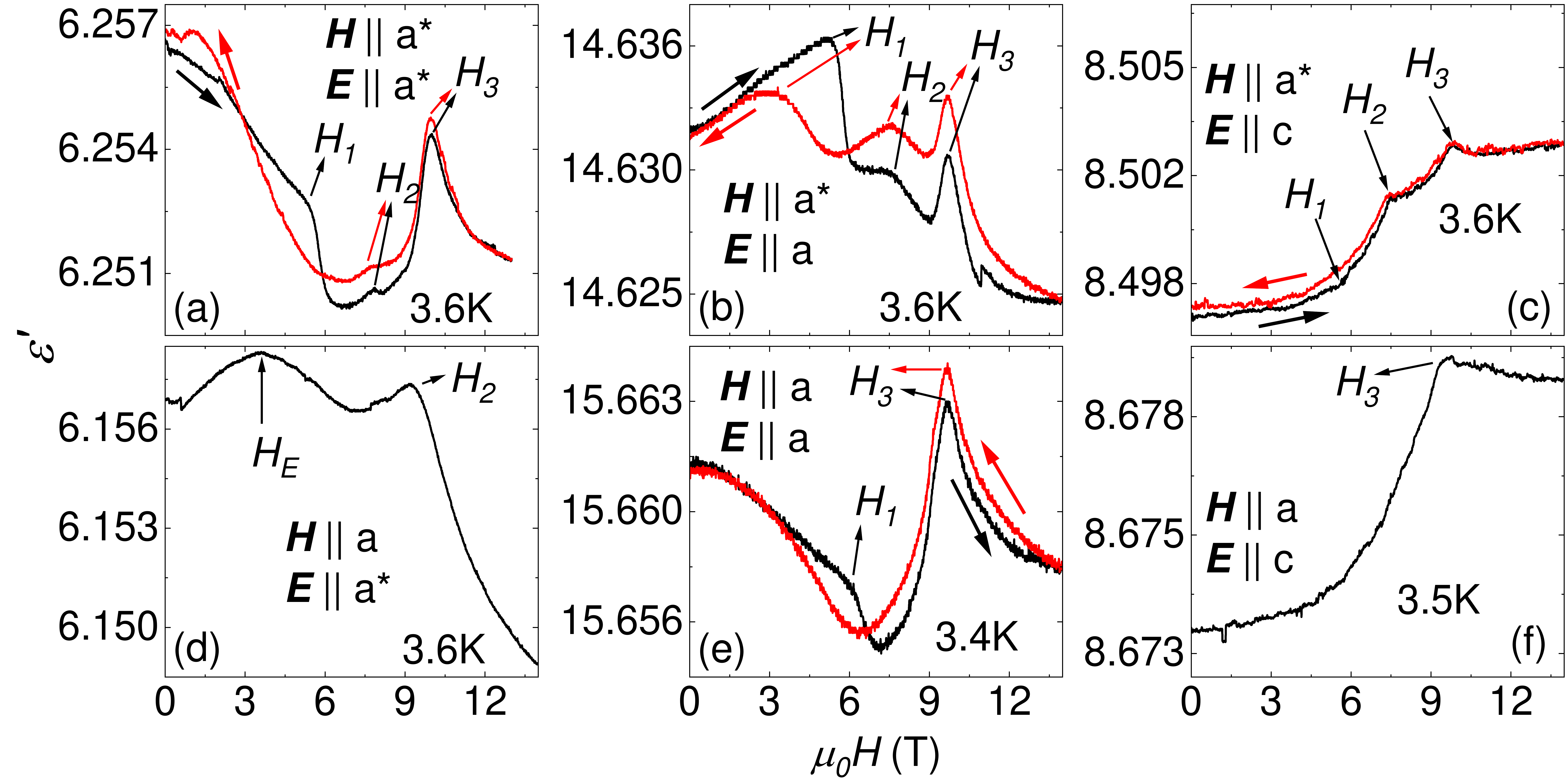}%
\caption{Dielectric constant $\varepsilon^{\prime} = \frac{\varepsilon}{\varepsilon_0}$ along all crystallographic orientations as a function of ${\bf H} \parallel a$ ((a)-(c)) and $a^*$ ((d)-(f)), taken in a superconducting magnet. Here $\varepsilon$ is the dielectric constant of the sample and $\varepsilon_0$ the vacuum. ${\bf H} \uparrow$ and ${\bf H}\downarrow$ are the up and down field-sweep, respectively. $H_{\text{1,2,3}}$ are the fields recorded in the phase diagram in Fig. \ref{NCTO-PD}. $H_{\text{1}}$ in panel (c) is the kink at which the slope of the curve suddenly increases. The complete data sets are available in the S.I.~\cite{supp}. \label{Capacitance}}
\end{figure*}

Next, we investigate the electrical properties of \NCTO{} by measuring the dielectric constant as a function of magnetic field ($\varepsilon^{\prime}(H)$) for various electric and magnetic field directions as shown in Fig.~\ref{Capacitance}. We note that the dielectric constant measurement has also been used to determine the phase boundaries of \RuCl{} that match well with phase boundaries obtained from other techniques \cite{mi2021stacking,zheng2018dielectric}. When the magnetic field is applied along $a^*$-axis with electric field applied along $a^*$, $a$ as shown in Fig.~\ref{Capacitance} (a) and (b) respectively, a similar hysteresis behavior as in the magnetization measurement shown in Fig.~\ref{MvsH-astar} is observed. Three phase boundaries $H_{\text{1}}, H_{\text{2}}$ and $H_{\text{3}}$ are clearly visible as peaks in the dielectric constant, and their positions match well with those found in the magnetization. When the electric field is applied along $c$-axis with ${\bf H}\parallel a^*$ (Fig.~\ref{Capacitance} (c)), the peaks corresponding to $H_{\text{1}}, H_{\text{2}}$ and $H_{\text{3}}$ are smaller, though still sharp.

Fig.~\ref{Capacitance} (d)-(f) shows $\varepsilon^{\prime}(H)$ for ${\bf H} \parallel a$. An additional hump ($H_{\text{E}}$) is observed when electric field $E$ is applied along $a^*$-axis. The origin of this feature needs further investigation as it does not overlap with any observations in other measurements. In contrast to the magnetization measurement, hysteretic behavior was observed for $\varepsilon^{\prime}(H)$ for ${\bf H} \parallel {\bf E} \parallel a$ as shown in panel (e). Since the dielectric constant strongly depends on the magnetic field and reflects the magnetic phase transition, \NCTO{} possesses a fairly strong magnetoelectric coupling. We later discuss the underlying mechanism based on the possible spin structures. For magnetic fields applied along $c$-axis we observed negligible field dependence in the dielectric constant as shown in the S.I.~\cite{supp}.

The temperature-dependent dielectric constant and dissipation measured at various frequencies are also collected as shown in Fig. S8 of the S.I. \cite{supp}. The three pronounced peaks are observed in the dissipative part of the dielectric constant, whose temperatures are dependent on frequency. These do not match any feature seen in the magnetization. Considering that the dielectric constant (electric capacitance) measurement reveals the dynamics of electric dipoles, a peak/hump is expected where they undergo strong fluctuation. Therefore, a speculation is that these humps could indicate the freezing of Na$^+$ positions as temperature decreases. Further studies are necessary to clarify these features.

\begin{figure}
\includegraphics[width=\linewidth]{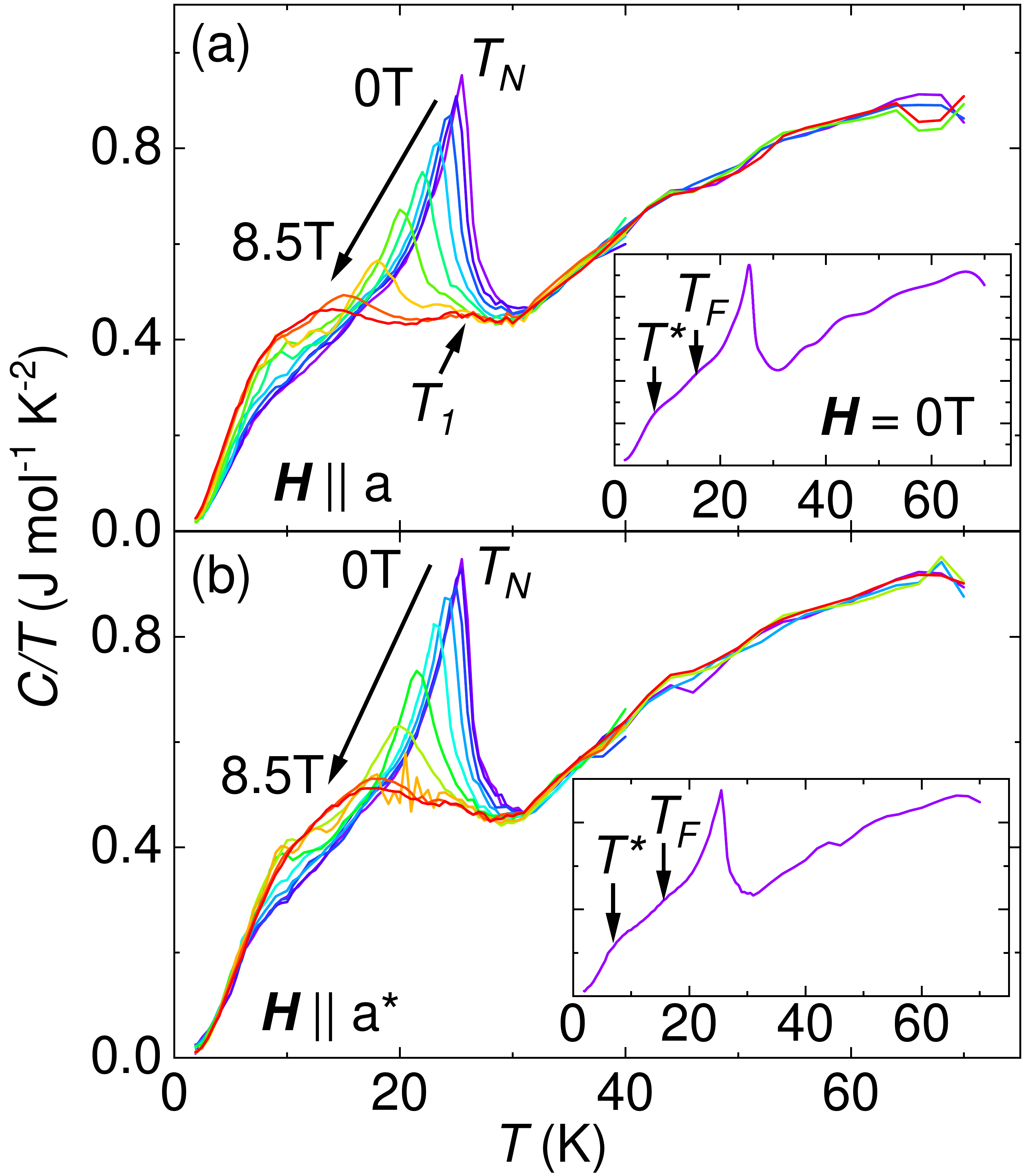}%
\caption{Specific heat divided by temperature ($C/T$) as a function of $T$ at various $H$ up to 8.5 T for (a) ${\bf H} \parallel a$ and (b) $\bf{H} \parallel$ $a^*$. The phase transition temperatures are labeled and corresponding features are indicated by the black arrows. The inset depicts the specific heat data at 0 T for a clearer view of $T_{\text{F}}$ and $T^*$. \label{SH}}
\end{figure}

The thermodynamic properties of \NCTO{} were also investigated and the specific heat divided by temperature ($C/T$) data are shown in Fig.~\ref{SH} at various $H$ up to 8.5 T. There is no significant difference between $\bf{H} \parallel$ $a$- and $a^*$-axes. For both directions, three phase transitions are observed, consistent with our magnetization measurement and previous reports \cite{lin2021field,yao2020ferrimagnetism}. The $T_{\text{N}}$ peak and $T^*$ hump are observed up to 8.5 T whereas the $T_{\text{F}}$ hump is difficult to extract above 6 T. With increasing magnetic field, the peak at $T_{\text{N}}$ gets weakened and suppressed to lower temperatures whereas the $T^*$ feature is robust against magnetic field. The $T_{\text{F}}$ feature when $\bf{H} \parallel$ $a$ is hard to identify at several magnetic field strengths but the rest of them show a slight decreasing trend of $T_{\text{F}}$ with increasing magnetic field. When $\bf{H} \parallel$ $a^*$, $T_{\text{F}}$ is largely independent of applied magnetic field.

\begin{figure*}
\includegraphics[width=\textwidth]{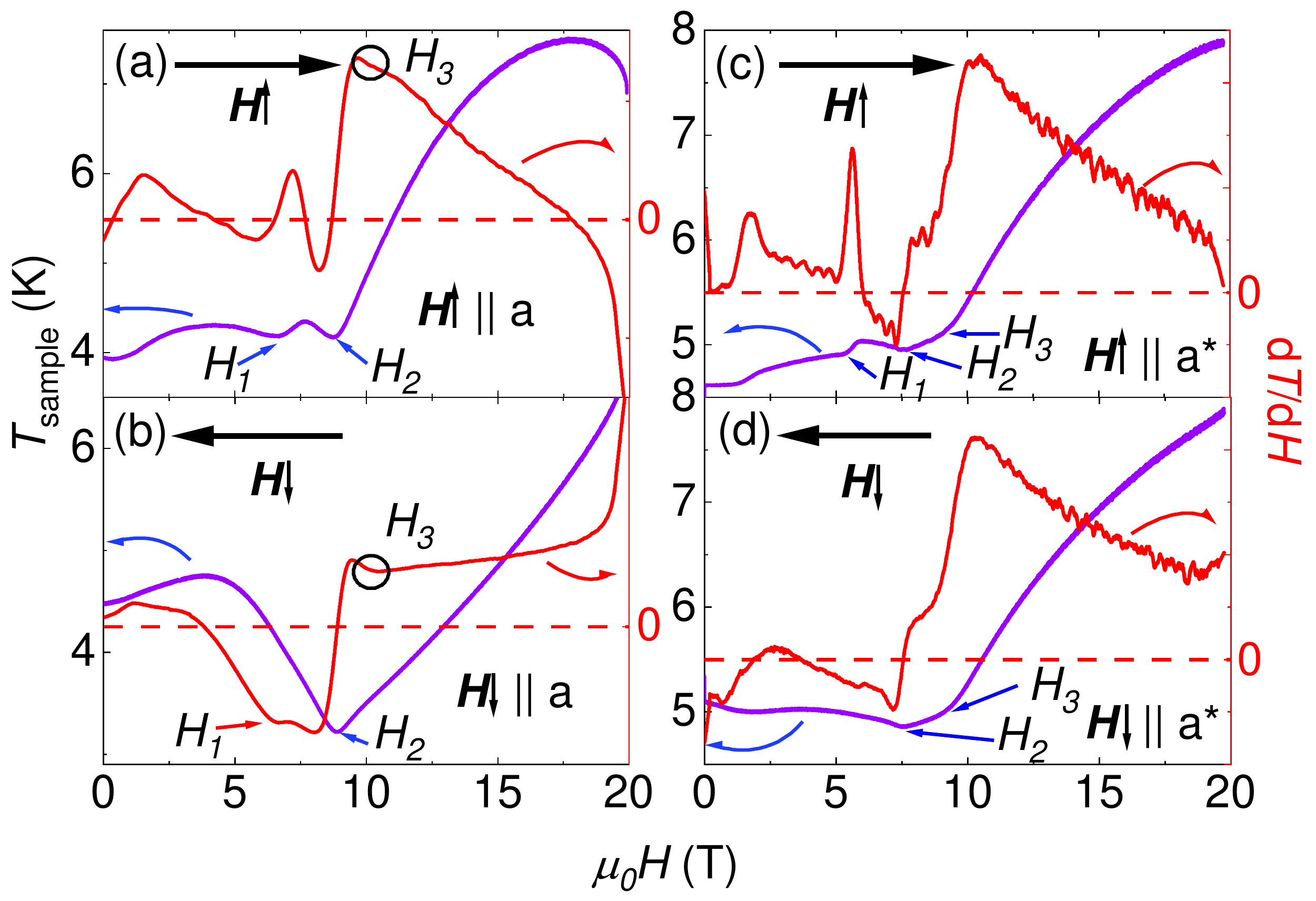}%
\caption{Magnetocaloric effect measurements data in millisecond pulsed magnetic fields. Sample temperature $T_{\rm sample}$ is plotted as a function of ${\bf H}$ applied along both $a$-axis ((a),(b)) and $a^*$-axis ((c),(d)). The purple lines are the sample temperatures and the red lines are the derivatives ($\frac{dT}{dH}$). Phase boundaries revealed in Fig.~\ref{NCTO-PD} are labeled and indicated by arrows and circles. The dashed red lines are indications of $\frac{dT}{dH} = 0$. The field sweep directions are further illustrated at the top of each plot. \label{MC}}
\end{figure*}

Fig.~\ref{MC} shows the magnetocaloric effect in quasi-adiabatic conditions in a 65 T pulsed magnet, pulsed to a maximum field of 20 T.  The sample temperature as a function of magnetic field is shown on the left axis and its derivative (d{\it T}/d{\it H}) on the right axis. The magnetic field vs. time profile is shown in the inset of Fig. S14 \cite{supp}. The 20 T peak field is chosen so the fast part of the pulse occurs in the region of interest up to 12 T, and the less adiabatic behavior that emerges as the sweep rate slows down and the field turns around where the sweep rate becomes zero occurs at fields above the region of interest.

We observe hysteresis in $T$($H$) originating from the 1$^{st}$ order phase transition at $H_{\text{1}}$ that was also seen in the other properties, as well as some thermal relaxation occurring at the highest fields where the field sweep rate slows down and passes through zero, causing the adiabatic behavior to become quasi-adiabatic. We note that with increasing field sweep speed, the hysteresis of 1$^{st}$ order phase transitions generally broadens due to the finite time needed to nucleate and grow the new phase. Thus the hysteresis in $H_{\text{1}}$ can be expected to open up significantly in these pulsed measurements. In some cases, 1$^{st}$ order phase transitions can be avoided altogether at fast sweep rates due to lack of time for the new phase to nucleate and grow (``supercooling"/``superfielding"). On the other hand, the non-hysteretic 2$^{nd}$ order-like phase transitions at $H_{\text{2}}$ and $H_{\text{3}}$ can be observed at similar fields as in dc magnetization measurements.

\begin{figure}
\includegraphics[width=\linewidth]{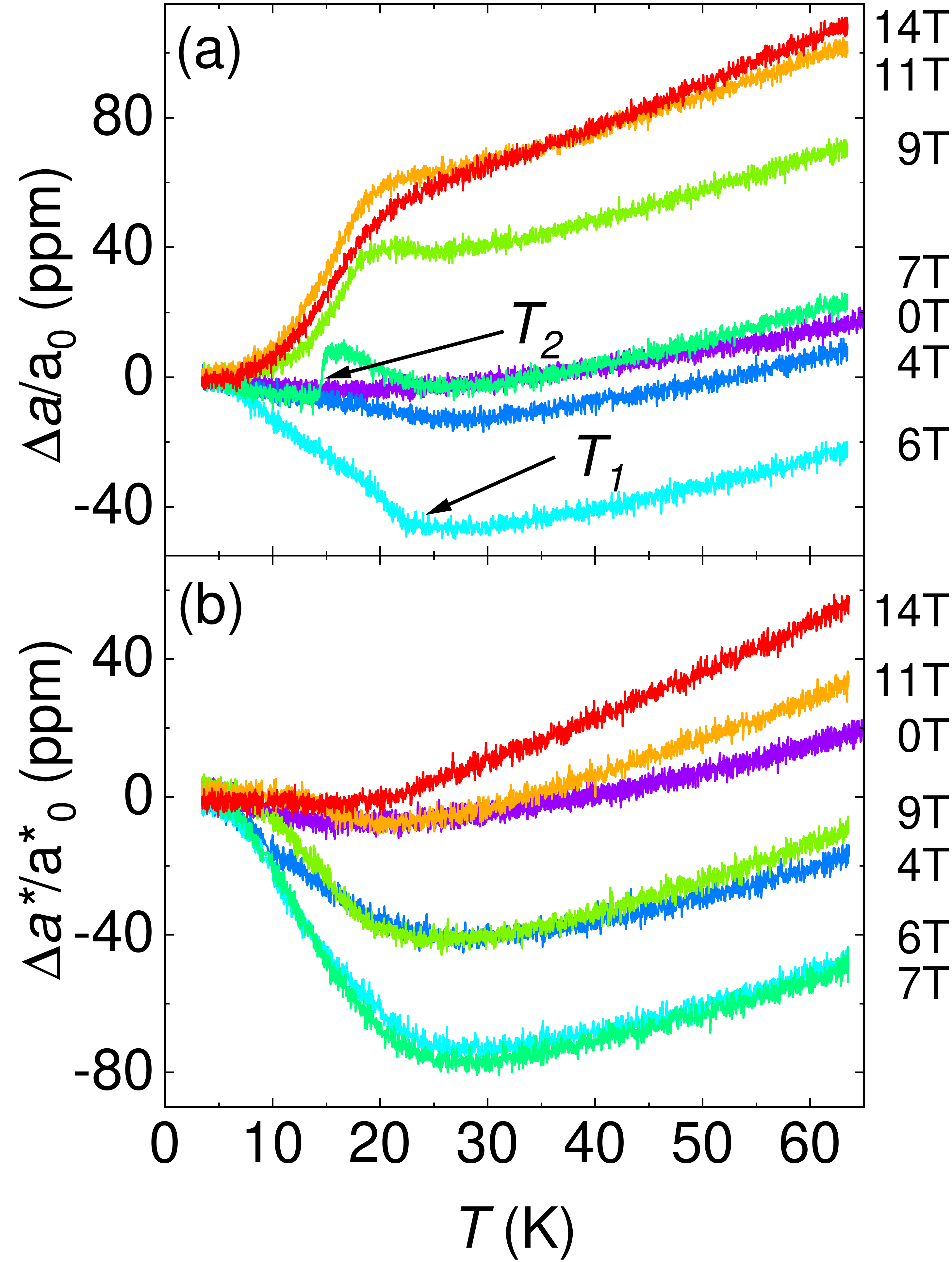}%
\caption{Thermal expansion as a function of temperature $T$ in (a) $a$-axis ($\Delta a/a_0$) and (b) $a^*$-axis ($\Delta a^*/a^*_0$) with various magnetic fields ${\bf H}$ applied along the same direction as the length change. 
$T_{\text{1,2}}$ represent the phase boundaries/crossovers recorded in the phase diagram. All data curves are normalized to the corresponding lowest-temperature thermal expansion values. The complete data sets are available in the S.I.~\cite{supp}. \label{MSvsT}}
\end{figure}

In the $T(H)$ data in Fig.~\ref{MC} we indicate the phase transitions $H_{\text{1}}$, $H_{\text{2}}$ and $H_{\text{3}}$ with arrows. These phase transitions appear as minima. This is consistent with increased spin disorder when approaching a phase transition, which forces the thermal entropy to drop to compensate. When \textbf{H} $\parallel a$, $H_{\text{3}}$ is difficult to observe due to a large background increase in temperature approaching saturation, but can be resolved as a wiggle in $dT/dH$. These observations are similar to those made for magnetocaloric effect data in $\alpha$-RuCl$_3$ \cite{bachus2020thermodynamic, schonemann2020thermal}. Above $H_{\text{3}}$, the temperature increases rapidly. This temperature increase reflects the spin gap that opens above magnetic saturation \cite{hong2021strongly}. The increase in thermal entropy compensates for the drop in spin entropy as the magnetization saturates and a spin gap opens and increases with increasing magnetic field.

When the magnetic field is parallel with $a^*$, all three phase boundaries are clearly resolved in $T(H)$ as well as $dT/dH$ on the up-sweep. We miss seeing $H_{\text{1}}$ in the down field-sweep as described above. All of $H_{\text{1}}$, $H_{\text{2}}$, and $H_{\text{3}}$ are observed as dips and kinks in the $T$($H$) curve or its first derivative.

We now move to thermal expansion and magnetostriction, i.e., length changes of the sample with temperature and field. Shown in Fig.~\ref{MSvsT} are the thermal expansion data of \NCTO{} as a function of $T$ with ${\bf H} \parallel a$ and $a^*$  up to 14 T. Unlike $\alpha$-RuCl$_3$ \cite{schonemann2020thermal}, along both $a$- and $a^*$-axes, the thermal expansion ($\Delta a(a^*)/a_0(a^*_0)$) shows very little temperature dependence at zero magnetic field, consistent with previous studies showing no structural transition \cite{lefranccois2016magnetic, bera2017zigzag, xiao2019crystal}. However, with increasing field, along the $a$-axis, a kink at $T_{\text{1}}$ develops, indicating an onset of slope change. This becomes more and more pronounced with increasing field until 6 T, above which the shape of the thermal expansion \textit{abruptly} changes and a sharp drop appears at $T_{\text{2}}$. At even higher fields, the $T_{\text{2}}$ feature broadens and eventually becomes a gradual decrease. Both features are observed almost always outside the antiferromagnetic phase of \NCTO{} and are independent of magnetic field strength, as illustrated in Fig.~\ref{NCTO-PD}. The origin of $T_{\text{1}}$ and $T_{\text{2}}$ are not yet determined and need further experimental input from structural-sensitive measurements such as X-ray or neutron diffraction. On the other hand, along $a^*$-axis, all features are broad and we do not identify any phase transitions.

\begin{figure*}
\includegraphics[width=\textwidth]{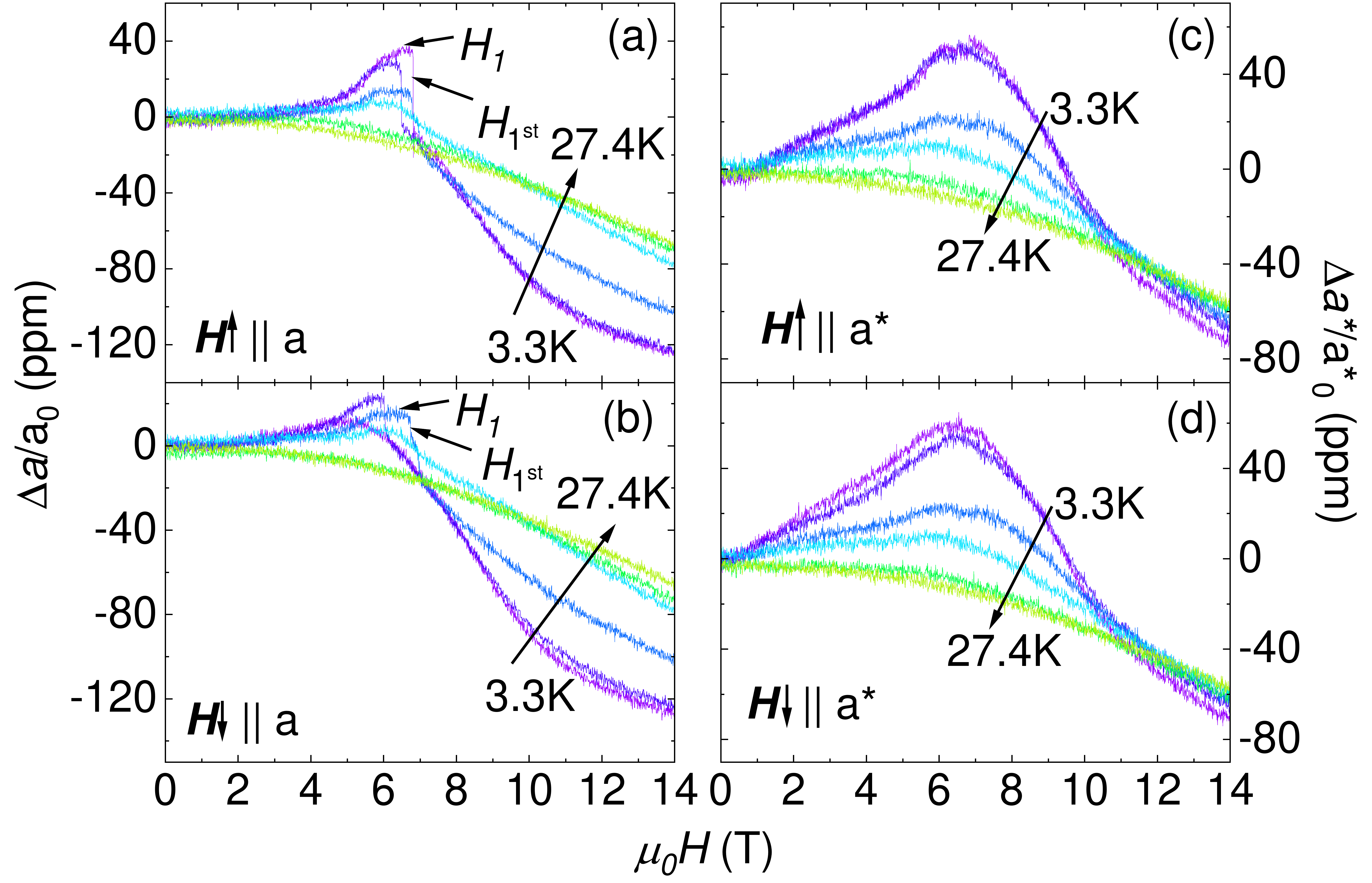}%
\caption{Magnetostriction as a function of applied magnetic field ${\bf H}$ in a superconducting magnet taken at various temperatures as indicated for both $a$-axis ((a), (b)) and $a^*$-axis ((c), (d)). ${\bf H}\uparrow$ and ${\bf H}\downarrow$ are the up and down field sweeps, respectively. $H_{\text{0}}$ and $H_{1^{\text{st}}}$ are the phase boundaries/crossovers recorded in the phase diagram. All data curves are normalized to the corresponding zero field thermal expansion values. The complete data sets are available in the S.I. \cite{supp}. \label{MSvsH}}
\end{figure*}

The magnetostriction data are shown in {Fig.~\ref{MSvsH}}. When \textbf{H} $\parallel a$, a peak is observed in the magnetostriction at 6 - 7 T, corresponding to $H_{\text{1}}$, followed by a discontinuous and hysteretic jump at $H_{1^{st}}$. In the up-sweeps, the amplitude of this jump decreases with increasing temperature until $T_{\text{F}}$, above which it disappears. But in the down-sweeps, the amplitude of this drop feature increases with increasing $T$ until $T_{\text{F}}$ at which the up- and down-sweep curves overlap with each other. It then suddenly becomes much weaker at higher temperatures and eventually becomes invisible above 22 K. On the other hand, along $a^*$-axis, only a broad maximum is observed which do not correspond to any phase transitions.

\subsection{Discussion}
Magnetic field versus temperature phase diagrams were built by combining the data from the Results subsection. In Fig.~\ref{NCTO-PD} we show the phase diagrams for the magnetic field in the plane. The four phase diagram correspond to up and down field sweeps for ${\bf H} \parallel a$ and $a^*$. (In the S.I. \cite{supp} we also show the phase diagram for ${\bf H} \parallel c$.)

The phase diagrams in Fig.~\ref{NCTO-PD} show four phases as a function of magnetic field. These phases are denoted as I, II, III and IV,
in addition to the high temperature paramagnetic phase. These four phases contrast with some of the previous studies where only three phases were observed \cite{lin2021field, xiao2021magnetic}, probably due to the limited number of measured quantities or field range in those studies. Recently a study of torque magnetometry and inelastic neutron scattering also showed some evidence of all four phases \cite{lin2022evidence}.

For ${\bf H} \parallel a$, we also observe apparent phase transitions $T_{\text{1}}$ and $T_{\text{2}}$ as a function of temperature in the thermal expansion and specific heat at high fields that were not previous reported.

Our comprehensive phase diagrams reveal a couple of interesting features of those phases. First, the phase boundaries $H_{\text{1}}$ and $H_{\text{2}}$ are nearly independent of temperature. Successive temperature-independent phase boundaries are not often observed except for in frustrated magnets \cite{2014LeeVVV, sears2017phase, schonemann2020thermal, huang2022successive, chern2017kitaev}. Therefore, this observation supports the existence of magnetic frustration in \NCTO{}, which is expected from Kitaev interactions as well as off-diagonal symmetric anisotropy, $\Gamma$ terms \cite{Janssen2016fieldPD,Janssen_2019}.  Among these phase transitions, the $H_{\text{1}}$ boundary along the $a^*$-axis is clearly 1$^{st}$ order, showing hysteresis between up and down field sweeps and metamagnetic behavior with a sudden change in magnetization (Fig.~\ref{MvsH-astar}). Along $a$-axis, we find that $H_{\text{1}}$ involves a rather large and discontinuous lattice distortion, as illustrated in the magnetostriction measurement. Considering that the spins are aligned along the $a$-axis in \NCTO{} {\cite{bera2017zigzag}}, this is different from a simple spin-flop or spin-flip phase transitions.

Recent inelastic neutron scattering and thermal conductivity measurements \cite{hong2021strongly, lin2022evidence} suggest that the phase between $H_{\text{2}}$ and $H_{\text{3}}$ could possibly be a Kitaev quantum spin liquid phase. It shows a restoration to an approximately six-fold symmetry in the hexagonal plane, and a broad region of low-lying excitations. Further studies are needed to distinguish KQSL from disordered or other possible phases.

We start by discussing the properties of the low-temperature, low-field phase I. The zero-field magnetic long-range ordering occurs below 27 K ($T_{\text{N}}$) and two subsequent phase transitions at 15 K ($T_{\text{F}}$) and 5 K ($T^{*}$) are observed, consistent with other literature and therein their natures are discussed \cite{lin2021field, xiao2021magnetic}. When the magnetic field is applied along $a$-axis in phase I, the 1$^{st}$ order $H_{\text{1}}$ phase transition to likely another antiferromagnetic phase (phase II) is clearly revealed in ac magnetic susceptibility (Fig.~\ref{MvsH-a}), dielectric constant (Fig.~\ref{Capacitance}), magnetocaloric effect (Fig.~\ref{MC}), and magnetostriction (Fig.~\ref{MSvsH}) in up sweeps. The transition involves a strong lattice contraction along the $a$-axis. Similar behaviors in ac magnetic susceptibility, thermal conductivity, magnetostriction and magnetocaloric effect have been observed in \RuCl{} \cite{schonemann2020thermal, Balz2019magnetostriction, balz2021field}.

Moving on to ${\bf H} \parallel a^*$ for $H_{\text{1}}$, the phase boundary is observed in dc magnetization and ac magnetic susceptibility both with a hysteresis loop (Fig.~\ref{MvsH-astar},  S5 in the S.I. \cite{supp}), dielectric constant (Fig.~\ref{Capacitance}), and magnetocaloric effect measurements (Fig.~\ref{MC}). One obvious difference compared to ${\bf H} \parallel a$ is that for ${\bf H} \parallel a^*$, the $H_{\text{1}}$ feature is more pronounced in dc magnetization measurements. This is not unreasonable as antiferromagnetic transitions are not always revealed in magnetization measurements. Another noticeable difference between $a$- and $a^*$-axes is from the dielectric constant. $H_{\text{1}}$ is observed regardless of electric field directions when ${\bf H} \parallel~a^*$ but only with ${\bf E} \parallel$ $a$ when ${\bf H} \parallel$ $a$. We note that phase II has different lattice constants from the zigzag antiferromagnetic ground state due to the 1$^{st}$ order phase transition at $H_{1}$. Further investigation such as neutron diffraction and X-ray measurements in magnetic field are necessary to investigate this.

The $H_{\text{2}}$ phase boundary is similar for ${\bf H} \parallel a$ and $a^*$ directions. It is observed in dc magnetization, ac susceptibility (Fig.~\ref{MvsH-a}, \ref{MvsH-astar}), dielectric constant (Fig.~\ref{Capacitance}), and magnetocaloric effect (Fig.~\ref{MC}) measurements. It manifests as a dip ($\bf{H} \parallel$ $a$) or a kink ($\bf{H} \parallel$ $a^*$) in $T$($H$) curves where the sample temperature starts to increase monotonically with increasing field. Such increase in lattice entropy in turn indicates a decrease in spin entropy in phase III under an quasi-adiabatic condition. Phase III has been proposed recently as a candidate KQSL phase according to the apparent resumption of the hexagonal symmetry in magnetic torque measurements, inelastic neutron diffraction measurements \cite{lin2022evidence}, and low-lying magnetic excitations in thermal conductivity \cite{hong2021strongly}. It is counter-intuitive that the spin entropy would decrease upon entering a spin liquid phase. However, we note that two types of KQSL phases exist: one is gappless and the other is gapped \cite{trebst2022kitaev}. For the former, gappless excitations contribute to the spin entropy at finite temperatures and thus, sample temperature would decrease within this phase. For the latter, energy gap protects the ground state and the spin entropy would decrease within this phase. Hence an increase of sample temperature is expected. Therefore, assuming phase III is indeed a KQSL phase, our results are more consistent with gapped KQSL scenario. In addition, we note that for KQSL-candidate $\alpha$-RuCl$_3$ the same entropy decrease was observed \cite{schonemann2020thermal}.

As reported in the literature \cite{yao2020ferrimagnetism, hong2021strongly}, the $H_{\text{3}}$ phase boundary appears like a continuation of the $T_{\text{N}}$ boundary from magnetization and specific heat measurements, with a mean-field-like shape in field-temperature. However, this phase boundary encompasses multiple magnetic phases I, II and III. Note that when $\bf{H} \parallel$ $a$, $H_{\text{3}}$ is only clearly observed in dc magnetization, ac magnetic susceptibility and dielectric constant measurements, but for $\bf{H} \parallel$ $a^*$, its feature is also very pronounced in the magnetocaloric effect.

Above $H_{\text{3}}$, the dc magnetization vs magnetic field becomes convex, appearing to saturate, as explained in the previous section. The sample temperature from the magnetocaloric effect continuously increases, consistent with a spin gap opening with increasing magnetic field. These two features support that phase IV is the spin polarized phase albeit with a magnetization that continues to slightly increase linearly up to at least 60 T due to a Van Vleck effect. A peak in the dielectric constant is usually associated with a phase transition involving electric dipole moments. As our electric polarization measurement did not yield any net electric polarization in this compound, such an electric ordering would have to be an antiferroelectric or disordered arrangement. The idea that Majorana excitations out of the KQSL phase create electrical patterns has been proposed for $\alpha$-RuCl$_3$, though there the electric patterns have no net dipole as they are radially symmetric \cite{Pereira20}. In general, magnetic spin configurations are known to produce electric polarization when the magnetism in conjunction with the lattice creates a polar axis, or alternatively magnetic configurations can create local dipoles that cancel each other in the bulk preventing a net electric polarization. Here, studying the possibility that a putative KQSL phase or its excitations could carry electric dipoles in Na$_2$Co$_2$TeO$_6$ will be an interesting future work.

Finally we note that for ${\bf H} \parallel a^*$ there is an apparent tricritical point where $T_{\text{N}}$, $H_{\text{2}}$ and $H_{\text{3}}$ meet. If these are all 2$^{nd}$ order phase transitions, such a tricritical point is not allowed by symmetry \cite{Landau65} or free energy continuity arguments \cite{Yip91}. One possibility is one of these phase boundaries is not a 2$^{nd}$ order phase transition. Another possibility is that this tricritical point merges with the first-order $H_{\text{1}}$ phase boundary, though this is not fully supported by our data. For ${\bf H} \parallel a$, there is no observed tricritical point of 2$^{nd}$ order phase transitions due to the addition of the $T_{\text{2}}$ phase line. We note also that where $T_{\text{1}}$ joins $T_{\text{N}}$ there may be a tricritical point, but more likely $T_{\text{1}}$ joins $T_{\text{N}}$ at $H = 0$, removing that conundrum.

\section{Conclusion}
In this work, we established a comprehensive ${\bf T-H}$ phase diagram of \NCTO{} based on its magnetic, electric, thermodynamic, and elastic properties. Three successive field-induced magnetic phases (I, II, III) are observed before magnetic saturation (IV), and the phase boundaries ($H_{\text{1}}$, $H_{\text{2}}$) are largely independent of temperature. This suggests the existence of magnetic frustration. Moreover, the dielectric constant is heavily dependent on magnetic field and it reveals all of the magnetic phase transitions, indicating a strong magnetoelectric coupling in \NCTO{} though without a measurable net electric polarization. Of the two proposed spin structures for phase I at zero magnetic field, the zigzag state fits our data better. By symmetry the zigzag state should not show an electric polarization in zero or applied magnetic fields, consistent with our measurements. The microscopic nature of phase II and III are still under investigation. But our work indicates that phase II has a different lattice constants compared to those in phase I, and phase III has lower spin entropy than phase II. At even higher fields, \NCTO{} enters the spin polarized phase (IV) where a spin gap opens. Strong peaks in the dielectric constant at the boundary between phase III and phase IV are consistent with an antiferroelectric or disordered-electric phase transition in conjunction with the magnetic one.

\section{Acknowledgement}
This work was principally lead by and supported by the U.S. Department of Energy, Office of Science, National Quantum Information Sciences Research Centers, Quantum Science Center. The facilities of the National High Magnetic Field Laboratory are supported by the National Science Foundation Cooperative Agreement No. DMR-1644779, and the State of Florida and the U.S. Department of Energy. Q. H. and H. D. Z. grew the samples with support from the National Science Foundation grant DMR-2003117. M. L. and S. M. T. acknowledge the LDRD program at Los Alamos National Laboratory for some initial work they did before joining the Quantum Science Center.\


\begin{thebibliography}{2}%
\makeatletter
\providecommand \@ifxundefined [1]{%
 \@ifx{#1\undefined}
}%
\providecommand \@ifnum [1]{%
 \ifnum #1\expandafter \@firstoftwo
 \else \expandafter \@secondoftwo
 \fi
}%
\providecommand \@ifx [1]{%
 \ifx #1\expandafter \@firstoftwo
 \else \expandafter \@secondoftwo
 \fi
}%
\providecommand \natexlab [1]{#1}%
\providecommand \enquote  [1]{``#1''}%
\providecommand \bibnamefont  [1]{#1}%
\providecommand \bibfnamefont [1]{#1}%
\providecommand \citenamefont [1]{#1}%
\providecommand \href@noop [0]{\@secondoftwo}%
\providecommand \href [0]{\begingroup \@sanitize@url \@href}%
\providecommand \@href[1]{\@@startlink{#1}\@@href}%
\providecommand \@@href[1]{\endgroup#1\@@endlink}%
\providecommand \@sanitize@url [0]{\catcode `\\12\catcode `\$12\catcode
  `\&12\catcode `\#12\catcode `\^12\catcode `\_12\catcode `\%12\relax}%
\providecommand \@@startlink[1]{}%
\providecommand \@@endlink[0]{}%
\providecommand \url  [0]{\begingroup\@sanitize@url \@url }%
\providecommand \@url [1]{\endgroup\@href {#1}{\urlprefix }}%
\providecommand \urlprefix  [0]{URL }%
\providecommand \Eprint [0]{\href }%
\providecommand \doibase [0]{https://doi.org/}%
\providecommand \selectlanguage [0]{\@gobble}%
\providecommand \bibinfo  [0]{\@secondoftwo}%
\providecommand \bibfield  [0]{\@secondoftwo}%
\providecommand \translation [1]{[#1]}%
\providecommand \BibitemOpen [0]{}%
\providecommand \bibitemStop [0]{}%
\providecommand \bibitemNoStop [0]{.\EOS\space}%
\providecommand \EOS [0]{\spacefactor3000\relax}%
\providecommand \BibitemShut  [1]{\csname bibitem#1\endcsname}%
\let\auto@bib@innerbib\@empty
\bibitem [{\citenamefont {Lee}\ \emph {et~al.}(2014)\citenamefont {Lee},
  \citenamefont {Hwang}, \citenamefont {Choi}, \citenamefont {Ma},
  \citenamefont {Dela~Cruz}, \citenamefont {Zhu}, \citenamefont {Ke},
  \citenamefont {Dun},\ and\ \citenamefont {Zhou}}]{Lee2014VV}%
  \BibitemOpen
  \bibfield  {author} {\bibinfo {author} {\bibfnamefont {M.}~\bibnamefont
  {Lee}}, \bibinfo {author} {\bibfnamefont {J.}~\bibnamefont {Hwang}}, \bibinfo
  {author} {\bibfnamefont {E.~S.}\ \bibnamefont {Choi}}, \bibinfo {author}
  {\bibfnamefont {J.}~\bibnamefont {Ma}}, \bibinfo {author} {\bibfnamefont
  {C.~R.}\ \bibnamefont {Dela~Cruz}}, \bibinfo {author} {\bibfnamefont
  {M.}~\bibnamefont {Zhu}}, \bibinfo {author} {\bibfnamefont {X.}~\bibnamefont
  {Ke}}, \bibinfo {author} {\bibfnamefont {Z.~L.}\ \bibnamefont {Dun}},\ and\
  \bibinfo {author} {\bibfnamefont {H.~D.}\ \bibnamefont {Zhou}},\ }\bibfield
  {title} {\bibinfo {title} {Series of phase transitions and multiferroicity in
  the quasi-two-dimensional spin-$\frac{1}{2}$ triangular-lattice
  antiferromagnet {B}a$_3${C}o{N}b$_2${O}$_9$},\ }\href
  {https://doi.org/10.1103/PhysRevB.89.104420} {\bibfield  {journal} {\bibinfo
  {journal} {Phys. Rev. B}\ }\textbf {\bibinfo {volume} {89}},\ \bibinfo
  {pages} {104420} (\bibinfo {year} {2014})}\BibitemShut {NoStop}%
\bibitem [{\citenamefont {Yao}\ and\ \citenamefont
  {Li}(2020)}]{yao2020ferrimagnetism}%
  \BibitemOpen
  \bibfield  {author} {\bibinfo {author} {\bibfnamefont {W.}~\bibnamefont
  {Yao}}\ and\ \bibinfo {author} {\bibfnamefont {Y.}~\bibnamefont {Li}},\
  }\bibfield  {title} {\bibinfo {title} {Ferrimagnetism and anisotropic phase
  tunability by magnetic fields in {N}a$_2${C}o$_2${T}e{O}$_6$},\ }\href@noop
  {} {\bibfield  {journal} {\bibinfo  {journal} {Physical Review B}\ }\textbf
  {\bibinfo {volume} {101}},\ \bibinfo {pages} {085120} (\bibinfo {year}
  {2020})}\BibitemShut {NoStop}%
\end{thebibliography}%


\begin{thebibliography}{66}%
\makeatletter
\providecommand \@ifxundefined [1]{%
 \@ifx{#1\undefined}
}%
\providecommand \@ifnum [1]{%
 \ifnum #1\expandafter \@firstoftwo
 \else \expandafter \@secondoftwo
 \fi
}%
\providecommand \@ifx [1]{%
 \ifx #1\expandafter \@firstoftwo
 \else \expandafter \@secondoftwo
 \fi
}%
\providecommand \natexlab [1]{#1}%
\providecommand \enquote  [1]{``#1''}%
\providecommand \bibnamefont  [1]{#1}%
\providecommand \bibfnamefont [1]{#1}%
\providecommand \citenamefont [1]{#1}%
\providecommand \href@noop [0]{\@secondoftwo}%
\providecommand \href [0]{\begingroup \@sanitize@url \@href}%
\providecommand \@href[1]{\@@startlink{#1}\@@href}%
\providecommand \@@href[1]{\endgroup#1\@@endlink}%
\providecommand \@sanitize@url [0]{\catcode `\\12\catcode `\$12\catcode
  `\&12\catcode `\#12\catcode `\^12\catcode `\_12\catcode `\%12\relax}%
\providecommand \@@startlink[1]{}%
\providecommand \@@endlink[0]{}%
\providecommand \url  [0]{\begingroup\@sanitize@url \@url }%
\providecommand \@url [1]{\endgroup\@href {#1}{\urlprefix }}%
\providecommand \urlprefix  [0]{URL }%
\providecommand \Eprint [0]{\href }%
\providecommand \doibase [0]{https://doi.org/}%
\providecommand \selectlanguage [0]{\@gobble}%
\providecommand \bibinfo  [0]{\@secondoftwo}%
\providecommand \bibfield  [0]{\@secondoftwo}%
\providecommand \translation [1]{[#1]}%
\providecommand \BibitemOpen [0]{}%
\providecommand \bibitemStop [0]{}%
\providecommand \bibitemNoStop [0]{.\EOS\space}%
\providecommand \EOS [0]{\spacefactor3000\relax}%
\providecommand \BibitemShut  [1]{\csname bibitem#1\endcsname}%
\let\auto@bib@innerbib\@empty
\bibitem [{\citenamefont {Kitaev}(2006)}]{kitaev2006anyons}%
  \BibitemOpen
  \bibfield  {author} {\bibinfo {author} {\bibfnamefont {A.}~\bibnamefont
  {Kitaev}},\ }\bibfield  {title} {\bibinfo {title} {Anyons in an exactly
  solved model and beyond},\ }\href@noop {} {\bibfield  {journal} {\bibinfo
  {journal} {Annals of Physics}\ }\textbf {\bibinfo {volume} {321}},\ \bibinfo
  {pages} {2} (\bibinfo {year} {2006})}\BibitemShut {NoStop}%
\bibitem [{\citenamefont {Nayak}\ \emph {et~al.}(2008)\citenamefont {Nayak},
  \citenamefont {Simon}, \citenamefont {Stern}, \citenamefont {Freedman},\ and\
  \citenamefont {Sarma}}]{nayak2008non}%
  \BibitemOpen
  \bibfield  {author} {\bibinfo {author} {\bibfnamefont {C.}~\bibnamefont
  {Nayak}}, \bibinfo {author} {\bibfnamefont {S.~H.}\ \bibnamefont {Simon}},
  \bibinfo {author} {\bibfnamefont {A.}~\bibnamefont {Stern}}, \bibinfo
  {author} {\bibfnamefont {M.}~\bibnamefont {Freedman}},\ and\ \bibinfo
  {author} {\bibfnamefont {S.~D.}\ \bibnamefont {Sarma}},\ }\bibfield  {title}
  {\bibinfo {title} {Non-{A}belian anyons and topological quantum
  computation},\ }\href@noop {} {\bibfield  {journal} {\bibinfo  {journal}
  {Reviews of Modern Physics}\ }\textbf {\bibinfo {volume} {80}},\ \bibinfo
  {pages} {1083} (\bibinfo {year} {2008})}\BibitemShut {NoStop}%
\bibitem [{\citenamefont {Kasahara}\ \emph {et~al.}(2018)\citenamefont
  {Kasahara}, \citenamefont {Ohnishi}, \citenamefont {Mizukami}, \citenamefont
  {Tanaka}, \citenamefont {Ma}, \citenamefont {Sugii}, \citenamefont {Kurita},
  \citenamefont {Tanaka}, \citenamefont {Nasu}, \citenamefont {Motome},
  \citenamefont {Shibauchi},\ and\ \citenamefont
  {Matsuda}}]{Kasahara2018half-quantized}%
  \BibitemOpen
  \bibfield  {author} {\bibinfo {author} {\bibfnamefont {Y.}~\bibnamefont
  {Kasahara}}, \bibinfo {author} {\bibfnamefont {T.}~\bibnamefont {Ohnishi}},
  \bibinfo {author} {\bibfnamefont {Y.}~\bibnamefont {Mizukami}}, \bibinfo
  {author} {\bibfnamefont {O.}~\bibnamefont {Tanaka}}, \bibinfo {author}
  {\bibfnamefont {S.}~\bibnamefont {Ma}}, \bibinfo {author} {\bibfnamefont
  {K.}~\bibnamefont {Sugii}}, \bibinfo {author} {\bibfnamefont
  {N.}~\bibnamefont {Kurita}}, \bibinfo {author} {\bibfnamefont
  {H.}~\bibnamefont {Tanaka}}, \bibinfo {author} {\bibfnamefont
  {J.}~\bibnamefont {Nasu}}, \bibinfo {author} {\bibfnamefont {Y.}~\bibnamefont
  {Motome}}, \bibinfo {author} {\bibfnamefont {T.}~\bibnamefont {Shibauchi}},\
  and\ \bibinfo {author} {\bibfnamefont {Y.}~\bibnamefont {Matsuda}},\
  }\bibfield  {title} {\bibinfo {title} {Majorana quantization and half-integer
  thermal quantum hall effect in a kitaev spin liquid},\ }\href@noop {}
  {\bibfield  {journal} {\bibinfo  {journal} {Nature}\ }\textbf {\bibinfo
  {volume} {559}},\ \bibinfo {pages} {227} (\bibinfo {year}
  {2018})}\BibitemShut {NoStop}%
\bibitem [{\citenamefont {Kasahara}\ \emph {et~al.}(2022)\citenamefont
  {Kasahara}, \citenamefont {Suetsugu}, \citenamefont {Asaba}, \citenamefont
  {Kasahara}, \citenamefont {Shibauchi}, \citenamefont {Kurita}, \citenamefont
  {Tanaka},\ and\ \citenamefont {Matsuda}}]{kasahara2022quantized}%
  \BibitemOpen
  \bibfield  {author} {\bibinfo {author} {\bibfnamefont {Y.}~\bibnamefont
  {Kasahara}}, \bibinfo {author} {\bibfnamefont {S.}~\bibnamefont {Suetsugu}},
  \bibinfo {author} {\bibfnamefont {T.}~\bibnamefont {Asaba}}, \bibinfo
  {author} {\bibfnamefont {S.}~\bibnamefont {Kasahara}}, \bibinfo {author}
  {\bibfnamefont {T.}~\bibnamefont {Shibauchi}}, \bibinfo {author}
  {\bibfnamefont {N.}~\bibnamefont {Kurita}}, \bibinfo {author} {\bibfnamefont
  {H.}~\bibnamefont {Tanaka}},\ and\ \bibinfo {author} {\bibfnamefont
  {Y.}~\bibnamefont {Matsuda}},\ }\bibfield  {title} {\bibinfo {title}
  {Quantized and unquantized thermal hall conductance of the kitaev spin liquid
  candidate $\alpha$- rucl 3},\ }\href@noop {} {\bibfield  {journal} {\bibinfo
  {journal} {Physical Review B}\ }\textbf {\bibinfo {volume} {106}},\ \bibinfo
  {pages} {L060410} (\bibinfo {year} {2022})}\BibitemShut {NoStop}%
\bibitem [{\citenamefont {Jackeli}\ and\ \citenamefont
  {Khaliullin}(2009)}]{jackeli2009mott}%
  \BibitemOpen
  \bibfield  {author} {\bibinfo {author} {\bibfnamefont {G.}~\bibnamefont
  {Jackeli}}\ and\ \bibinfo {author} {\bibfnamefont {G.}~\bibnamefont
  {Khaliullin}},\ }\bibfield  {title} {\bibinfo {title} {Mott insulators in the
  strong spin-orbit coupling limit: from {H}eisenberg to a quantum compass and
  {K}itaev models},\ }\href@noop {} {\bibfield  {journal} {\bibinfo  {journal}
  {Physical review letters}\ }\textbf {\bibinfo {volume} {102}},\ \bibinfo
  {pages} {017205} (\bibinfo {year} {2009})}\BibitemShut {NoStop}%
\bibitem [{\citenamefont {Chaloupka}\ \emph {et~al.}(2010)\citenamefont
  {Chaloupka}, \citenamefont {Jackeli},\ and\ \citenamefont
  {Khaliullin}}]{chaloupka2010kitaev}%
  \BibitemOpen
  \bibfield  {author} {\bibinfo {author} {\bibfnamefont {J.}~\bibnamefont
  {Chaloupka}}, \bibinfo {author} {\bibfnamefont {G.}~\bibnamefont {Jackeli}},\
  and\ \bibinfo {author} {\bibfnamefont {G.}~\bibnamefont {Khaliullin}},\
  }\bibfield  {title} {\bibinfo {title} {Kitaev-{H}eisenberg model on a
  honeycomb lattice: possible exotic phases in iridium oxides
  {A}$_2${I}r{O}$_3$},\ }\href@noop {} {\bibfield  {journal} {\bibinfo
  {journal} {Physical review letters}\ }\textbf {\bibinfo {volume} {105}},\
  \bibinfo {pages} {027204} (\bibinfo {year} {2010})}\BibitemShut {NoStop}%
\bibitem [{\citenamefont {Kitagawa}\ \emph {et~al.}(2018)\citenamefont
  {Kitagawa}, \citenamefont {Takayama}, \citenamefont {Matsumoto},
  \citenamefont {Kato}, \citenamefont {Takano}, \citenamefont {Kishimoto},
  \citenamefont {Bette}, \citenamefont {Dinnebier}, \citenamefont {Jackeli},\
  and\ \citenamefont {Takagi}}]{kitagawa2018spin}%
  \BibitemOpen
  \bibfield  {author} {\bibinfo {author} {\bibfnamefont {K.}~\bibnamefont
  {Kitagawa}}, \bibinfo {author} {\bibfnamefont {T.}~\bibnamefont {Takayama}},
  \bibinfo {author} {\bibfnamefont {Y.}~\bibnamefont {Matsumoto}}, \bibinfo
  {author} {\bibfnamefont {A.}~\bibnamefont {Kato}}, \bibinfo {author}
  {\bibfnamefont {R.}~\bibnamefont {Takano}}, \bibinfo {author} {\bibfnamefont
  {Y.}~\bibnamefont {Kishimoto}}, \bibinfo {author} {\bibfnamefont
  {S.}~\bibnamefont {Bette}}, \bibinfo {author} {\bibfnamefont
  {R.}~\bibnamefont {Dinnebier}}, \bibinfo {author} {\bibfnamefont
  {G.}~\bibnamefont {Jackeli}},\ and\ \bibinfo {author} {\bibfnamefont
  {H.}~\bibnamefont {Takagi}},\ }\bibfield  {title} {\bibinfo {title} {A
  spin--orbital-entangled quantum liquid on a honeycomb lattice},\ }\href@noop
  {} {\bibfield  {journal} {\bibinfo  {journal} {Nature}\ }\textbf {\bibinfo
  {volume} {554}},\ \bibinfo {pages} {341} (\bibinfo {year}
  {2018})}\BibitemShut {NoStop}%
\bibitem [{\citenamefont {Bahrami}\ \emph {et~al.}(2019)\citenamefont
  {Bahrami}, \citenamefont {Lafargue-Dit-Hauret}, \citenamefont {Lebedev},
  \citenamefont {Movshovich}, \citenamefont {Yang}, \citenamefont {Broido},
  \citenamefont {Rocquefelte},\ and\ \citenamefont
  {Tafti}}]{bahrami2019thermodynamic}%
  \BibitemOpen
  \bibfield  {author} {\bibinfo {author} {\bibfnamefont {F.}~\bibnamefont
  {Bahrami}}, \bibinfo {author} {\bibfnamefont {W.}~\bibnamefont
  {Lafargue-Dit-Hauret}}, \bibinfo {author} {\bibfnamefont {O.~I.}\
  \bibnamefont {Lebedev}}, \bibinfo {author} {\bibfnamefont {R.}~\bibnamefont
  {Movshovich}}, \bibinfo {author} {\bibfnamefont {H.-Y.}\ \bibnamefont
  {Yang}}, \bibinfo {author} {\bibfnamefont {D.}~\bibnamefont {Broido}},
  \bibinfo {author} {\bibfnamefont {X.}~\bibnamefont {Rocquefelte}},\ and\
  \bibinfo {author} {\bibfnamefont {F.}~\bibnamefont {Tafti}},\ }\bibfield
  {title} {\bibinfo {title} {Thermodynamic evidence of proximity to a kitaev
  spin liquid in {A}g$_3${L}i{I}r$_2${O}$_6$},\ }\href@noop {} {\bibfield
  {journal} {\bibinfo  {journal} {Physical Review Letters}\ }\textbf {\bibinfo
  {volume} {123}},\ \bibinfo {pages} {237203} (\bibinfo {year}
  {2019})}\BibitemShut {NoStop}%
\bibitem [{\citenamefont {Chakraborty}\ \emph {et~al.}(2021)\citenamefont
  {Chakraborty}, \citenamefont {Kumar}, \citenamefont {Bachhar}, \citenamefont
  {B{\"u}ttgen}, \citenamefont {Yokoyama}, \citenamefont {Biswas},
  \citenamefont {Siruguri}, \citenamefont {Pujari}, \citenamefont {Dasgupta},\
  and\ \citenamefont {Mahajan}}]{chakraborty2021unusual}%
  \BibitemOpen
  \bibfield  {author} {\bibinfo {author} {\bibfnamefont {A.}~\bibnamefont
  {Chakraborty}}, \bibinfo {author} {\bibfnamefont {V.}~\bibnamefont {Kumar}},
  \bibinfo {author} {\bibfnamefont {S.}~\bibnamefont {Bachhar}}, \bibinfo
  {author} {\bibfnamefont {N.}~\bibnamefont {B{\"u}ttgen}}, \bibinfo {author}
  {\bibfnamefont {K.}~\bibnamefont {Yokoyama}}, \bibinfo {author}
  {\bibfnamefont {P.~K.}\ \bibnamefont {Biswas}}, \bibinfo {author}
  {\bibfnamefont {V.}~\bibnamefont {Siruguri}}, \bibinfo {author}
  {\bibfnamefont {S.}~\bibnamefont {Pujari}}, \bibinfo {author} {\bibfnamefont
  {I.}~\bibnamefont {Dasgupta}},\ and\ \bibinfo {author} {\bibfnamefont
  {A.~V.}\ \bibnamefont {Mahajan}},\ }\bibfield  {title} {\bibinfo {title}
  {Unusual spin dynamics in the low-temperature magnetically ordered state of
  {A}g$_3${L}i{I}r$_2${O}$_6$},\ }\href@noop {} {\bibfield  {journal} {\bibinfo
   {journal} {Physical Review B}\ }\textbf {\bibinfo {volume} {104}},\ \bibinfo
  {pages} {115106} (\bibinfo {year} {2021})}\BibitemShut {NoStop}%
\bibitem [{\citenamefont {Bahrami}\ \emph {et~al.}(2021)\citenamefont
  {Bahrami}, \citenamefont {Kenney}, \citenamefont {Wang}, \citenamefont
  {Berlie}, \citenamefont {Lebedev}, \citenamefont {Graf},\ and\ \citenamefont
  {Tafti}}]{Bahrami2021}%
  \BibitemOpen
  \bibfield  {author} {\bibinfo {author} {\bibfnamefont {F.}~\bibnamefont
  {Bahrami}}, \bibinfo {author} {\bibfnamefont {E.~M.}\ \bibnamefont {Kenney}},
  \bibinfo {author} {\bibfnamefont {C.}~\bibnamefont {Wang}}, \bibinfo {author}
  {\bibfnamefont {A.}~\bibnamefont {Berlie}}, \bibinfo {author} {\bibfnamefont
  {O.~I.}\ \bibnamefont {Lebedev}}, \bibinfo {author} {\bibfnamefont {M.~J.}\
  \bibnamefont {Graf}},\ and\ \bibinfo {author} {\bibfnamefont
  {F.}~\bibnamefont {Tafti}},\ }\bibfield  {title} {\bibinfo {title} {Effect of
  structural disorder on the kitaev magnet
  ${\mathrm{ag}}_{3}{\mathrm{liir}}_{2}{\mathrm{o}}_{6}$},\ }\href
  {https://doi.org/10.1103/PhysRevB.103.094427} {\bibfield  {journal} {\bibinfo
   {journal} {Phys. Rev. B}\ }\textbf {\bibinfo {volume} {103}},\ \bibinfo
  {pages} {094427} (\bibinfo {year} {2021})}\BibitemShut {NoStop}%
\bibitem [{\citenamefont {Li}\ and\ \citenamefont
  {Valent\'{\i}}(2022)}]{LiValenti2022}%
  \BibitemOpen
  \bibfield  {author} {\bibinfo {author} {\bibfnamefont {Y.}~\bibnamefont
  {Li}}\ and\ \bibinfo {author} {\bibfnamefont {R.}~\bibnamefont
  {Valent\'{\i}}},\ }\bibfield  {title} {\bibinfo {title} {Role of disorder in
  electronic and magnetic properties of
  ${\mathrm{ag}}_{3}{\mathrm{liir}}_{2}{\mathrm{o}}_{6}$},\ }\href
  {https://doi.org/10.1103/PhysRevB.105.115123} {\bibfield  {journal} {\bibinfo
   {journal} {Phys. Rev. B}\ }\textbf {\bibinfo {volume} {105}},\ \bibinfo
  {pages} {115123} (\bibinfo {year} {2022})}\BibitemShut {NoStop}%
\bibitem [{\citenamefont {Plumb}\ \emph {et~al.}(2014)\citenamefont {Plumb},
  \citenamefont {Clancy}, \citenamefont {Sandilands}, \citenamefont {Shankar},
  \citenamefont {Hu}, \citenamefont {Burch}, \citenamefont {Kee},\ and\
  \citenamefont {Kim}}]{plumb2014alpha}%
  \BibitemOpen
  \bibfield  {author} {\bibinfo {author} {\bibfnamefont {K.}~\bibnamefont
  {Plumb}}, \bibinfo {author} {\bibfnamefont {J.}~\bibnamefont {Clancy}},
  \bibinfo {author} {\bibfnamefont {L.}~\bibnamefont {Sandilands}}, \bibinfo
  {author} {\bibfnamefont {V.~V.}\ \bibnamefont {Shankar}}, \bibinfo {author}
  {\bibfnamefont {Y.}~\bibnamefont {Hu}}, \bibinfo {author} {\bibfnamefont
  {K.}~\bibnamefont {Burch}}, \bibinfo {author} {\bibfnamefont {H.-Y.}\
  \bibnamefont {Kee}},\ and\ \bibinfo {author} {\bibfnamefont {Y.-J.}\
  \bibnamefont {Kim}},\ }\bibfield  {title} {\bibinfo {title} {$\alpha$-
  {R}u{C}l$_3$: {A} spin-orbit assisted mott insulator on a honeycomb
  lattice},\ }\href@noop {} {\bibfield  {journal} {\bibinfo  {journal}
  {Physical Review B}\ }\textbf {\bibinfo {volume} {90}},\ \bibinfo {pages}
  {041112} (\bibinfo {year} {2014})}\BibitemShut {NoStop}%
\bibitem [{\citenamefont {Banerjee}\ \emph {et~al.}(2016)\citenamefont
  {Banerjee}, \citenamefont {Bridges}, \citenamefont {Yan}, \citenamefont
  {Aczel}, \citenamefont {Li}, \citenamefont {Stone}, \citenamefont {Granroth},
  \citenamefont {Lumsden}, \citenamefont {Yiu}, \citenamefont {Knolle} \emph
  {et~al.}}]{banerjee2016proximate}%
  \BibitemOpen
  \bibfield  {author} {\bibinfo {author} {\bibfnamefont {A.}~\bibnamefont
  {Banerjee}}, \bibinfo {author} {\bibfnamefont {C.}~\bibnamefont {Bridges}},
  \bibinfo {author} {\bibfnamefont {J.-Q.}\ \bibnamefont {Yan}}, \bibinfo
  {author} {\bibfnamefont {A.}~\bibnamefont {Aczel}}, \bibinfo {author}
  {\bibfnamefont {L.}~\bibnamefont {Li}}, \bibinfo {author} {\bibfnamefont
  {M.}~\bibnamefont {Stone}}, \bibinfo {author} {\bibfnamefont
  {G.}~\bibnamefont {Granroth}}, \bibinfo {author} {\bibfnamefont
  {M.}~\bibnamefont {Lumsden}}, \bibinfo {author} {\bibfnamefont
  {Y.}~\bibnamefont {Yiu}}, \bibinfo {author} {\bibfnamefont {J.}~\bibnamefont
  {Knolle}}, \emph {et~al.},\ }\bibfield  {title} {\bibinfo {title} {Proximate
  kitaev quantum spin liquid behaviour in a honeycomb magnet},\ }\href@noop {}
  {\bibfield  {journal} {\bibinfo  {journal} {Nature materials}\ }\textbf
  {\bibinfo {volume} {15}},\ \bibinfo {pages} {733} (\bibinfo {year}
  {2016})}\BibitemShut {NoStop}%
\bibitem [{\citenamefont {Do}\ \emph {et~al.}(2017)\citenamefont {Do},
  \citenamefont {Park}, \citenamefont {Yoshitake}, \citenamefont {Nasu},
  \citenamefont {Motome}, \citenamefont {Kwon}, \citenamefont {Adroja},
  \citenamefont {Voneshen}, \citenamefont {Kim}, \citenamefont {Jang},
  \citenamefont {Park}, \citenamefont {Choi},\ and\ \citenamefont
  {Ji}}]{do2017RuClINS}%
  \BibitemOpen
  \bibfield  {author} {\bibinfo {author} {\bibfnamefont {S.-H.}\ \bibnamefont
  {Do}}, \bibinfo {author} {\bibfnamefont {S.-Y.}\ \bibnamefont {Park}},
  \bibinfo {author} {\bibfnamefont {J.}~\bibnamefont {Yoshitake}}, \bibinfo
  {author} {\bibfnamefont {J.}~\bibnamefont {Nasu}}, \bibinfo {author}
  {\bibfnamefont {Y.}~\bibnamefont {Motome}}, \bibinfo {author} {\bibfnamefont
  {Y.~S.}\ \bibnamefont {Kwon}}, \bibinfo {author} {\bibfnamefont {D.~T.}\
  \bibnamefont {Adroja}}, \bibinfo {author} {\bibfnamefont {D.~J.}\
  \bibnamefont {Voneshen}}, \bibinfo {author} {\bibfnamefont {K.}~\bibnamefont
  {Kim}}, \bibinfo {author} {\bibfnamefont {T.-H.}\ \bibnamefont {Jang}},
  \bibinfo {author} {\bibfnamefont {J.-H.}\ \bibnamefont {Park}}, \bibinfo
  {author} {\bibfnamefont {K.-Y.}\ \bibnamefont {Choi}},\ and\ \bibinfo
  {author} {\bibfnamefont {S.}~\bibnamefont {Ji}},\ }\bibfield  {title}
  {\bibinfo {title} {Majorana fermions in the kitaev quantum spin system
  $\alpha$-rucl$_{3}$},\ }\href
  {https://doi.org/https://doi.org/10.1038/nphys4264} {\bibfield  {journal}
  {\bibinfo  {journal} {Nature Physics}\ }\textbf {\bibinfo {volume} {13}},\
  \bibinfo {pages} {1079} (\bibinfo {year} {2017})}\BibitemShut {NoStop}%
\bibitem [{\citenamefont {Czajka}\ \emph {et~al.}(2021)\citenamefont {Czajka},
  \citenamefont {Gao}, \citenamefont {Hirschberger}, \citenamefont
  {Lampen-Kelley}, \citenamefont {Banerjee}, \citenamefont {Yan}, \citenamefont
  {Mandrus}, \citenamefont {Nagler},\ and\ \citenamefont
  {Ong}}]{czajka2021oscillations}%
  \BibitemOpen
  \bibfield  {author} {\bibinfo {author} {\bibfnamefont {P.}~\bibnamefont
  {Czajka}}, \bibinfo {author} {\bibfnamefont {T.}~\bibnamefont {Gao}},
  \bibinfo {author} {\bibfnamefont {M.}~\bibnamefont {Hirschberger}}, \bibinfo
  {author} {\bibfnamefont {P.}~\bibnamefont {Lampen-Kelley}}, \bibinfo {author}
  {\bibfnamefont {A.}~\bibnamefont {Banerjee}}, \bibinfo {author}
  {\bibfnamefont {J.}~\bibnamefont {Yan}}, \bibinfo {author} {\bibfnamefont
  {D.~G.}\ \bibnamefont {Mandrus}}, \bibinfo {author} {\bibfnamefont {S.~E.}\
  \bibnamefont {Nagler}},\ and\ \bibinfo {author} {\bibfnamefont
  {N.}~\bibnamefont {Ong}},\ }\bibfield  {title} {\bibinfo {title}
  {Oscillations of the thermal conductivity in the spin-liquid state of
  $\alpha$-{R}u{C}l$_3$},\ }\href@noop {} {\bibfield  {journal} {\bibinfo
  {journal} {Nature Physics}\ }\textbf {\bibinfo {volume} {17}},\ \bibinfo
  {pages} {915} (\bibinfo {year} {2021})}\BibitemShut {NoStop}%
\bibitem [{\citenamefont {Bruin}\ \emph
  {et~al.}(2022{\natexlab{a}})\citenamefont {Bruin}, \citenamefont {Claus},
  \citenamefont {Matsumoto}, \citenamefont {Nuss}, \citenamefont {Laha},
  \citenamefont {Lotsch}, \citenamefont {Kurita}, \citenamefont {Tanaka},\ and\
  \citenamefont {Takagi}}]{bruin2022origin}%
  \BibitemOpen
  \bibfield  {author} {\bibinfo {author} {\bibfnamefont {J.}~\bibnamefont
  {Bruin}}, \bibinfo {author} {\bibfnamefont {R.}~\bibnamefont {Claus}},
  \bibinfo {author} {\bibfnamefont {Y.}~\bibnamefont {Matsumoto}}, \bibinfo
  {author} {\bibfnamefont {J.}~\bibnamefont {Nuss}}, \bibinfo {author}
  {\bibfnamefont {S.}~\bibnamefont {Laha}}, \bibinfo {author} {\bibfnamefont
  {B.}~\bibnamefont {Lotsch}}, \bibinfo {author} {\bibfnamefont
  {N.}~\bibnamefont {Kurita}}, \bibinfo {author} {\bibfnamefont
  {H.}~\bibnamefont {Tanaka}},\ and\ \bibinfo {author} {\bibfnamefont
  {H.}~\bibnamefont {Takagi}},\ }\bibfield  {title} {\bibinfo {title} {Origin
  of oscillatory structures in the magnetothermal conductivity of the putative
  kitaev magnet $\alpha$-{R}u{C}l$_3$},\ }\href@noop {} {\bibfield  {journal}
  {\bibinfo  {journal} {APL Materials}\ }\textbf {\bibinfo {volume} {10}},\
  \bibinfo {pages} {090703} (\bibinfo {year} {2022}{\natexlab{a}})}\BibitemShut
  {NoStop}%
\bibitem [{\citenamefont {Bruin}\ \emph
  {et~al.}(2022{\natexlab{b}})\citenamefont {Bruin}, \citenamefont {Claus},
  \citenamefont {Matsumoto}, \citenamefont {Kurita}, \citenamefont {Tanaka},\
  and\ \citenamefont {Takagi}}]{bruin2022robustness}%
  \BibitemOpen
  \bibfield  {author} {\bibinfo {author} {\bibfnamefont {J.}~\bibnamefont
  {Bruin}}, \bibinfo {author} {\bibfnamefont {R.}~\bibnamefont {Claus}},
  \bibinfo {author} {\bibfnamefont {Y.}~\bibnamefont {Matsumoto}}, \bibinfo
  {author} {\bibfnamefont {N.}~\bibnamefont {Kurita}}, \bibinfo {author}
  {\bibfnamefont {H.}~\bibnamefont {Tanaka}},\ and\ \bibinfo {author}
  {\bibfnamefont {H.}~\bibnamefont {Takagi}},\ }\bibfield  {title} {\bibinfo
  {title} {Robustness of the thermal hall effect close to half-quantization in
  $\alpha$-{R}u{C}l$_3$},\ }\href@noop {} {\bibfield  {journal} {\bibinfo
  {journal} {Nature Physics}\ }\textbf {\bibinfo {volume} {18}},\ \bibinfo
  {pages} {401} (\bibinfo {year} {2022}{\natexlab{b}})}\BibitemShut {NoStop}%
\bibitem [{\citenamefont {Liu}\ and\ \citenamefont
  {Khaliullin}(2018)}]{liu2018pseudospin}%
  \BibitemOpen
  \bibfield  {author} {\bibinfo {author} {\bibfnamefont {H.}~\bibnamefont
  {Liu}}\ and\ \bibinfo {author} {\bibfnamefont {G.}~\bibnamefont
  {Khaliullin}},\ }\bibfield  {title} {\bibinfo {title} {Pseudospin exchange
  interactions in d$^7$ cobalt compounds: possible realization of the {K}itaev
  model},\ }\href@noop {} {\bibfield  {journal} {\bibinfo  {journal} {Physical
  Review B}\ }\textbf {\bibinfo {volume} {97}},\ \bibinfo {pages} {014407}
  (\bibinfo {year} {2018})}\BibitemShut {NoStop}%
\bibitem [{\citenamefont {Sano}\ \emph {et~al.}(2018)\citenamefont {Sano},
  \citenamefont {Kato},\ and\ \citenamefont {Motome}}]{Sano18}%
  \BibitemOpen
  \bibfield  {author} {\bibinfo {author} {\bibfnamefont {R.}~\bibnamefont
  {Sano}}, \bibinfo {author} {\bibfnamefont {Y.}~\bibnamefont {Kato}},\ and\
  \bibinfo {author} {\bibfnamefont {Y.}~\bibnamefont {Motome}},\ }\bibfield
  {title} {\bibinfo {title} {Kitaev-heisenberg hamiltonian for high-spin
  ${d}^{7}$ mott insulators},\ }\href
  {https://doi.org/10.1103/PhysRevB.97.014408} {\bibfield  {journal} {\bibinfo
  {journal} {Phys. Rev. B}\ }\textbf {\bibinfo {volume} {97}},\ \bibinfo
  {pages} {014408} (\bibinfo {year} {2018})}\BibitemShut {NoStop}%
\bibitem [{\citenamefont {Liu}(2021)}]{liu2021towards}%
  \BibitemOpen
  \bibfield  {author} {\bibinfo {author} {\bibfnamefont {H.}~\bibnamefont
  {Liu}},\ }\bibfield  {title} {\bibinfo {title} {Towards kitaev spin liquid in
  3d transition metal compounds},\ }\href@noop {} {\bibfield  {journal}
  {\bibinfo  {journal} {International Journal of Modern Physics B}\ }\textbf
  {\bibinfo {volume} {35}},\ \bibinfo {pages} {2130006} (\bibinfo {year}
  {2021})}\BibitemShut {NoStop}%
\bibitem [{\citenamefont {Kim}\ \emph {et~al.}(2021)\citenamefont {Kim},
  \citenamefont {Jeong}, \citenamefont {Lin}, \citenamefont {Park},
  \citenamefont {Masuda}, \citenamefont {Asai}, \citenamefont {Itoh},
  \citenamefont {Kim}, \citenamefont {Zhou}, \citenamefont {Ma} \emph
  {et~al.}}]{kim2021antiferromagnetic}%
  \BibitemOpen
  \bibfield  {author} {\bibinfo {author} {\bibfnamefont {C.}~\bibnamefont
  {Kim}}, \bibinfo {author} {\bibfnamefont {J.}~\bibnamefont {Jeong}}, \bibinfo
  {author} {\bibfnamefont {G.}~\bibnamefont {Lin}}, \bibinfo {author}
  {\bibfnamefont {P.}~\bibnamefont {Park}}, \bibinfo {author} {\bibfnamefont
  {T.}~\bibnamefont {Masuda}}, \bibinfo {author} {\bibfnamefont
  {S.}~\bibnamefont {Asai}}, \bibinfo {author} {\bibfnamefont {S.}~\bibnamefont
  {Itoh}}, \bibinfo {author} {\bibfnamefont {H.-S.}\ \bibnamefont {Kim}},
  \bibinfo {author} {\bibfnamefont {H.}~\bibnamefont {Zhou}}, \bibinfo {author}
  {\bibfnamefont {J.}~\bibnamefont {Ma}}, \emph {et~al.},\ }\bibfield  {title}
  {\bibinfo {title} {Antiferromagnetic kitaev interaction in {J}$_{eff}$= 1/2
  cobalt honeycomb materials {N}a$_3${C}o$_2${S}b{O}$_6$ and
  {N}a$_2${C}o$_2${T}e{O}$_6$},\ }\href@noop {} {\bibfield  {journal} {\bibinfo
   {journal} {Journal of Physics: Condensed Matter}\ }\textbf {\bibinfo
  {volume} {34}},\ \bibinfo {pages} {045802} (\bibinfo {year}
  {2021})}\BibitemShut {NoStop}%
\bibitem [{\citenamefont {Lin}\ \emph {et~al.}(2021)\citenamefont {Lin},
  \citenamefont {Jeong}, \citenamefont {Kim}, \citenamefont {Wang},
  \citenamefont {Huang}, \citenamefont {Masuda}, \citenamefont {Asai},
  \citenamefont {Itoh}, \citenamefont {G{\"u}nther}, \citenamefont {Russina}
  \emph {et~al.}}]{lin2021field}%
  \BibitemOpen
  \bibfield  {author} {\bibinfo {author} {\bibfnamefont {G.}~\bibnamefont
  {Lin}}, \bibinfo {author} {\bibfnamefont {J.}~\bibnamefont {Jeong}}, \bibinfo
  {author} {\bibfnamefont {C.}~\bibnamefont {Kim}}, \bibinfo {author}
  {\bibfnamefont {Y.}~\bibnamefont {Wang}}, \bibinfo {author} {\bibfnamefont
  {Q.}~\bibnamefont {Huang}}, \bibinfo {author} {\bibfnamefont
  {T.}~\bibnamefont {Masuda}}, \bibinfo {author} {\bibfnamefont
  {S.}~\bibnamefont {Asai}}, \bibinfo {author} {\bibfnamefont {S.}~\bibnamefont
  {Itoh}}, \bibinfo {author} {\bibfnamefont {G.}~\bibnamefont {G{\"u}nther}},
  \bibinfo {author} {\bibfnamefont {M.}~\bibnamefont {Russina}}, \emph
  {et~al.},\ }\bibfield  {title} {\bibinfo {title} {Field-induced quantum spin
  disordered state in spin-1/2 honeycomb magnet {N}a$_2${C}o$_2${T}e{O}$_6$},\
  }\href@noop {} {\bibfield  {journal} {\bibinfo  {journal} {Nature
  communications}\ }\textbf {\bibinfo {volume} {12}},\ \bibinfo {pages} {1}
  (\bibinfo {year} {2021})}\BibitemShut {NoStop}%
\bibitem [{\citenamefont {Sanders}\ \emph {et~al.}(2022)\citenamefont
  {Sanders}, \citenamefont {Mole}, \citenamefont {Liu}, \citenamefont {Brown},
  \citenamefont {Yu}, \citenamefont {Ling},\ and\ \citenamefont
  {Rachel}}]{sanders2022dominant}%
  \BibitemOpen
  \bibfield  {author} {\bibinfo {author} {\bibfnamefont {A.~L.}\ \bibnamefont
  {Sanders}}, \bibinfo {author} {\bibfnamefont {R.~A.}\ \bibnamefont {Mole}},
  \bibinfo {author} {\bibfnamefont {J.}~\bibnamefont {Liu}}, \bibinfo {author}
  {\bibfnamefont {A.~J.}\ \bibnamefont {Brown}}, \bibinfo {author}
  {\bibfnamefont {D.}~\bibnamefont {Yu}}, \bibinfo {author} {\bibfnamefont
  {C.~D.}\ \bibnamefont {Ling}},\ and\ \bibinfo {author} {\bibfnamefont
  {S.}~\bibnamefont {Rachel}},\ }\bibfield  {title} {\bibinfo {title} {Dominant
  kitaev interactions in the honeycomb materials {N}a$_3${C}o$_2${S}b{O}$_6$
  and {N}a$_2${C}o$_2${T}e{O}$_6$},\ }\href@noop {} {\bibfield  {journal}
  {\bibinfo  {journal} {Physical Review B}\ }\textbf {\bibinfo {volume}
  {106}},\ \bibinfo {pages} {014413} (\bibinfo {year} {2022})}\BibitemShut
  {NoStop}%
\bibitem [{\citenamefont {Yao}\ \emph {et~al.}(2022)\citenamefont {Yao},
  \citenamefont {Iida}, \citenamefont {Kamazawa},\ and\ \citenamefont
  {Li}}]{yao2022excitations}%
  \BibitemOpen
  \bibfield  {author} {\bibinfo {author} {\bibfnamefont {W.}~\bibnamefont
  {Yao}}, \bibinfo {author} {\bibfnamefont {K.}~\bibnamefont {Iida}}, \bibinfo
  {author} {\bibfnamefont {K.}~\bibnamefont {Kamazawa}},\ and\ \bibinfo
  {author} {\bibfnamefont {Y.}~\bibnamefont {Li}},\ }\bibfield  {title}
  {\bibinfo {title} {Excitations in the ordered and paramagnetic states of
  honeycomb magnet {N}a$_2${C}o$_2${T}e{O}$_6$},\ }\href
  {https://doi.org/10.1103/PhysRevLett.129.147202} {\bibfield  {journal}
  {\bibinfo  {journal} {Phys. Rev. Lett.}\ }\textbf {\bibinfo {volume} {129}},\
  \bibinfo {pages} {147202} (\bibinfo {year} {2022})}\BibitemShut {NoStop}%
\bibitem [{\citenamefont {Viciu}\ \emph {et~al.}(2007)\citenamefont {Viciu},
  \citenamefont {Huang}, \citenamefont {Morosan}, \citenamefont {Zandbergen},
  \citenamefont {Greenbaum}, \citenamefont {McQueen},\ and\ \citenamefont
  {Cava}}]{viciu2007structure}%
  \BibitemOpen
  \bibfield  {author} {\bibinfo {author} {\bibfnamefont {L.}~\bibnamefont
  {Viciu}}, \bibinfo {author} {\bibfnamefont {Q.}~\bibnamefont {Huang}},
  \bibinfo {author} {\bibfnamefont {E.}~\bibnamefont {Morosan}}, \bibinfo
  {author} {\bibfnamefont {H.}~\bibnamefont {Zandbergen}}, \bibinfo {author}
  {\bibfnamefont {N.}~\bibnamefont {Greenbaum}}, \bibinfo {author}
  {\bibfnamefont {T.}~\bibnamefont {McQueen}},\ and\ \bibinfo {author}
  {\bibfnamefont {R.}~\bibnamefont {Cava}},\ }\bibfield  {title} {\bibinfo
  {title} {Structure and basic magnetic properties of the honeycomb lattice
  compounds {N}a$_2${C}o$_2${T}e{O}$_6$ and {N}a$_3${C}o$_2${S}b{O}$_6$},\
  }\href@noop {} {\bibfield  {journal} {\bibinfo  {journal} {Journal of Solid
  State Chemistry}\ }\textbf {\bibinfo {volume} {180}},\ \bibinfo {pages}
  {1060} (\bibinfo {year} {2007})}\BibitemShut {NoStop}%
\bibitem [{\citenamefont {Winter}(2022)}]{winter2022magnetic}%
  \BibitemOpen
  \bibfield  {author} {\bibinfo {author} {\bibfnamefont {S.~M.}\ \bibnamefont
  {Winter}},\ }\bibfield  {title} {\bibinfo {title} {Magnetic couplings in
  edge-sharing high-spin $d^7$ compounds},\ }\href@noop {} {\bibfield
  {journal} {\bibinfo  {journal} {J. Phys. Mater.}\ }\textbf {\bibinfo {volume}
  {5}},\ \bibinfo {pages} {045003} (\bibinfo {year} {2022})}\BibitemShut
  {NoStop}%
\bibitem [{\citenamefont {Bera}\ \emph {et~al.}(2017)\citenamefont {Bera},
  \citenamefont {Yusuf}, \citenamefont {Kumar},\ and\ \citenamefont
  {Ritter}}]{bera2017zigzag}%
  \BibitemOpen
  \bibfield  {author} {\bibinfo {author} {\bibfnamefont {A.}~\bibnamefont
  {Bera}}, \bibinfo {author} {\bibfnamefont {S.}~\bibnamefont {Yusuf}},
  \bibinfo {author} {\bibfnamefont {A.}~\bibnamefont {Kumar}},\ and\ \bibinfo
  {author} {\bibfnamefont {C.}~\bibnamefont {Ritter}},\ }\bibfield  {title}
  {\bibinfo {title} {Zigzag antiferromagnetic ground state with anisotropic
  correlation lengths in the quasi-two-dimensional honeycomb lattice compound
  {N}a$_2${C}o$_2${T}e{O}$_6$},\ }\href@noop {} {\bibfield  {journal} {\bibinfo
   {journal} {Physical Review B}\ }\textbf {\bibinfo {volume} {95}},\ \bibinfo
  {pages} {094424} (\bibinfo {year} {2017})}\BibitemShut {NoStop}%
\bibitem [{\citenamefont {Hong}\ \emph {et~al.}(2021)\citenamefont {Hong},
  \citenamefont {Gillig}, \citenamefont {Hentrich}, \citenamefont {Yao},
  \citenamefont {Kocsis}, \citenamefont {Witte}, \citenamefont {Schreiner},
  \citenamefont {Baumann}, \citenamefont {P{\'e}rez}, \citenamefont {Wolter}
  \emph {et~al.}}]{hong2021strongly}%
  \BibitemOpen
  \bibfield  {author} {\bibinfo {author} {\bibfnamefont {X.}~\bibnamefont
  {Hong}}, \bibinfo {author} {\bibfnamefont {M.}~\bibnamefont {Gillig}},
  \bibinfo {author} {\bibfnamefont {R.}~\bibnamefont {Hentrich}}, \bibinfo
  {author} {\bibfnamefont {W.}~\bibnamefont {Yao}}, \bibinfo {author}
  {\bibfnamefont {V.}~\bibnamefont {Kocsis}}, \bibinfo {author} {\bibfnamefont
  {A.~R.}\ \bibnamefont {Witte}}, \bibinfo {author} {\bibfnamefont
  {T.}~\bibnamefont {Schreiner}}, \bibinfo {author} {\bibfnamefont
  {D.}~\bibnamefont {Baumann}}, \bibinfo {author} {\bibfnamefont
  {N.}~\bibnamefont {P{\'e}rez}}, \bibinfo {author} {\bibfnamefont {A.~U.}\
  \bibnamefont {Wolter}}, \emph {et~al.},\ }\bibfield  {title} {\bibinfo
  {title} {Strongly scattered phonon heat transport of the candidate kitaev
  material {N}a$_2${C}o$_2${T}e{O}$_6$},\ }\href@noop {} {\bibfield  {journal}
  {\bibinfo  {journal} {Physical Review B}\ }\textbf {\bibinfo {volume}
  {104}},\ \bibinfo {pages} {144426} (\bibinfo {year} {2021})}\BibitemShut
  {NoStop}%
\bibitem [{\citenamefont {Xiao}\ \emph {et~al.}(2021)\citenamefont {Xiao},
  \citenamefont {Xia}, \citenamefont {Song},\ and\ \citenamefont
  {Xiao}}]{xiao2021magnetic}%
  \BibitemOpen
  \bibfield  {author} {\bibinfo {author} {\bibfnamefont {G.}~\bibnamefont
  {Xiao}}, \bibinfo {author} {\bibfnamefont {Z.}~\bibnamefont {Xia}}, \bibinfo
  {author} {\bibfnamefont {Y.}~\bibnamefont {Song}},\ and\ \bibinfo {author}
  {\bibfnamefont {L.}~\bibnamefont {Xiao}},\ }\bibfield  {title} {\bibinfo
  {title} {Magnetic properties and phase diagram of quasi-two-dimensional
  {N}a$_2${C}o$_2${T}e{O}$_6$ single crystal under high magnetic field},\
  }\href@noop {} {\bibfield  {journal} {\bibinfo  {journal} {Journal of
  Physics: Condensed Matter}\ }\textbf {\bibinfo {volume} {34}},\ \bibinfo
  {pages} {075801} (\bibinfo {year} {2021})}\BibitemShut {NoStop}%
\bibitem [{\citenamefont {Lin}\ \emph {et~al.}(2022)\citenamefont {Lin},
  \citenamefont {Zhao}, \citenamefont {Li}, \citenamefont {Shu}, \citenamefont
  {Ma}, \citenamefont {Jiao}, \citenamefont {Huang}, \citenamefont {Sheng},
  \citenamefont {Kolesnikov}, \citenamefont {Li} \emph
  {et~al.}}]{lin2022evidence}%
  \BibitemOpen
  \bibfield  {author} {\bibinfo {author} {\bibfnamefont {G.}~\bibnamefont
  {Lin}}, \bibinfo {author} {\bibfnamefont {Q.}~\bibnamefont {Zhao}}, \bibinfo
  {author} {\bibfnamefont {G.}~\bibnamefont {Li}}, \bibinfo {author}
  {\bibfnamefont {M.}~\bibnamefont {Shu}}, \bibinfo {author} {\bibfnamefont
  {Y.}~\bibnamefont {Ma}}, \bibinfo {author} {\bibfnamefont {J.}~\bibnamefont
  {Jiao}}, \bibinfo {author} {\bibfnamefont {Q.}~\bibnamefont {Huang}},
  \bibinfo {author} {\bibfnamefont {J.}~\bibnamefont {Sheng}}, \bibinfo
  {author} {\bibfnamefont {A.}~\bibnamefont {Kolesnikov}}, \bibinfo {author}
  {\bibfnamefont {L.}~\bibnamefont {Li}}, \emph {et~al.},\ }\bibfield  {title}
  {\bibinfo {title} {Evidence for field induced quantum spin liquid behavior in
  a spin-1/2 honeycomb magnet},\ }\href@noop {} {\bibfield  {journal} {\bibinfo
   {journal} {Research Square [https://doi.org/10.21203/rs.3.rs-2034295/v1]}\ }
  (\bibinfo {year} {2022})}\BibitemShut {NoStop}%
\bibitem [{\citenamefont {Lefran{\c{c}}ois}\ \emph {et~al.}(2016)\citenamefont
  {Lefran{\c{c}}ois}, \citenamefont {Songvilay}, \citenamefont {Robert},
  \citenamefont {Nataf}, \citenamefont {Jordan}, \citenamefont {Chaix},
  \citenamefont {Colin}, \citenamefont {Lejay}, \citenamefont {Hadj-Azzem},
  \citenamefont {Ballou} \emph {et~al.}}]{lefranccois2016magnetic}%
  \BibitemOpen
  \bibfield  {author} {\bibinfo {author} {\bibfnamefont {E.}~\bibnamefont
  {Lefran{\c{c}}ois}}, \bibinfo {author} {\bibfnamefont {M.}~\bibnamefont
  {Songvilay}}, \bibinfo {author} {\bibfnamefont {J.}~\bibnamefont {Robert}},
  \bibinfo {author} {\bibfnamefont {G.}~\bibnamefont {Nataf}}, \bibinfo
  {author} {\bibfnamefont {E.}~\bibnamefont {Jordan}}, \bibinfo {author}
  {\bibfnamefont {L.}~\bibnamefont {Chaix}}, \bibinfo {author} {\bibfnamefont
  {C.}~\bibnamefont {Colin}}, \bibinfo {author} {\bibfnamefont
  {P.}~\bibnamefont {Lejay}}, \bibinfo {author} {\bibfnamefont
  {A.}~\bibnamefont {Hadj-Azzem}}, \bibinfo {author} {\bibfnamefont
  {R.}~\bibnamefont {Ballou}}, \emph {et~al.},\ }\bibfield  {title} {\bibinfo
  {title} {Magnetic properties of the honeycomb oxide
  {N}a$_2${C}o$_2${T}e{O}$_6$},\ }\href@noop {} {\bibfield  {journal} {\bibinfo
   {journal} {Physical Review B}\ }\textbf {\bibinfo {volume} {94}},\ \bibinfo
  {pages} {214416} (\bibinfo {year} {2016})}\BibitemShut {NoStop}%
\bibitem [{\citenamefont {Johnson}\ \emph {et~al.}(2015)\citenamefont
  {Johnson}, \citenamefont {Williams}, \citenamefont {Haghighirad},
  \citenamefont {Singleton}, \citenamefont {Zapf}, \citenamefont {Manuel},
  \citenamefont {Mazin}, \citenamefont {Li}, \citenamefont {Jeschke},
  \citenamefont {Valent{\'\i}} \emph {et~al.}}]{johnson2015monoclinic}%
  \BibitemOpen
  \bibfield  {author} {\bibinfo {author} {\bibfnamefont {R.~D.}\ \bibnamefont
  {Johnson}}, \bibinfo {author} {\bibfnamefont {S.}~\bibnamefont {Williams}},
  \bibinfo {author} {\bibfnamefont {A.}~\bibnamefont {Haghighirad}}, \bibinfo
  {author} {\bibfnamefont {J.}~\bibnamefont {Singleton}}, \bibinfo {author}
  {\bibfnamefont {V.}~\bibnamefont {Zapf}}, \bibinfo {author} {\bibfnamefont
  {P.}~\bibnamefont {Manuel}}, \bibinfo {author} {\bibfnamefont
  {I.}~\bibnamefont {Mazin}}, \bibinfo {author} {\bibfnamefont
  {Y.}~\bibnamefont {Li}}, \bibinfo {author} {\bibfnamefont {H.~O.}\
  \bibnamefont {Jeschke}}, \bibinfo {author} {\bibfnamefont {R.}~\bibnamefont
  {Valent{\'\i}}}, \emph {et~al.},\ }\bibfield  {title} {\bibinfo {title}
  {Monoclinic crystal structure of $\alpha$- {R}u{C}l$_3$ and the zigzag
  antiferromagnetic ground state},\ }\href@noop {} {\bibfield  {journal}
  {\bibinfo  {journal} {Physical Review B}\ }\textbf {\bibinfo {volume} {92}},\
  \bibinfo {pages} {235119} (\bibinfo {year} {2015})}\BibitemShut {NoStop}%
\bibitem [{\citenamefont {Songvilay}\ \emph {et~al.}(2020)\citenamefont
  {Songvilay}, \citenamefont {Robert}, \citenamefont {Petit}, \citenamefont
  {Rodriguez-Rivera}, \citenamefont {Ratcliff}, \citenamefont {Damay},
  \citenamefont {Bal{\'e}dent}, \citenamefont {Jim{\'e}nez-Ruiz}, \citenamefont
  {Lejay}, \citenamefont {Pachoud} \emph {et~al.}}]{songvilay2020kitaev}%
  \BibitemOpen
  \bibfield  {author} {\bibinfo {author} {\bibfnamefont {M.}~\bibnamefont
  {Songvilay}}, \bibinfo {author} {\bibfnamefont {J.}~\bibnamefont {Robert}},
  \bibinfo {author} {\bibfnamefont {S.}~\bibnamefont {Petit}}, \bibinfo
  {author} {\bibfnamefont {J.}~\bibnamefont {Rodriguez-Rivera}}, \bibinfo
  {author} {\bibfnamefont {W.}~\bibnamefont {Ratcliff}}, \bibinfo {author}
  {\bibfnamefont {F.}~\bibnamefont {Damay}}, \bibinfo {author} {\bibfnamefont
  {V.}~\bibnamefont {Bal{\'e}dent}}, \bibinfo {author} {\bibfnamefont
  {M.}~\bibnamefont {Jim{\'e}nez-Ruiz}}, \bibinfo {author} {\bibfnamefont
  {P.}~\bibnamefont {Lejay}}, \bibinfo {author} {\bibfnamefont
  {E.}~\bibnamefont {Pachoud}}, \emph {et~al.},\ }\bibfield  {title} {\bibinfo
  {title} {Kitaev interactions in the {C}o honeycomb antiferromagnets
  {N}a$_3${C}o$_2${S}b{O}$_6$ and {N}a$_2${C}o$_2${T}e{O}$_6$},\ }\href@noop {}
  {\bibfield  {journal} {\bibinfo  {journal} {Physical Review B}\ }\textbf
  {\bibinfo {volume} {102}},\ \bibinfo {pages} {224429} (\bibinfo {year}
  {2020})}\BibitemShut {NoStop}%
\bibitem [{\citenamefont {Samarakoon}\ \emph {et~al.}(2021)\citenamefont
  {Samarakoon}, \citenamefont {Chen}, \citenamefont {Zhou},\ and\ \citenamefont
  {Garlea}}]{samarakoon2021static}%
  \BibitemOpen
  \bibfield  {author} {\bibinfo {author} {\bibfnamefont {A.~M.}\ \bibnamefont
  {Samarakoon}}, \bibinfo {author} {\bibfnamefont {Q.}~\bibnamefont {Chen}},
  \bibinfo {author} {\bibfnamefont {H.}~\bibnamefont {Zhou}},\ and\ \bibinfo
  {author} {\bibfnamefont {V.~O.}\ \bibnamefont {Garlea}},\ }\bibfield  {title}
  {\bibinfo {title} {Static and dynamic magnetic properties of honeycomb
  lattice antiferromagnets {N}a$_2${M}$_2${T}e{O}$_6$, {M} = {C}o and {N}i},\
  }\href@noop {} {\bibfield  {journal} {\bibinfo  {journal} {Physical Review
  B}\ }\textbf {\bibinfo {volume} {104}},\ \bibinfo {pages} {184415} (\bibinfo
  {year} {2021})}\BibitemShut {NoStop}%
\bibitem [{\citenamefont {Chen}\ \emph {et~al.}(2021)\citenamefont {Chen},
  \citenamefont {Li}, \citenamefont {Hu}, \citenamefont {Hu}, \citenamefont
  {Yue}, \citenamefont {Sutarto}, \citenamefont {He}, \citenamefont {Iida},
  \citenamefont {Kamazawa}, \citenamefont {Yu} \emph {et~al.}}]{chen2021spin}%
  \BibitemOpen
  \bibfield  {author} {\bibinfo {author} {\bibfnamefont {W.}~\bibnamefont
  {Chen}}, \bibinfo {author} {\bibfnamefont {X.}~\bibnamefont {Li}}, \bibinfo
  {author} {\bibfnamefont {Z.}~\bibnamefont {Hu}}, \bibinfo {author}
  {\bibfnamefont {Z.}~\bibnamefont {Hu}}, \bibinfo {author} {\bibfnamefont
  {L.}~\bibnamefont {Yue}}, \bibinfo {author} {\bibfnamefont {R.}~\bibnamefont
  {Sutarto}}, \bibinfo {author} {\bibfnamefont {F.}~\bibnamefont {He}},
  \bibinfo {author} {\bibfnamefont {K.}~\bibnamefont {Iida}}, \bibinfo {author}
  {\bibfnamefont {K.}~\bibnamefont {Kamazawa}}, \bibinfo {author}
  {\bibfnamefont {W.}~\bibnamefont {Yu}}, \emph {et~al.},\ }\bibfield  {title}
  {\bibinfo {title} {Spin-orbit phase behavior of {N}a$_2${C}o$_2${T}e{O}$_6$
  at low temperatures},\ }\href@noop {} {\bibfield  {journal} {\bibinfo
  {journal} {Physical Review B}\ }\textbf {\bibinfo {volume} {103}},\ \bibinfo
  {pages} {L180404} (\bibinfo {year} {2021})}\BibitemShut {NoStop}%
\bibitem [{\citenamefont {Lee}\ \emph {et~al.}(2021)\citenamefont {Lee},
  \citenamefont {Lee}, \citenamefont {Choi}, \citenamefont {Jang},
  \citenamefont {Kalaivanan}, \citenamefont {Sankar},\ and\ \citenamefont
  {Choi}}]{lee2021multistage}%
  \BibitemOpen
  \bibfield  {author} {\bibinfo {author} {\bibfnamefont {C.}~\bibnamefont
  {Lee}}, \bibinfo {author} {\bibfnamefont {S.}~\bibnamefont {Lee}}, \bibinfo
  {author} {\bibfnamefont {Y.}~\bibnamefont {Choi}}, \bibinfo {author}
  {\bibfnamefont {Z.}~\bibnamefont {Jang}}, \bibinfo {author} {\bibfnamefont
  {R.}~\bibnamefont {Kalaivanan}}, \bibinfo {author} {\bibfnamefont
  {R.}~\bibnamefont {Sankar}},\ and\ \bibinfo {author} {\bibfnamefont {K.-Y.}\
  \bibnamefont {Choi}},\ }\bibfield  {title} {\bibinfo {title} {Multistage
  development of anisotropic magnetic correlations in the {C}o-based honeycomb
  lattice {N}a$_2${C}o$_2${T}e{O}$_6$},\ }\href@noop {} {\bibfield  {journal}
  {\bibinfo  {journal} {Physical Review B}\ }\textbf {\bibinfo {volume}
  {103}},\ \bibinfo {pages} {214447} (\bibinfo {year} {2021})}\BibitemShut
  {NoStop}%
\bibitem [{\citenamefont {Yao}\ and\ \citenamefont
  {Li}(2020)}]{yao2020ferrimagnetism}%
  \BibitemOpen
  \bibfield  {author} {\bibinfo {author} {\bibfnamefont {W.}~\bibnamefont
  {Yao}}\ and\ \bibinfo {author} {\bibfnamefont {Y.}~\bibnamefont {Li}},\
  }\bibfield  {title} {\bibinfo {title} {Ferrimagnetism and anisotropic phase
  tunability by magnetic fields in {N}a$_2${C}o$_2${T}e{O}$_6$},\ }\href@noop
  {} {\bibfield  {journal} {\bibinfo  {journal} {Physical Review B}\ }\textbf
  {\bibinfo {volume} {101}},\ \bibinfo {pages} {085120} (\bibinfo {year}
  {2020})}\BibitemShut {NoStop}%
\bibitem [{\citenamefont {Zhang}\ \emph {et~al.}(2022)\citenamefont {Zhang},
  \citenamefont {Lee}, \citenamefont {Lee}, \citenamefont {Woods},
  \citenamefont {Thomas}, \citenamefont {Movshovich}, \citenamefont {Brosha},
  \citenamefont {Huang}, \citenamefont {Zhou},\ and\ \citenamefont
  {Zapf}}]{supp}%
  \BibitemOpen
  \bibfield  {author} {\bibinfo {author} {\bibfnamefont {S.}~\bibnamefont
  {Zhang}}, \bibinfo {author} {\bibfnamefont {M.}~\bibnamefont {Lee}}, \bibinfo
  {author} {\bibfnamefont {S.}~\bibnamefont {Lee}}, \bibinfo {author}
  {\bibfnamefont {A.~J.}\ \bibnamefont {Woods}}, \bibinfo {author}
  {\bibfnamefont {S.~M.}\ \bibnamefont {Thomas}}, \bibinfo {author}
  {\bibfnamefont {R.}~\bibnamefont {Movshovich}}, \bibinfo {author}
  {\bibfnamefont {E.}~\bibnamefont {Brosha}}, \bibinfo {author} {\bibfnamefont
  {Q.}~\bibnamefont {Huang}}, \bibinfo {author} {\bibfnamefont
  {H.}~\bibnamefont {Zhou}},\ and\ \bibinfo {author} {\bibfnamefont
  {V.}~\bibnamefont {Zapf}},\ }\href@noop {} {\bibinfo {title} {Supplemental
  material of electronic and magnetic phase diagrams of kitaev quantum spin
  liquid candidate {N}a$_2${C}o$_2${T}e{O}$_6$}},\ \bibinfo {howpublished}
  {\url{URL_will_be_inserted_by_publisher}} (\bibinfo {year}
  {2022})\BibitemShut {NoStop}%
\bibitem [{\citenamefont {Zapf}\ \emph {et~al.}(2010)\citenamefont {Zapf},
  \citenamefont {Kenzelmann}, \citenamefont {Wolff-Fabris}, \citenamefont
  {Balakirev},\ and\ \citenamefont {Chen}}]{Zapf2010}%
  \BibitemOpen
  \bibfield  {author} {\bibinfo {author} {\bibfnamefont {V.~S.}\ \bibnamefont
  {Zapf}}, \bibinfo {author} {\bibfnamefont {M.}~\bibnamefont {Kenzelmann}},
  \bibinfo {author} {\bibfnamefont {F.}~\bibnamefont {Wolff-Fabris}}, \bibinfo
  {author} {\bibfnamefont {F.}~\bibnamefont {Balakirev}},\ and\ \bibinfo
  {author} {\bibfnamefont {Y.}~\bibnamefont {Chen}},\ }\bibfield  {title}
  {\bibinfo {title} {Magnetically induced electric polarization in an
  organometallic magnet},\ }\href {https://doi.org/10.1103/PhysRevB.82.060402}
  {\bibfield  {journal} {\bibinfo  {journal} {Phys. Rev. B}\ }\textbf {\bibinfo
  {volume} {82}},\ \bibinfo {pages} {060402} (\bibinfo {year}
  {2010})}\BibitemShut {NoStop}%
\bibitem [{\citenamefont {Chai}\ \emph {et~al.}(2021)\citenamefont {Chai},
  \citenamefont {Cong}, \citenamefont {He}, \citenamefont {Su}, \citenamefont
  {Ding}, \citenamefont {Singleton}, \citenamefont {Zapf},\ and\ \citenamefont
  {Sun}}]{Zapf2021}%
  \BibitemOpen
  \bibfield  {author} {\bibinfo {author} {\bibfnamefont {Y.-S.}\ \bibnamefont
  {Chai}}, \bibinfo {author} {\bibfnamefont {J.-Z.}\ \bibnamefont {Cong}},
  \bibinfo {author} {\bibfnamefont {J.-C.}\ \bibnamefont {He}}, \bibinfo
  {author} {\bibfnamefont {D.}~\bibnamefont {Su}}, \bibinfo {author}
  {\bibfnamefont {X.-X.}\ \bibnamefont {Ding}}, \bibinfo {author}
  {\bibfnamefont {J.}~\bibnamefont {Singleton}}, \bibinfo {author}
  {\bibfnamefont {V.}~\bibnamefont {Zapf}},\ and\ \bibinfo {author}
  {\bibfnamefont {Y.}~\bibnamefont {Sun}},\ }\bibfield  {title} {\bibinfo
  {title} {Giant magnetostriction and nonsaturating electric polarization up to
  60 {T} in the polar magnet {C}a{B}a{C}o$_{4}${O}$_{7}$},\ }\href
  {https://doi.org/10.1103/PhysRevB.103.174433} {\bibfield  {journal} {\bibinfo
   {journal} {Phys. Rev. B}\ }\textbf {\bibinfo {volume} {103}},\ \bibinfo
  {pages} {174433} (\bibinfo {year} {2021})}\BibitemShut {NoStop}%
\bibitem [{\citenamefont {Jaime}\ \emph {et~al.}(2017)\citenamefont {Jaime},
  \citenamefont {Corval{\'a}n~Moya}, \citenamefont {Weickert}, \citenamefont
  {Zapf}, \citenamefont {Balakirev}, \citenamefont {Wartenbe}, \citenamefont
  {Rosa}, \citenamefont {Betts}, \citenamefont {Rodriguez}, \citenamefont
  {Crooker} \emph {et~al.}}]{jaime2017fiber}%
  \BibitemOpen
  \bibfield  {author} {\bibinfo {author} {\bibfnamefont {M.}~\bibnamefont
  {Jaime}}, \bibinfo {author} {\bibfnamefont {C.}~\bibnamefont
  {Corval{\'a}n~Moya}}, \bibinfo {author} {\bibfnamefont {F.}~\bibnamefont
  {Weickert}}, \bibinfo {author} {\bibfnamefont {V.}~\bibnamefont {Zapf}},
  \bibinfo {author} {\bibfnamefont {F.~F.}\ \bibnamefont {Balakirev}}, \bibinfo
  {author} {\bibfnamefont {M.}~\bibnamefont {Wartenbe}}, \bibinfo {author}
  {\bibfnamefont {P.~F.}\ \bibnamefont {Rosa}}, \bibinfo {author}
  {\bibfnamefont {J.~B.}\ \bibnamefont {Betts}}, \bibinfo {author}
  {\bibfnamefont {G.}~\bibnamefont {Rodriguez}}, \bibinfo {author}
  {\bibfnamefont {S.~A.}\ \bibnamefont {Crooker}}, \emph {et~al.},\ }\bibfield
  {title} {\bibinfo {title} {Fiber {B}ragg grating dilatometry in extreme
  magnetic field and cryogenic conditions},\ }\href@noop {} {\bibfield
  {journal} {\bibinfo  {journal} {Sensors}\ }\textbf {\bibinfo {volume} {17}},\
  \bibinfo {pages} {2572} (\bibinfo {year} {2017})}\BibitemShut {NoStop}%
\bibitem [{\citenamefont {Fiebig}(2005)}]{Fiebig2005}%
  \BibitemOpen
  \bibfield  {author} {\bibinfo {author} {\bibfnamefont {M.}~\bibnamefont
  {Fiebig}},\ }\bibfield  {title} {\bibinfo {title} {Revival of the
  magnetoelectric effect},\ }\href@noop {} {\bibfield  {journal} {\bibinfo
  {journal} {Journal of physics D: applied physics}\ }\textbf {\bibinfo
  {volume} {38}},\ \bibinfo {pages} {R123} (\bibinfo {year}
  {2005})}\BibitemShut {NoStop}%
\bibitem [{\citenamefont {Agyei}\ and\ \citenamefont
  {Birman}(1990)}]{agyei1990linear}%
  \BibitemOpen
  \bibfield  {author} {\bibinfo {author} {\bibfnamefont {A.}~\bibnamefont
  {Agyei}}\ and\ \bibinfo {author} {\bibfnamefont {J.~L.}\ \bibnamefont
  {Birman}},\ }\bibfield  {title} {\bibinfo {title} {On the linear
  magnetoelectric effect},\ }\href@noop {} {\bibfield  {journal} {\bibinfo
  {journal} {Journal of Physics: Condensed Matter}\ }\textbf {\bibinfo {volume}
  {2}},\ \bibinfo {pages} {3007} (\bibinfo {year} {1990})}\BibitemShut
  {NoStop}%
\bibitem [{\citenamefont {Rivera}(1994)}]{rivera1994linear}%
  \BibitemOpen
  \bibfield  {author} {\bibinfo {author} {\bibfnamefont {J.-P.}\ \bibnamefont
  {Rivera}},\ }\bibfield  {title} {\bibinfo {title} {The linear magnetoelectric
  effect in licopo4revisited},\ }\href@noop {} {\bibfield  {journal} {\bibinfo
  {journal} {Ferroelectrics}\ }\textbf {\bibinfo {volume} {161}},\ \bibinfo
  {pages} {147} (\bibinfo {year} {1994})}\BibitemShut {NoStop}%
\bibitem [{\citenamefont {Spaldin}\ and\ \citenamefont
  {Ramesh}(2019)}]{Spaldin19}%
  \BibitemOpen
  \bibfield  {author} {\bibinfo {author} {\bibfnamefont {N.~A.}\ \bibnamefont
  {Spaldin}}\ and\ \bibinfo {author} {\bibfnamefont {R.}~\bibnamefont
  {Ramesh}},\ }\bibfield  {title} {\bibinfo {title} {Advances in
  magnetoelectric multiferroics},\ }\href@noop {} {\bibfield  {journal}
  {\bibinfo  {journal} {Nature materials}\ }\textbf {\bibinfo {volume} {18}},\
  \bibinfo {pages} {203} (\bibinfo {year} {2019})}\BibitemShut {NoStop}%
\bibitem [{\citenamefont {Mukherjee}\ \emph {et~al.}(2022)\citenamefont
  {Mukherjee}, \citenamefont {Manna}, \citenamefont {Saha}, \citenamefont
  {Majumdar},\ and\ \citenamefont {Giri}}]{mukherjee2022ferroelectric}%
  \BibitemOpen
  \bibfield  {author} {\bibinfo {author} {\bibfnamefont {S.}~\bibnamefont
  {Mukherjee}}, \bibinfo {author} {\bibfnamefont {G.}~\bibnamefont {Manna}},
  \bibinfo {author} {\bibfnamefont {P.}~\bibnamefont {Saha}}, \bibinfo {author}
  {\bibfnamefont {S.}~\bibnamefont {Majumdar}},\ and\ \bibinfo {author}
  {\bibfnamefont {S.}~\bibnamefont {Giri}},\ }\bibfield  {title} {\bibinfo
  {title} {Ferroelectric order with a linear high-field magnetoelectric
  coupling in {N}a$_2${C}o$_2${T}e{O}$_6$: A proposed kitaev compound},\
  }\href@noop {} {\bibfield  {journal} {\bibinfo  {journal} {Physical Review
  Materials}\ }\textbf {\bibinfo {volume} {6}},\ \bibinfo {pages} {054407}
  (\bibinfo {year} {2022})}\BibitemShut {NoStop}%
\bibitem [{\citenamefont {Xiao}\ \emph {et~al.}(2019)\citenamefont {Xiao},
  \citenamefont {Xia}, \citenamefont {Zhang}, \citenamefont {Yue},
  \citenamefont {Huang}, \citenamefont {Zhang}, \citenamefont {Yang},
  \citenamefont {Song}, \citenamefont {Wei}, \citenamefont {Deng} \emph
  {et~al.}}]{xiao2019crystal}%
  \BibitemOpen
  \bibfield  {author} {\bibinfo {author} {\bibfnamefont {G.}~\bibnamefont
  {Xiao}}, \bibinfo {author} {\bibfnamefont {Z.}~\bibnamefont {Xia}}, \bibinfo
  {author} {\bibfnamefont {W.}~\bibnamefont {Zhang}}, \bibinfo {author}
  {\bibfnamefont {X.}~\bibnamefont {Yue}}, \bibinfo {author} {\bibfnamefont
  {S.}~\bibnamefont {Huang}}, \bibinfo {author} {\bibfnamefont
  {X.}~\bibnamefont {Zhang}}, \bibinfo {author} {\bibfnamefont
  {F.}~\bibnamefont {Yang}}, \bibinfo {author} {\bibfnamefont {Y.}~\bibnamefont
  {Song}}, \bibinfo {author} {\bibfnamefont {M.}~\bibnamefont {Wei}}, \bibinfo
  {author} {\bibfnamefont {H.}~\bibnamefont {Deng}}, \emph {et~al.},\
  }\bibfield  {title} {\bibinfo {title} {Crystal growth and the magnetic
  properties of {N}a$_2${C}o$_2${T}e{O}$_6$ with quasi-two-dimensional
  honeycomb lattice},\ }\href@noop {} {\bibfield  {journal} {\bibinfo
  {journal} {Crystal Growth \& Design}\ }\textbf {\bibinfo {volume} {19}},\
  \bibinfo {pages} {2658} (\bibinfo {year} {2019})}\BibitemShut {NoStop}%
\bibitem [{\citenamefont {Spaldin}\ \emph {et~al.}(2008)\citenamefont
  {Spaldin}, \citenamefont {Fiebig},\ and\ \citenamefont
  {Mostovoy}}]{Spaldin2008toroidal}%
  \BibitemOpen
  \bibfield  {author} {\bibinfo {author} {\bibfnamefont {N.~A.}\ \bibnamefont
  {Spaldin}}, \bibinfo {author} {\bibfnamefont {M.}~\bibnamefont {Fiebig}},\
  and\ \bibinfo {author} {\bibfnamefont {M.}~\bibnamefont {Mostovoy}},\
  }\bibfield  {title} {\bibinfo {title} {The toroidal moment in
  condensed-matter physics and its relation to the magnetoelectric effect},\
  }\href {https://doi.org/10.1088/0953-8984/20/43/434203} {\bibfield  {journal}
  {\bibinfo  {journal} {Journal of Physics: Condensed Matter}\ }\textbf
  {\bibinfo {volume} {20}},\ \bibinfo {pages} {434203} (\bibinfo {year}
  {2008})}\BibitemShut {NoStop}%
\bibitem [{\citenamefont {Yao}\ \emph {et~al.}(2023)\citenamefont {Yao},
  \citenamefont {Zhao}, \citenamefont {Qiu}, \citenamefont {Balz},
  \citenamefont {Stewart}, \citenamefont {Lynn},\ and\ \citenamefont
  {Li}}]{yao2022magnetic}%
  \BibitemOpen
  \bibfield  {author} {\bibinfo {author} {\bibfnamefont {W.}~\bibnamefont
  {Yao}}, \bibinfo {author} {\bibfnamefont {Y.}~\bibnamefont {Zhao}}, \bibinfo
  {author} {\bibfnamefont {Y.}~\bibnamefont {Qiu}}, \bibinfo {author}
  {\bibfnamefont {C.}~\bibnamefont {Balz}}, \bibinfo {author} {\bibfnamefont
  {J.~R.}\ \bibnamefont {Stewart}}, \bibinfo {author} {\bibfnamefont {J.~W.}\
  \bibnamefont {Lynn}},\ and\ \bibinfo {author} {\bibfnamefont
  {Y.}~\bibnamefont {Li}},\ }\bibfield  {title} {\bibinfo {title} {Magnetic
  ground state of the kitaev {N}a$_{2}${C}o$_{2}${T}e{O}$_{6}$ spin liquid
  candidate},\ }\href@noop {} {\bibfield  {journal} {\bibinfo  {journal}
  {Physical Review Research}\ }\textbf {\bibinfo {volume} {5}},\ \bibinfo
  {pages} {L022045} (\bibinfo {year} {2023})}\BibitemShut {NoStop}%
\bibitem [{\citenamefont {Lee}\ \emph {et~al.}(2014{\natexlab{a}})\citenamefont
  {Lee}, \citenamefont {Hwang}, \citenamefont {Choi}, \citenamefont {Ma},
  \citenamefont {Dela~Cruz}, \citenamefont {Zhu}, \citenamefont {Ke},
  \citenamefont {Dun},\ and\ \citenamefont {Zhou}}]{Lee2014VV}%
  \BibitemOpen
  \bibfield  {author} {\bibinfo {author} {\bibfnamefont {M.}~\bibnamefont
  {Lee}}, \bibinfo {author} {\bibfnamefont {J.}~\bibnamefont {Hwang}}, \bibinfo
  {author} {\bibfnamefont {E.~S.}\ \bibnamefont {Choi}}, \bibinfo {author}
  {\bibfnamefont {J.}~\bibnamefont {Ma}}, \bibinfo {author} {\bibfnamefont
  {C.~R.}\ \bibnamefont {Dela~Cruz}}, \bibinfo {author} {\bibfnamefont
  {M.}~\bibnamefont {Zhu}}, \bibinfo {author} {\bibfnamefont {X.}~\bibnamefont
  {Ke}}, \bibinfo {author} {\bibfnamefont {Z.~L.}\ \bibnamefont {Dun}},\ and\
  \bibinfo {author} {\bibfnamefont {H.~D.}\ \bibnamefont {Zhou}},\ }\bibfield
  {title} {\bibinfo {title} {Series of phase transitions and multiferroicity in
  the quasi-two-dimensional spin-$\frac{1}{2}$ triangular-lattice
  antiferromagnet {B}a$_3${C}o{N}b$_2${O}$_9$},\ }\href
  {https://doi.org/10.1103/PhysRevB.89.104420} {\bibfield  {journal} {\bibinfo
  {journal} {Phys. Rev. B}\ }\textbf {\bibinfo {volume} {89}},\ \bibinfo
  {pages} {104420} (\bibinfo {year} {2014}{\natexlab{a}})}\BibitemShut
  {NoStop}%
\bibitem [{\citenamefont {Mi}\ \emph {et~al.}(2021)\citenamefont {Mi},
  \citenamefont {Wang}, \citenamefont {Gui}, \citenamefont {Pi}, \citenamefont
  {Zheng}, \citenamefont {Yang}, \citenamefont {Gan}, \citenamefont {Wang},
  \citenamefont {Li}, \citenamefont {Wang} \emph {et~al.}}]{mi2021stacking}%
  \BibitemOpen
  \bibfield  {author} {\bibinfo {author} {\bibfnamefont {X.}~\bibnamefont
  {Mi}}, \bibinfo {author} {\bibfnamefont {X.}~\bibnamefont {Wang}}, \bibinfo
  {author} {\bibfnamefont {H.}~\bibnamefont {Gui}}, \bibinfo {author}
  {\bibfnamefont {M.}~\bibnamefont {Pi}}, \bibinfo {author} {\bibfnamefont
  {T.}~\bibnamefont {Zheng}}, \bibinfo {author} {\bibfnamefont
  {K.}~\bibnamefont {Yang}}, \bibinfo {author} {\bibfnamefont {Y.}~\bibnamefont
  {Gan}}, \bibinfo {author} {\bibfnamefont {P.}~\bibnamefont {Wang}}, \bibinfo
  {author} {\bibfnamefont {A.}~\bibnamefont {Li}}, \bibinfo {author}
  {\bibfnamefont {A.}~\bibnamefont {Wang}}, \emph {et~al.},\ }\bibfield
  {title} {\bibinfo {title} {Stacking faults in $\alpha$- {R}u{C}l$_3$ revealed
  by local electric polarization},\ }\href@noop {} {\bibfield  {journal}
  {\bibinfo  {journal} {Physical Review B}\ }\textbf {\bibinfo {volume}
  {103}},\ \bibinfo {pages} {174413} (\bibinfo {year} {2021})}\BibitemShut
  {NoStop}%
\bibitem [{\citenamefont {Zheng}\ \emph {et~al.}(2018)\citenamefont {Zheng},
  \citenamefont {Cui}, \citenamefont {Li}, \citenamefont {Ran}, \citenamefont
  {Wen},\ and\ \citenamefont {Yu}}]{zheng2018dielectric}%
  \BibitemOpen
  \bibfield  {author} {\bibinfo {author} {\bibfnamefont {J.-C.}\ \bibnamefont
  {Zheng}}, \bibinfo {author} {\bibfnamefont {Y.}~\bibnamefont {Cui}}, \bibinfo
  {author} {\bibfnamefont {T.-R.}\ \bibnamefont {Li}}, \bibinfo {author}
  {\bibfnamefont {K.-J.}\ \bibnamefont {Ran}}, \bibinfo {author} {\bibfnamefont
  {J.}~\bibnamefont {Wen}},\ and\ \bibinfo {author} {\bibfnamefont
  {W.}~\bibnamefont {Yu}},\ }\bibfield  {title} {\bibinfo {title} {Dielectric
  evidence for possible type-{II} multiferroicity in $\alpha$-{R}u{C}l$_3$},\
  }\href {https://doi.org/https://doi.org/10.1007/s11433-017-9166-1} {\bibfield
   {journal} {\bibinfo  {journal} {Science China Physics, Mechanics and
  Astronomy}\ }\textbf {\bibinfo {volume} {61}},\ \bibinfo {pages} {057021}
  (\bibinfo {year} {2018})}\BibitemShut {NoStop}%
\bibitem [{\citenamefont {Bachus}\ \emph {et~al.}(2020)\citenamefont {Bachus},
  \citenamefont {Kaib}, \citenamefont {Tokiwa}, \citenamefont {Jesche},
  \citenamefont {Tsurkan}, \citenamefont {Loidl}, \citenamefont {Winter},
  \citenamefont {Tsirlin}, \citenamefont {Valenti},\ and\ \citenamefont
  {Gegenwart}}]{bachus2020thermodynamic}%
  \BibitemOpen
  \bibfield  {author} {\bibinfo {author} {\bibfnamefont {S.}~\bibnamefont
  {Bachus}}, \bibinfo {author} {\bibfnamefont {D.~A.}\ \bibnamefont {Kaib}},
  \bibinfo {author} {\bibfnamefont {Y.}~\bibnamefont {Tokiwa}}, \bibinfo
  {author} {\bibfnamefont {A.}~\bibnamefont {Jesche}}, \bibinfo {author}
  {\bibfnamefont {V.}~\bibnamefont {Tsurkan}}, \bibinfo {author} {\bibfnamefont
  {A.}~\bibnamefont {Loidl}}, \bibinfo {author} {\bibfnamefont {S.~M.}\
  \bibnamefont {Winter}}, \bibinfo {author} {\bibfnamefont {A.~A.}\
  \bibnamefont {Tsirlin}}, \bibinfo {author} {\bibfnamefont {R.}~\bibnamefont
  {Valenti}},\ and\ \bibinfo {author} {\bibfnamefont {P.}~\bibnamefont
  {Gegenwart}},\ }\bibfield  {title} {\bibinfo {title} {Thermodynamic
  perspective on field-induced behavior of $\alpha$-{R}u{C}l$_3$},\ }\href@noop
  {} {\bibfield  {journal} {\bibinfo  {journal} {Physical Review Letters}\
  }\textbf {\bibinfo {volume} {125}},\ \bibinfo {pages} {097203} (\bibinfo
  {year} {2020})}\BibitemShut {NoStop}%
\bibitem [{\citenamefont {Sch{\"o}nemann}\ \emph {et~al.}(2020)\citenamefont
  {Sch{\"o}nemann}, \citenamefont {Imajo}, \citenamefont {Weickert},
  \citenamefont {Yan}, \citenamefont {Mandrus}, \citenamefont {Takano},
  \citenamefont {Brosha}, \citenamefont {Rosa}, \citenamefont {Nagler},
  \citenamefont {Kindo} \emph {et~al.}}]{schonemann2020thermal}%
  \BibitemOpen
  \bibfield  {author} {\bibinfo {author} {\bibfnamefont {R.}~\bibnamefont
  {Sch{\"o}nemann}}, \bibinfo {author} {\bibfnamefont {S.}~\bibnamefont
  {Imajo}}, \bibinfo {author} {\bibfnamefont {F.}~\bibnamefont {Weickert}},
  \bibinfo {author} {\bibfnamefont {J.}~\bibnamefont {Yan}}, \bibinfo {author}
  {\bibfnamefont {D.~G.}\ \bibnamefont {Mandrus}}, \bibinfo {author}
  {\bibfnamefont {Y.}~\bibnamefont {Takano}}, \bibinfo {author} {\bibfnamefont
  {E.~L.}\ \bibnamefont {Brosha}}, \bibinfo {author} {\bibfnamefont {P.~F.}\
  \bibnamefont {Rosa}}, \bibinfo {author} {\bibfnamefont {S.~E.}\ \bibnamefont
  {Nagler}}, \bibinfo {author} {\bibfnamefont {K.}~\bibnamefont {Kindo}}, \emph
  {et~al.},\ }\bibfield  {title} {\bibinfo {title} {Thermal and magnetoelastic
  properties of $\alpha$- {R}u{C}l$_3$ in the field-induced low-temperature
  states},\ }\href@noop {} {\bibfield  {journal} {\bibinfo  {journal} {Physical
  Review B}\ }\textbf {\bibinfo {volume} {102}},\ \bibinfo {pages} {214432}
  (\bibinfo {year} {2020})}\BibitemShut {NoStop}%
\bibitem [{\citenamefont {Lee}\ \emph {et~al.}(2014{\natexlab{b}})\citenamefont
  {Lee}, \citenamefont {Choi}, \citenamefont {Huang}, \citenamefont {Ma},
  \citenamefont {Dela~Cruz}, \citenamefont {Matsuda}, \citenamefont {Tian},
  \citenamefont {Dun}, \citenamefont {Dong},\ and\ \citenamefont
  {Zhou}}]{2014LeeVVV}%
  \BibitemOpen
  \bibfield  {author} {\bibinfo {author} {\bibfnamefont {M.}~\bibnamefont
  {Lee}}, \bibinfo {author} {\bibfnamefont {E.~S.}\ \bibnamefont {Choi}},
  \bibinfo {author} {\bibfnamefont {X.}~\bibnamefont {Huang}}, \bibinfo
  {author} {\bibfnamefont {J.}~\bibnamefont {Ma}}, \bibinfo {author}
  {\bibfnamefont {C.~R.}\ \bibnamefont {Dela~Cruz}}, \bibinfo {author}
  {\bibfnamefont {M.}~\bibnamefont {Matsuda}}, \bibinfo {author} {\bibfnamefont
  {W.}~\bibnamefont {Tian}}, \bibinfo {author} {\bibfnamefont {Z.~L.}\
  \bibnamefont {Dun}}, \bibinfo {author} {\bibfnamefont {S.}~\bibnamefont
  {Dong}},\ and\ \bibinfo {author} {\bibfnamefont {H.~D.}\ \bibnamefont
  {Zhou}},\ }\bibfield  {title} {\bibinfo {title} {Magnetic phase diagram and
  multiferroicity of {B}a$_{3}${M}n{N}b$_{2}${O}$_{9}$: A spin-$\frac{5}{2}$
  triangular lattice antiferromagnet with weak easy-axis anisotropy},\ }\href
  {https://doi.org/10.1103/PhysRevB.90.224402} {\bibfield  {journal} {\bibinfo
  {journal} {Phys. Rev. B}\ }\textbf {\bibinfo {volume} {90}},\ \bibinfo
  {pages} {224402} (\bibinfo {year} {2014}{\natexlab{b}})}\BibitemShut
  {NoStop}%
\bibitem [{\citenamefont {Sears}\ \emph {et~al.}(2017)\citenamefont {Sears},
  \citenamefont {Zhao}, \citenamefont {Xu}, \citenamefont {Lynn},\ and\
  \citenamefont {Kim}}]{sears2017phase}%
  \BibitemOpen
  \bibfield  {author} {\bibinfo {author} {\bibfnamefont {J.~A.}\ \bibnamefont
  {Sears}}, \bibinfo {author} {\bibfnamefont {Y.}~\bibnamefont {Zhao}},
  \bibinfo {author} {\bibfnamefont {Z.}~\bibnamefont {Xu}}, \bibinfo {author}
  {\bibfnamefont {J.~W.}\ \bibnamefont {Lynn}},\ and\ \bibinfo {author}
  {\bibfnamefont {Y.-J.}\ \bibnamefont {Kim}},\ }\bibfield  {title} {\bibinfo
  {title} {Phase diagram of $\alpha$-{R}u{C}l$_3$ in an in-plane magnetic
  field},\ }\href@noop {} {\bibfield  {journal} {\bibinfo  {journal} {Physical
  Review B}\ }\textbf {\bibinfo {volume} {95}},\ \bibinfo {pages} {180411}
  (\bibinfo {year} {2017})}\BibitemShut {NoStop}%
\bibitem [{\citenamefont {Huang}\ \emph {et~al.}(2022)\citenamefont {Huang},
  \citenamefont {Lee}, \citenamefont {sang Choi}, \citenamefont {Ma},
  \citenamefont {Cruz},\ and\ \citenamefont {Zhou}}]{huang2022successive}%
  \BibitemOpen
  \bibfield  {author} {\bibinfo {author} {\bibfnamefont {Q.}~\bibnamefont
  {Huang}}, \bibinfo {author} {\bibfnamefont {M.}~\bibnamefont {Lee}}, \bibinfo
  {author} {\bibfnamefont {E.}~\bibnamefont {sang Choi}}, \bibinfo {author}
  {\bibfnamefont {J.}~\bibnamefont {Ma}}, \bibinfo {author} {\bibfnamefont
  {C.~D.}\ \bibnamefont {Cruz}},\ and\ \bibinfo {author} {\bibfnamefont
  {H.}~\bibnamefont {Zhou}},\ }\bibfield  {title} {\bibinfo {title} {Successive
  phase transitions and multiferroicity in deformed triangular-lattice
  antiferromagnets {C}a$_3${MN}b$_2${O}$_9$ ({M}= {C}o, {N}i) with spatial
  anisotropy},\ }\href@noop {} {\bibfield  {journal} {\bibinfo  {journal} {ECS
  Journal of Solid State Science and Technology}\ } (\bibinfo {year}
  {2022})}\BibitemShut {NoStop}%
\bibitem [{\citenamefont {Chern}\ \emph {et~al.}(2017)\citenamefont {Chern},
  \citenamefont {Sizyuk}, \citenamefont {Price},\ and\ \citenamefont
  {Perkins}}]{chern2017kitaev}%
  \BibitemOpen
  \bibfield  {author} {\bibinfo {author} {\bibfnamefont {G.-W.}\ \bibnamefont
  {Chern}}, \bibinfo {author} {\bibfnamefont {Y.}~\bibnamefont {Sizyuk}},
  \bibinfo {author} {\bibfnamefont {C.}~\bibnamefont {Price}},\ and\ \bibinfo
  {author} {\bibfnamefont {N.~B.}\ \bibnamefont {Perkins}},\ }\bibfield
  {title} {\bibinfo {title} {Kitaev-heisenberg model in a magnetic field:
  order-by-disorder and commensurate-incommensurate transitions},\ }\href@noop
  {} {\bibfield  {journal} {\bibinfo  {journal} {Physical Review B}\ }\textbf
  {\bibinfo {volume} {95}},\ \bibinfo {pages} {144427} (\bibinfo {year}
  {2017})}\BibitemShut {NoStop}%
\bibitem [{\citenamefont {Janssen}\ \emph {et~al.}(2016)\citenamefont
  {Janssen}, \citenamefont {Andrade},\ and\ \citenamefont
  {Vojta}}]{Janssen2016fieldPD}%
  \BibitemOpen
  \bibfield  {author} {\bibinfo {author} {\bibfnamefont {L.}~\bibnamefont
  {Janssen}}, \bibinfo {author} {\bibfnamefont {E.~C.}\ \bibnamefont
  {Andrade}},\ and\ \bibinfo {author} {\bibfnamefont {M.}~\bibnamefont
  {Vojta}},\ }\bibfield  {title} {\bibinfo {title} {Honeycomb-lattice
  heisenberg-kitaev model in a magnetic field: Spin canting, metamagnetism, and
  vortex crystals},\ }\href {https://doi.org/10.1103/PhysRevLett.117.277202}
  {\bibfield  {journal} {\bibinfo  {journal} {Phys. Rev. Lett.}\ }\textbf
  {\bibinfo {volume} {117}},\ \bibinfo {pages} {277202} (\bibinfo {year}
  {2016})}\BibitemShut {NoStop}%
\bibitem [{\citenamefont {Janssen}\ and\ \citenamefont
  {Vojta}(2019)}]{Janssen_2019}%
  \BibitemOpen
  \bibfield  {author} {\bibinfo {author} {\bibfnamefont {L.}~\bibnamefont
  {Janssen}}\ and\ \bibinfo {author} {\bibfnamefont {M.}~\bibnamefont
  {Vojta}},\ }\bibfield  {title} {\bibinfo {title} {Heisenberg–kitaev physics
  in magnetic fields},\ }\href {https://doi.org/10.1088/1361-648X/ab283e}
  {\bibfield  {journal} {\bibinfo  {journal} {Journal of Physics: Condensed
  Matter}\ }\textbf {\bibinfo {volume} {31}},\ \bibinfo {pages} {423002}
  (\bibinfo {year} {2019})}\BibitemShut {NoStop}%
\bibitem [{\citenamefont {Balz}\ \emph {et~al.}(2019)\citenamefont {Balz},
  \citenamefont {Lampen-Kelley}, \citenamefont {Banerjee}, \citenamefont {Yan},
  \citenamefont {Lu}, \citenamefont {Hu}, \citenamefont {Yadav}, \citenamefont
  {Takano}, \citenamefont {Liu}, \citenamefont {Tennant}, \citenamefont
  {Lumsden}, \citenamefont {Mandrus},\ and\ \citenamefont
  {Nagler}}]{Balz2019magnetostriction}%
  \BibitemOpen
  \bibfield  {author} {\bibinfo {author} {\bibfnamefont {C.}~\bibnamefont
  {Balz}}, \bibinfo {author} {\bibfnamefont {P.}~\bibnamefont {Lampen-Kelley}},
  \bibinfo {author} {\bibfnamefont {A.}~\bibnamefont {Banerjee}}, \bibinfo
  {author} {\bibfnamefont {J.}~\bibnamefont {Yan}}, \bibinfo {author}
  {\bibfnamefont {Z.}~\bibnamefont {Lu}}, \bibinfo {author} {\bibfnamefont
  {X.}~\bibnamefont {Hu}}, \bibinfo {author} {\bibfnamefont {S.~M.}\
  \bibnamefont {Yadav}}, \bibinfo {author} {\bibfnamefont {Y.}~\bibnamefont
  {Takano}}, \bibinfo {author} {\bibfnamefont {Y.}~\bibnamefont {Liu}},
  \bibinfo {author} {\bibfnamefont {D.~A.}\ \bibnamefont {Tennant}}, \bibinfo
  {author} {\bibfnamefont {M.~D.}\ \bibnamefont {Lumsden}}, \bibinfo {author}
  {\bibfnamefont {D.}~\bibnamefont {Mandrus}},\ and\ \bibinfo {author}
  {\bibfnamefont {S.~E.}\ \bibnamefont {Nagler}},\ }\bibfield  {title}
  {\bibinfo {title} {Finite field regime for a quantum spin liquid in
  $\alpha$-{R}u{C}l$_3$},\ }\href {https://doi.org/10.1103/PhysRevB.100.060405}
  {\bibfield  {journal} {\bibinfo  {journal} {Phys. Rev. B}\ }\textbf {\bibinfo
  {volume} {100}},\ \bibinfo {pages} {060405} (\bibinfo {year}
  {2019})}\BibitemShut {NoStop}%
\bibitem [{\citenamefont {Balz}\ \emph {et~al.}(2021)\citenamefont {Balz},
  \citenamefont {Janssen}, \citenamefont {Lampen-Kelley}, \citenamefont
  {Banerjee}, \citenamefont {Liu}, \citenamefont {Yan}, \citenamefont
  {Mandrus}, \citenamefont {Vojta},\ and\ \citenamefont
  {Nagler}}]{balz2021field}%
  \BibitemOpen
  \bibfield  {author} {\bibinfo {author} {\bibfnamefont {C.}~\bibnamefont
  {Balz}}, \bibinfo {author} {\bibfnamefont {L.}~\bibnamefont {Janssen}},
  \bibinfo {author} {\bibfnamefont {P.}~\bibnamefont {Lampen-Kelley}}, \bibinfo
  {author} {\bibfnamefont {A.}~\bibnamefont {Banerjee}}, \bibinfo {author}
  {\bibfnamefont {Y.}~\bibnamefont {Liu}}, \bibinfo {author} {\bibfnamefont
  {J.-Q.}\ \bibnamefont {Yan}}, \bibinfo {author} {\bibfnamefont
  {D.}~\bibnamefont {Mandrus}}, \bibinfo {author} {\bibfnamefont
  {M.}~\bibnamefont {Vojta}},\ and\ \bibinfo {author} {\bibfnamefont {S.~E.}\
  \bibnamefont {Nagler}},\ }\bibfield  {title} {\bibinfo {title} {Field-induced
  intermediate ordered phase and anisotropic interlayer interactions in
  $\alpha$- {R}u{C}l$_3$},\ }\href@noop {} {\bibfield  {journal} {\bibinfo
  {journal} {Physical Review B}\ }\textbf {\bibinfo {volume} {103}},\ \bibinfo
  {pages} {174417} (\bibinfo {year} {2021})}\BibitemShut {NoStop}%
\bibitem [{\citenamefont {Trebst}\ and\ \citenamefont
  {Hickey}(2022)}]{trebst2022kitaev}%
  \BibitemOpen
  \bibfield  {author} {\bibinfo {author} {\bibfnamefont {S.}~\bibnamefont
  {Trebst}}\ and\ \bibinfo {author} {\bibfnamefont {C.}~\bibnamefont
  {Hickey}},\ }\bibfield  {title} {\bibinfo {title} {Kitaev materials},\
  }\href@noop {} {\bibfield  {journal} {\bibinfo  {journal} {Physics Reports}\
  }\textbf {\bibinfo {volume} {950}},\ \bibinfo {pages} {1} (\bibinfo {year}
  {2022})}\BibitemShut {NoStop}%
\bibitem [{\citenamefont {Pereira}\ and\ \citenamefont
  {Egger}(2020)}]{Pereira20}%
  \BibitemOpen
  \bibfield  {author} {\bibinfo {author} {\bibfnamefont {R.~G.}\ \bibnamefont
  {Pereira}}\ and\ \bibinfo {author} {\bibfnamefont {R.}~\bibnamefont
  {Egger}},\ }\bibfield  {title} {\bibinfo {title} {Electrical access to ising
  anyons in kitaev spin liquids},\ }\href
  {https://doi.org/10.1103/PhysRevLett.125.227202} {\bibfield  {journal}
  {\bibinfo  {journal} {Phys. Rev. Lett.}\ }\textbf {\bibinfo {volume} {125}},\
  \bibinfo {pages} {227202} (\bibinfo {year} {2020})}\BibitemShut {NoStop}%
\bibitem [{\citenamefont {Landau}(1937)}]{Landau65}%
  \BibitemOpen
  \bibfield  {author} {\bibinfo {author} {\bibfnamefont {L.~D.}\ \bibnamefont
  {Landau}},\ }\href@noop {} {\bibfield  {journal} {\bibinfo  {journal} {Zh.
  Eksp. Teor. Fiz.}\ }\textbf {\bibinfo {volume} {7}},\ \bibinfo {pages} {627}
  (\bibinfo {year} {1937})},\ \bibinfo {note} {or, "On phase transitions II",
  by L. D. Landau, in "Collected Works of L. D. Landau", Gordon and Breach, N.
  Y., 1965)}\BibitemShut {NoStop}%
\bibitem [{\citenamefont {Yip}\ \emph {et~al.}(1991)\citenamefont {Yip},
  \citenamefont {Li},\ and\ \citenamefont {Kumar}}]{Yip91}%
  \BibitemOpen
  \bibfield  {author} {\bibinfo {author} {\bibfnamefont {S.~K.}\ \bibnamefont
  {Yip}}, \bibinfo {author} {\bibfnamefont {T.}~\bibnamefont {Li}},\ and\
  \bibinfo {author} {\bibfnamefont {P.}~\bibnamefont {Kumar}},\ }\bibfield
  {title} {\bibinfo {title} {Thermodynamic considerations and the phase diagram
  of superconducting {UP}t$_{3}$},\ }\href
  {https://doi.org/10.1103/PhysRevB.43.2742} {\bibfield  {journal} {\bibinfo
  {journal} {Phys. Rev. B}\ }\textbf {\bibinfo {volume} {43}},\ \bibinfo
  {pages} {2742} (\bibinfo {year} {1991})}\BibitemShut {NoStop}%
\end{thebibliography}
%

\end{document}



\title{Supplemental Material of ``Electronic and magnetic phase diagrams of Kitaev quantum spin liquid candidate Na$_2$Co$_2$TeO$_6$"}


\author{Shengzhi Zhang}
\email{shengzhi@lanl.gov}
\affiliation{National High Magnetic Field Laboratory, Los Alamos National Laboratory, Los Alamos, New Mexico 87545, USA.}
\author{Sangyun Lee}
\affiliation{MPA-Q, Los Alamos National Laboratory, Los Alamos, New Mexico 87545, USA.}
\author{Andrew J. Woods}
\affiliation{MPA-Q, Los Alamos National Laboratory, Los Alamos, New Mexico 87545, USA.}
\author{Sean M. Thomas}
\affiliation{MPA-Q, Los Alamos National Laboratory, Los Alamos, New Mexico 87545, USA.}
\author{Roman Movshovich}
\affiliation{MPA-Q, Los Alamos National Laboratory, Los Alamos, New Mexico 87545, USA.}
\author{Eric Brosha}
\affiliation{Los Alamos National Laboratory, Los Alamos, New Mexico 87545, USA.}
\author{Qing Huang}
\affiliation{Department of Physics, University of Tennessee, Knoxville, Tennessee 37996, USA.}
\author{Haidong Zhou}
\affiliation{Department of Physics, University of Tennessee, Knoxville, Tennessee 37996, USA.}
\author{Vivien S. Zapf}
\email{vzapf@lanl.gov}
\affiliation{National High Magnetic Field Laboratory, Los Alamos National Laboratory, Los Alamos, New Mexico 87545, USA.}
\author{Minseong Lee}
\email{ml10k@lanl.gov}
\affiliation{National High Magnetic Field Laboratory, Los Alamos National Laboratory, Los Alamos, New Mexico 87545, USA.}

\date{\today}

\begin{abstract}

\end{abstract}


\maketitle

\setcounter{equation}{0}
\setcounter{figure}{0}
\setcounter{table}{0}
\setcounter{page}{1}
\makeatletter
\renewcommand{\theequation}{S\arabic{equation}}
\renewcommand{\thefigure}{S\arabic{figure}}
\renewcommand{\thetable}{S\arabic{table}}
\renewcommand{\etal}{\textit{et al. }}
\renewcommand{\insitu}{\textit{in situ}}

\section{Experimental configurations}
\begin{figure}
\includegraphics[width=\linewidth]{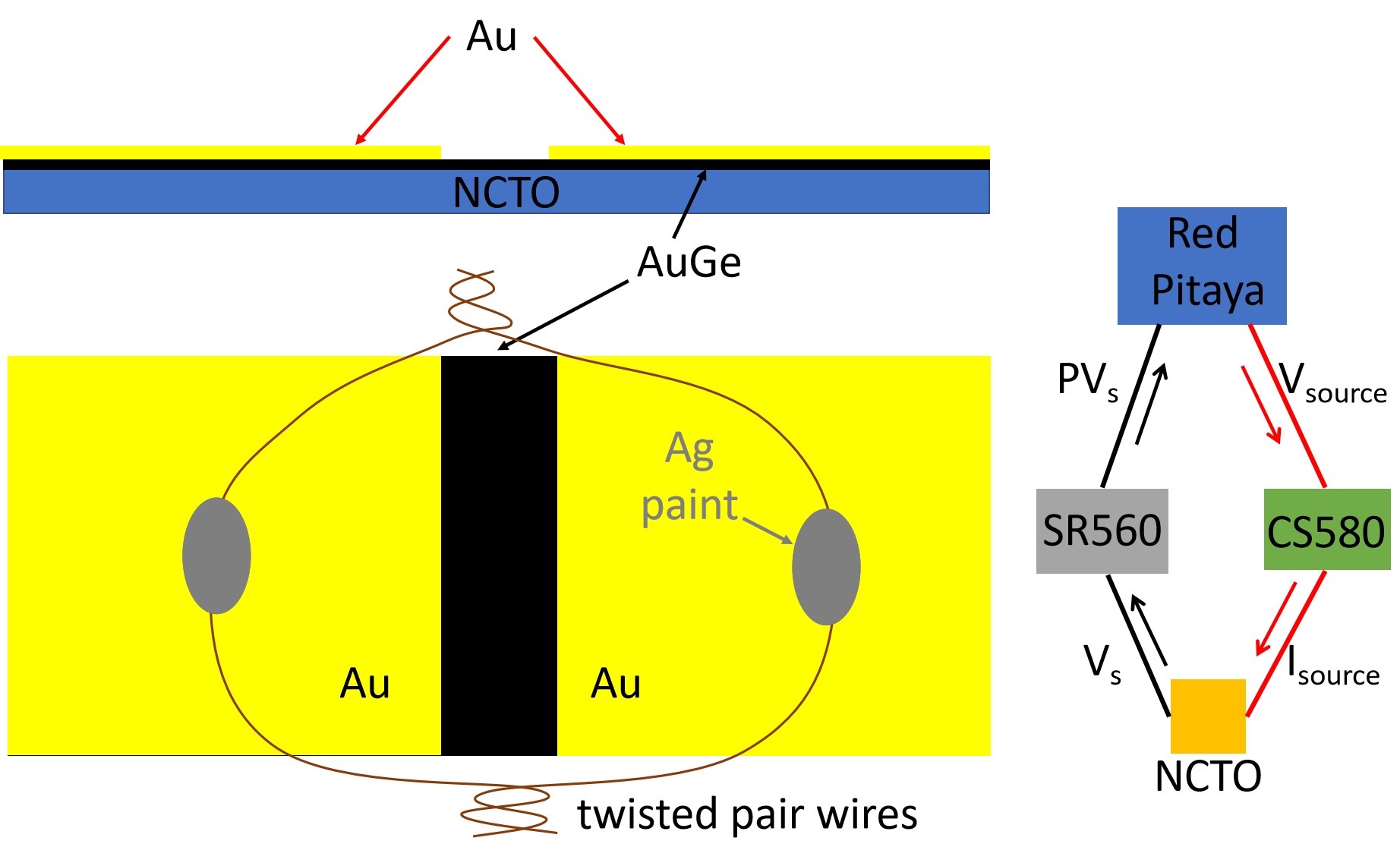}%
\caption {Top and bottom of the left panel respectively show the side and top views of the magnetocaloric effect measurement wiring configuration. The right hand side illustrates the standard four-point resistance measurement setup. $\text{V}_{\text{source}}$, $\text{I}_{\text{source}}$ are the sourcing voltage from the Red Pitaya and the converted current applied to the sample, respectively. $\text{V}_{\text{s}}$ is the induced voltage in the sample, and P is the amplification of the pre-amplifier (SR560). \label{SUPP-MC-cartoon}}
\end{figure}

This section illustrates the experimental setup of the magnetocaloric effect and magnetostriction measurements. Shown in the left-hand-side of Fig.~\ref{SUPP-MC-cartoon} is a schematic drawing of the wiring configuration for magnetocaloric effect measurements. A 10 nm AuGe thin film was deposited on the {\it ab}-plane of an as-grown \NCTO{} single crystal and two gold contact pads were then deposited on the thin film, leaving a thin stripe exposed. Twisted pair of manganin wires were attached to the contact pads using silver paint. Then a standard four-probe resistance measurement was carried out in the 65 T pulsed field magnet at Pulsed Field Facility at National High Magnetic Field Laboratory whose set-up is illustrated in the right-hand-side of Fig.~\ref{SUPP-MC-cartoon}. A sourcing voltage ($\text{V}_{\text{source}}$) is supplied from a Red Pitaya (https://redpitaya.com/) and converted into current ($\text{I}_{\text{source}}$) by a Stanford Research System (SRS) CS580 voltage-controlled current source. The induced voltage from the sample is amplified by an SRS SR560 voltage pre-amplifier ($\text{V}_{\text{s}} \rightarrow \text{PV}_{\text{s}}$) before being measured in the Red Pitaya. Magnetic field was applied within the plane so that heating from the eddy current was negligible.

\begin{figure}
\includegraphics[width=\linewidth]{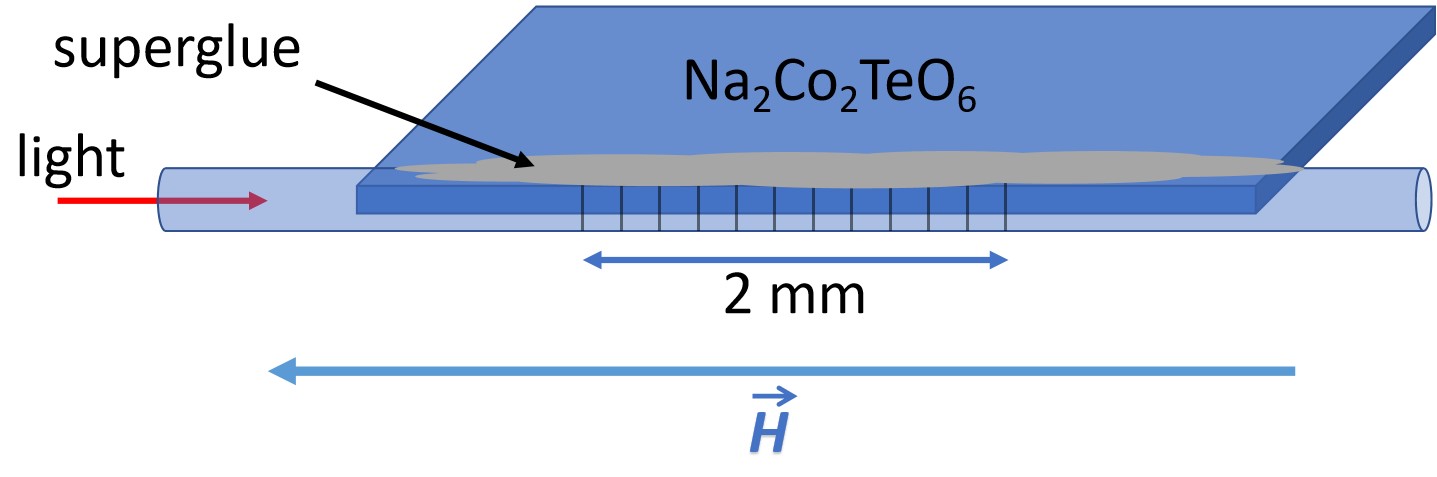}%
\caption {A cartoon illustrating the configuration of the FBG magnetostriction measurement. The thick rod and the set of vertical short lines represent the fiber and the Bragg grating, respectively. The grey amorphous shaped area represents the superglue. Magnetic field {\bf H} is applied along the fiber as indicated by the arrow. \label{SUPP-MS-cartoon}}
\end{figure}

Shown in Fig.~\ref{SUPP-MS-cartoon} is an illustration of the crystal attached to the optical fiber in the magnetostriction measurement using fiber Bragg grating (FBG) technique. Single crystalline \NCTO{} was glued to the 2 mm grates using Pattex ultra gel superglue along either $a$- or $a^*$-axis. The edge of the sample is much longer than the grating period to ensure full coverage. As demonstrated, the sample is thinner than the fiber so a certain amount of superglue was needed to ensure a secure bonding which may cause extra strain on the sample.

\section{Magneto-electrical poling effect}
To strengthen our argument on the spin structure of Phase I, we further eliminate the effect of electrical and magnetic domains by performing magnetization and electrical polarization measurements at low temperatures after \textit{magneto-electric poling}. That is to apply a 4 T magnetic field and 200 V electric field to a polycrystal at 100 K and during the cool-down process. This will align the magnetic moments and electric dipoles to form a monodomain to eliminate the effect of multidomain effects. Note that in these experiment, the applied magnetic field is always perpendicular to the electrical polarization direction, in line with the configuration demonstrated in Fig. 1 (a).

\begin{figure}
\includegraphics[width=\linewidth]{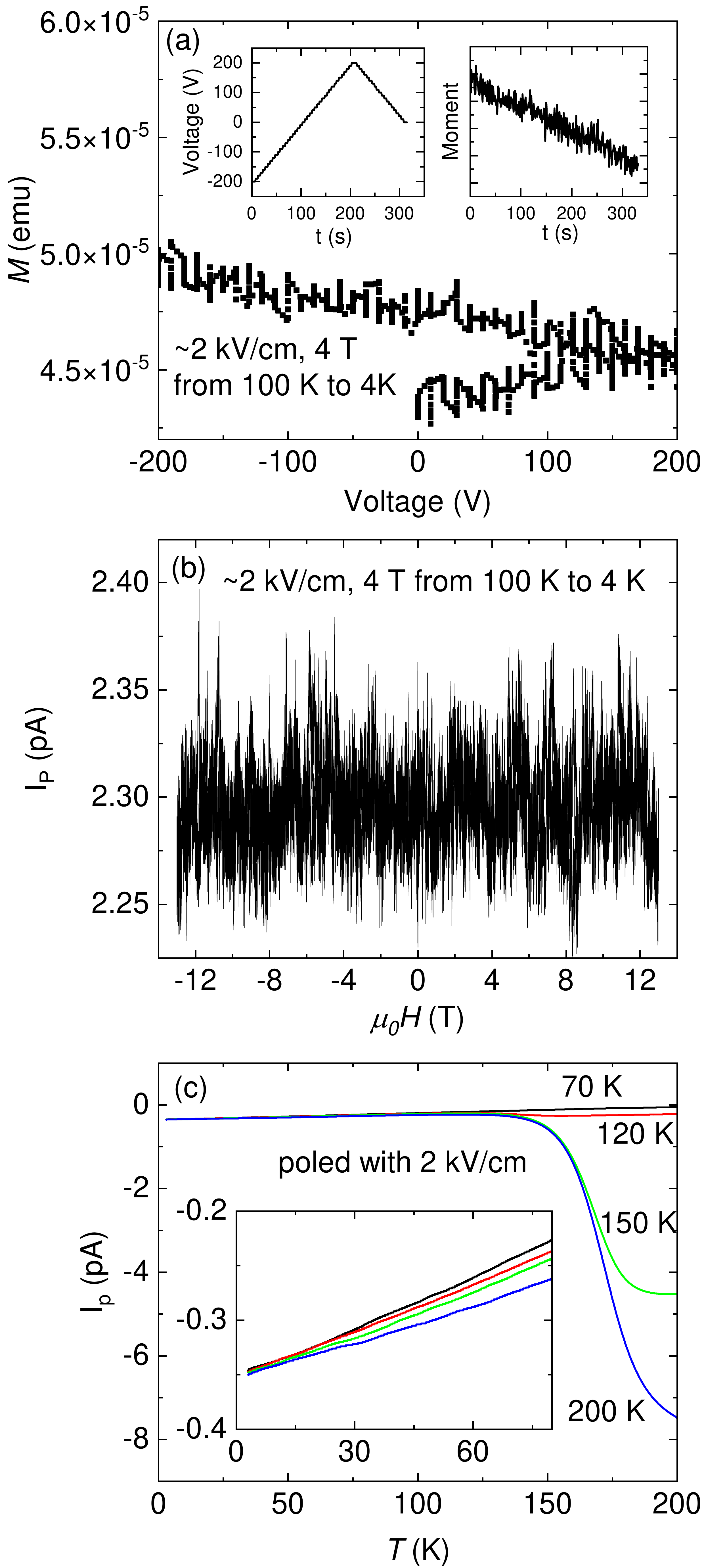}%
\caption {(a) Magnetization as a function of applied voltage across a polycrystalline \NCTO{}. The left and right inset shows the time-dependent applied voltage and magnetization. (b) Magnetoelectric current as a function of magnetic field of the same polycrystalline \NCTO{} as in panel (a). In both data-sets, a magnetic field of 4 T and an electric field of about 2kV/cm are applied from 100K to 4K at which the data were taken. (c) Temperature-dependent pyroelectric current of the same \NCTO{} polycrystal sample. Electric field of 2kV/cm was applied at 70 K, 120 K, 150 K, and 200 K, respectively, and during cooling down. The data were measured while warming up to 200 K with 0 V/cm electric field applied. \label{SUPP-MEpoling}}
\end{figure}

From our results as shown in Fig.~\ref{SUPP-MEpoling} (a) and (b), no linear magneto-electric coupling is observed, negating the triple-Q scenario, supporting our claim in the main text.

In addition, to examine the discrepancy between our measurement and the ones made by Mukherjee \textit {et al.}, we attempted to reproduce their observations but could not observe any pyroelectric current shown in their paper, as shown in Fig.~\ref{SUPP-MEpoling} (c). 

\section{Complete data sets}

\subsection{Magnetization}
Dc magnetization measurements up to 60 T were performed in the Pulsed Field Facility at National High Magnetic Field Laboratory by detecting the field-induced voltage across a pick-up coil made of high purity Cu. The samples were carefully aligned and put in a 1 mm diameter sample holder. For each temperature, two measurements were performed with the sample in and out of the pick-up coil. The signal difference between the two was taken as the sample signal which was then calibrated according to the PPMS dc magnetization data.

\begin{figure}
\includegraphics[width=\linewidth]{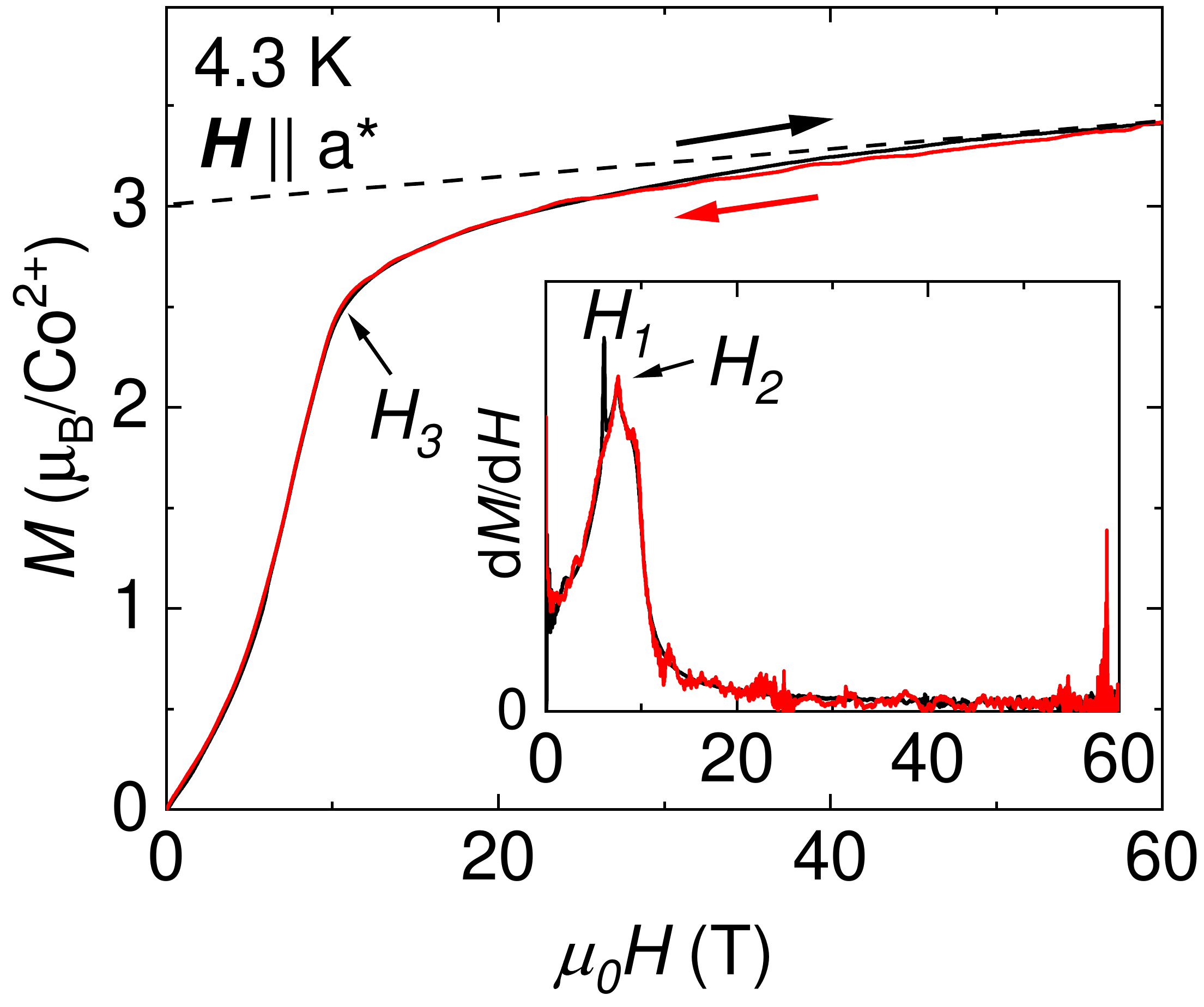}%
\caption{Magnetization $M$ as a function of magnetic field $H$ applied along a*-axis measured in millisecond pulsed fields to 60 T at 4.3 K. The black and red lines are up- and down-sweeps, respectively. The inset is the derivative $dM/dH$. \label{MvsH-astar-pulsed}}
\end{figure}

First we show the magnetization as a function of magnetic field up to 60 T measured in the pulsed field facility in Fig.~\ref{MvsH-astar-pulsed}. Only a linear Van-Vleck contribution \cite{Lee2014VV} is observed above $H_3$ as also mentioned in the main text.

\begin{figure*}
\includegraphics[width=\textwidth]{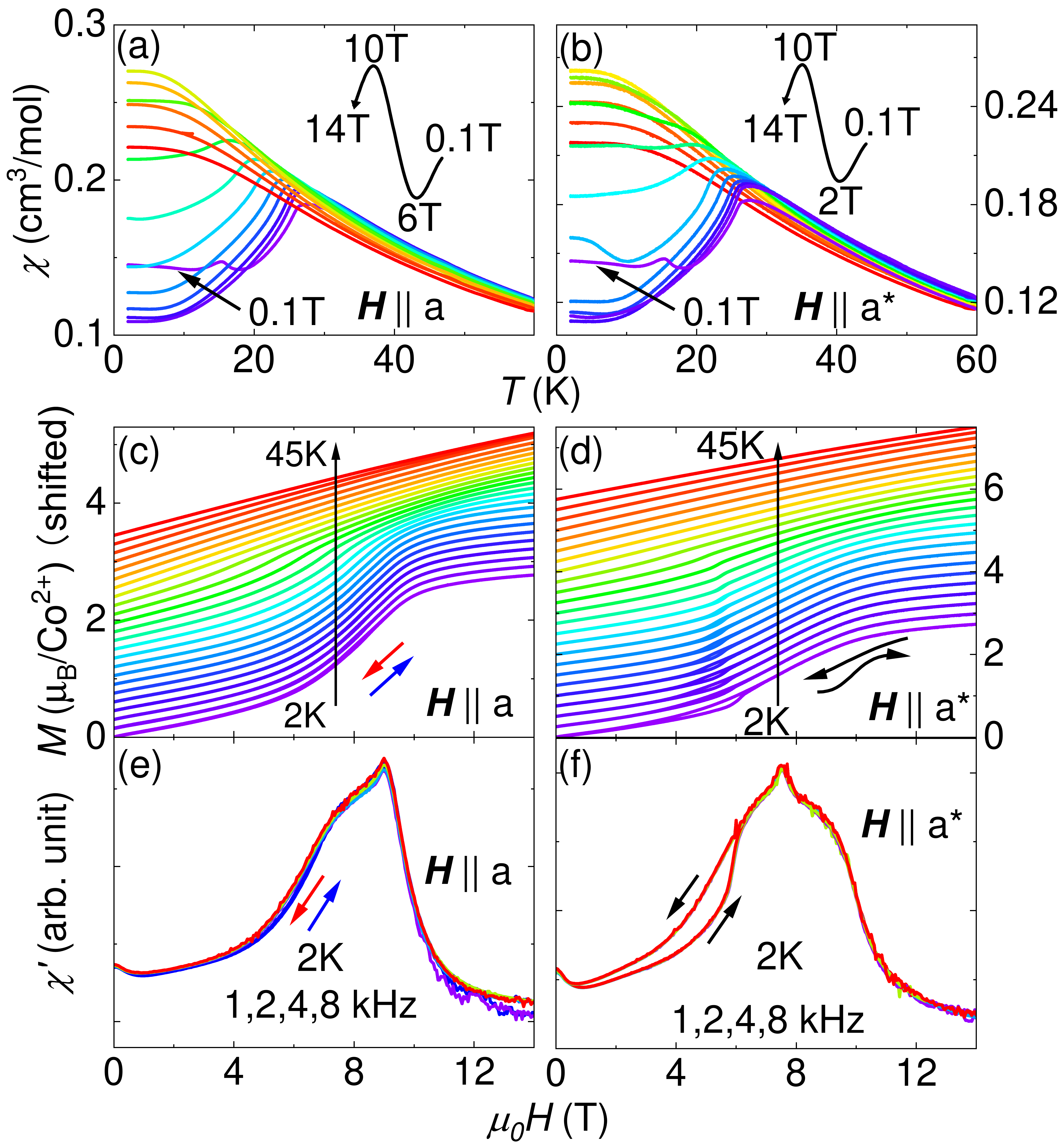}%
\caption {(a),(b) Complete data set of magnetic susceptibility as a function of temperature for ${\bf H} \parallel a$ and $a^{*}$, respectively. The s-shaped curves labeled with field strength illustrate the evolution of low temperature magnetic susceptibility with magnetic field. (c),(d) Complete data set of magnetization as a function of magnetic field for ${\bf H} \parallel$ a and a$^{*}$, respectively. Curves are shifted from 2 K data for clarity. (e),(f) ac magnetic susceptibility as a function of magnetic field along $a$, and $a^*$ axes, respectively, measured at 2 K with 1, 2, 4, 8 kHz. The arrows indicate the field sweep directions. \label{SUPP-Magnetization-inPlane}}
\end{figure*}

Fig.~\ref{SUPP-Magnetization-inPlane} (a)-(d) are the complete data sets of dc magnetization measurements from which Fig.~2 and Fig.~3 in the main text are selected. The hysteresis behavior with ${\bf H} \parallel a^{*}$ is more clearly shown in Fig.~\ref{SUPP-Magnetization-inPlane} (d). Panel (e) and (f) are the magnetic field dependent ac susceptibility illustrating the absence of frequency dependence. A pronounced hysteresis loop is again observed when ${\bf H} \parallel a^*$.

\begin{figure*}
\includegraphics[width=\textwidth]{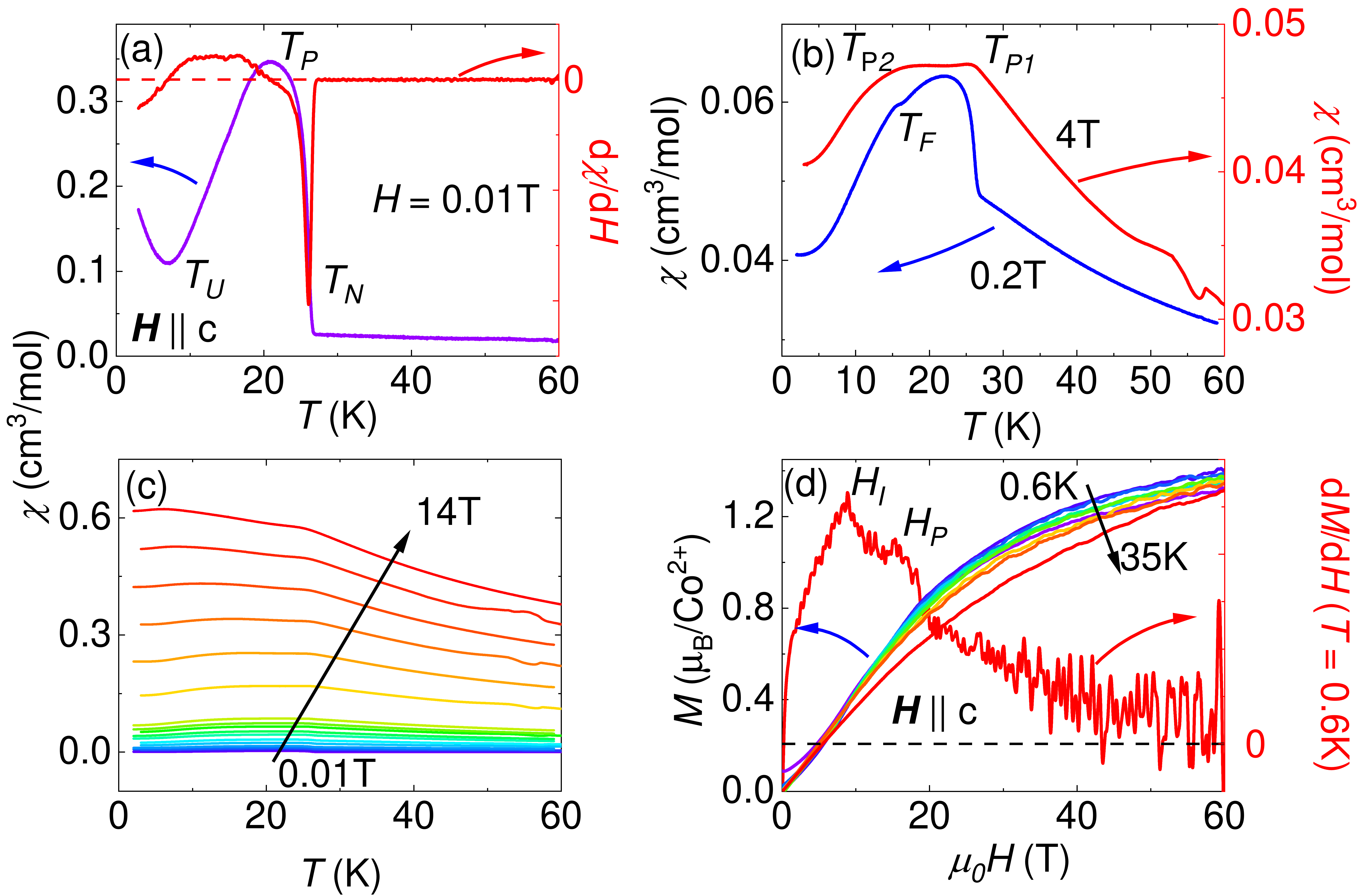}%
\caption {${\bf H} \parallel c$ (a),(b) Magnetic susceptibility ($\chi$) as a function of temperature taken at 0.01 T, 0.2 T and 2 T to show the critical temperatures in Fig.~\ref{SUPP-PD-c}. (c) A complete data set of temperature dependent $\chi$ from 0.01 T to 14 T. (d) Magnetic moment as a function of applied magnetic field in a temperature range of 0.6 K to 35 K. This is measured in a 65 T pulsed field facility. The right axis is the derivative $dM/dH$ at 0.6 K. Critical fields used in Fig.~\ref{SUPP-PD-c} are also labeled. \label{SUPP-Magnetization-c}}
\end{figure*}

The $c$-axis magnetic susceptibility data are displayed in Fig.~\ref{SUPP-Magnetization-c} (a) and (b). Similarly to the in-plane measurements presented in the main text, the antiferromagnetic transition at $T_{\text{N}} \approx$ 27 K as well as the spin canting at $T_{\text{F}} \approx$ 15 K are both observed. Observation of $T_{\text{F}}$ is consistent with the scenario in which the spins are slightly canted towards $c$-axis. On the other hand, observation of $T_{\text{N}}$ along $c$-axis as a sharp increase in dc magnetization may be explained by ferrimagnetism in the system \cite{yao2020ferrimagnetism}. However, it shows up as a sharp $increase$ in magnetic susceptibility, drastically different from the in-plane behavior. With increasing magnetic field, both $T_{\text{N}}$ and $T_{\text{F}}$ increase slightly until 2 T above which they diminish. In addition to the critical temperatures observed in {\it ab}-plane, a broad peak is observed in the magnetic susceptibility at $T_P$ (panel (a)) which splits into two peaks ($T_{\text{P1}}$ and $T_{\text{P2}}$) at higher fields (panel (b)). $T_{\text{P2}}$ gets suppressed by increasing magnetic field whereas $T_{\text{P2}}$ remains largely constant even at 14 T. At even lower temperatures, an upturn appears at $T_U$ which is quickly suppressed by very small magnetic field. Fig.~\ref{SUPP-Magnetization-c} (d) shows the magnetic moment as a function of magnetic field up to 60 T. From their first derivatives, we can extract two critical fields defined by the major peak (inflection point $H_{\text{I}}$) and a second peak in the first derivative ($H_{\text{P}}$). 

\begin{figure}
\includegraphics[width=\linewidth]{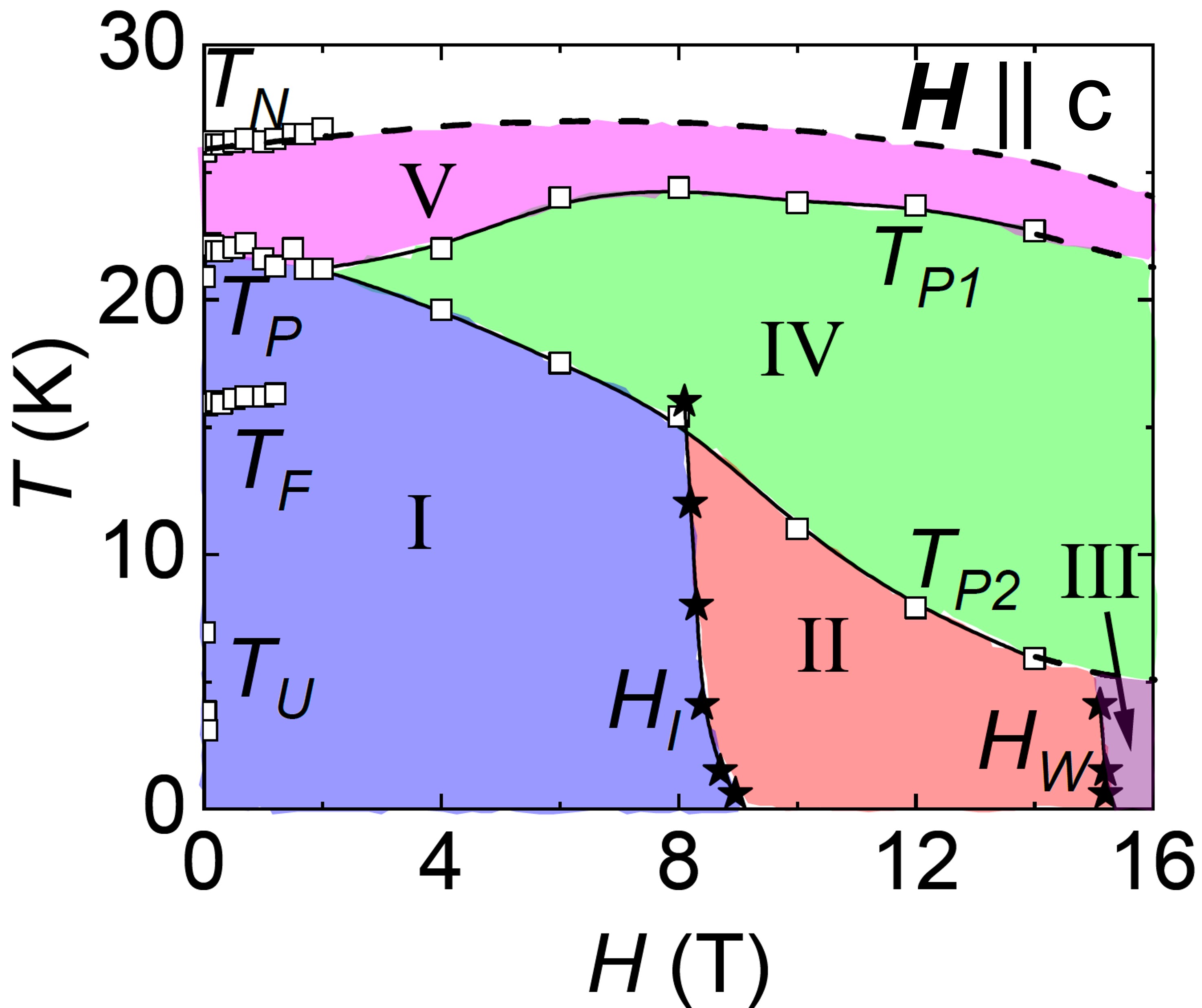}%
\caption {${\bf H} \parallel c$ temperature-field ($\bf{T}$-$\bf{H}$) phase diagram constructed from Fig.~\ref{SUPP-Magnetization-c}. The empty squares are from temperature sweeps and the solid stars are from magnetic field sweeps. The phase regions are labeled from I-V. \label{SUPP-PD-c}.}
\end{figure}

The $c$-axis temperature-field (${\bf T-H}$) phase diagram is plotted in Fig.~\ref{SUPP-PD-c}. Similar to the in-plane phase diagram Fig.~1 in the main text, phase boundaries with little temperature dependence are observed ($H_{\text{I}}$, $H_{\text{P}}$). More data points are needed to complete this phase diagram.

\subsection{Dielectric constant}

\begin{figure}
\includegraphics[width=\linewidth]{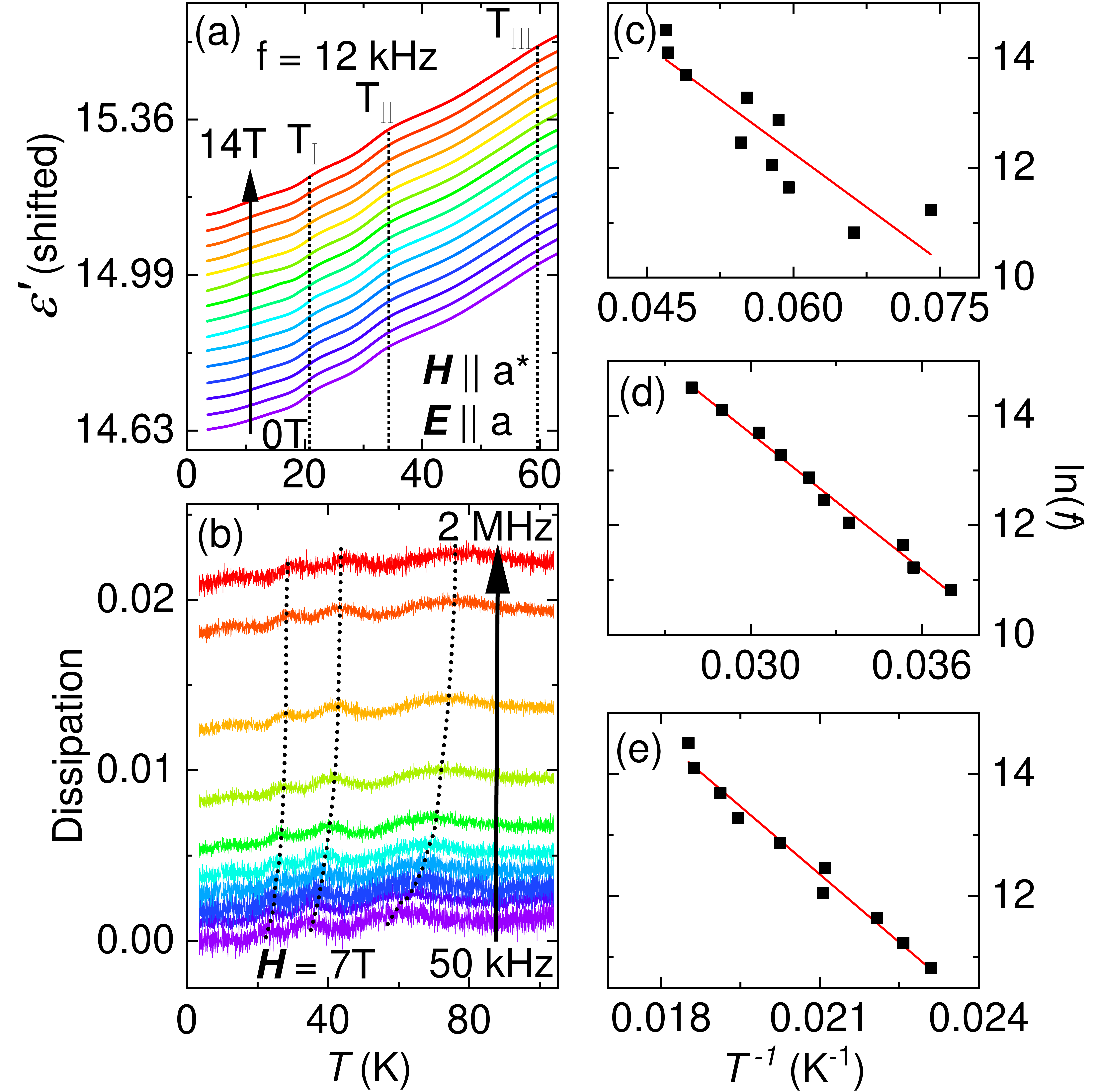}%
\caption{(a) Dielectric constant ($\varepsilon^{\prime}$) as a function of temperature under various magnetic field up to 14 T when magnetic field is applied along a*-axis and electric field along a-axis. The labels $T_{\text{I}}$, $T_{\text{II}}$, and $T_{\text{III}}$ indicate the temperatures at which three humps are observed. The dotted lines are guide to the eyes showing $T_{\text{I,II,III}}$ are independent of magnetic field. The other curves are shifted from ${\bf H}$ = 0 T curve for clarity. (b) Dissipation as a function of temperature measured with various frequencies. The dashed lines are guide to the eyes showing $T_{\text{I,II,III}}$ are frequency dependent. (c) - (f) $\ln$($f$) of each curve as a function of the locations of hump I to hump III in terms of $T^{-1}$, respectively. The solid red lines are fittings to the Arrhenius expression $\ln$ ($f$) = $\ln (A) - \frac{E_a}{k_B} \frac{1}{T}$. \label{SUPP-DC-Freq}}
\end{figure}

The temperature dependent dielectric constant data measured at magnetic fields up to 14 T are shown in Fig.~\ref{SUPP-DC-Freq} (a) with magnetic field applied along $a^*$-axis and electric field along $a$-axis. Three weak humps at $T_{\text{I}}$ through $T_{\text{III}}$ are observed as indicated by the dotted lines. These hump features are better observed in the dissipation as pronounced peaks. Therefore, we show in Fig.~\ref{SUPP-DC-Freq} (b) the frequency dependent dissipation measured at $H =$ 7 T. This is a good representation of all magnetic fields because the features showed no dependence on magnetic field strength as illustrated in panel Fig.~\ref{SUPP-DC-Freq} (a). As indicated by the dotted lines in panel (b), $T_{\text{I,II,III}}$ all increase with increasing frequency from 50 kHz to 2 MHz. By plotting the Arrhenius plot for all three features (Figs.~\ref{SUPP-DC-Freq}(c) - (e)), the activation energies of $T_I$ to $T_{III}$ features are extracted to be 11.27(1.9) meV, 35.68(1.2) meV, and 64.06(3.1) meV, respectively. In addition to their lacking magnetic field dependence, it should also be pointed out that none of $T_{\text{I,II,III}}$ matches the magnetic transitions plotted in Fig.~1 in the main text, indicating that they are not associated with any magnetic orderings.

\begin{figure*}
\includegraphics[width=\textwidth]{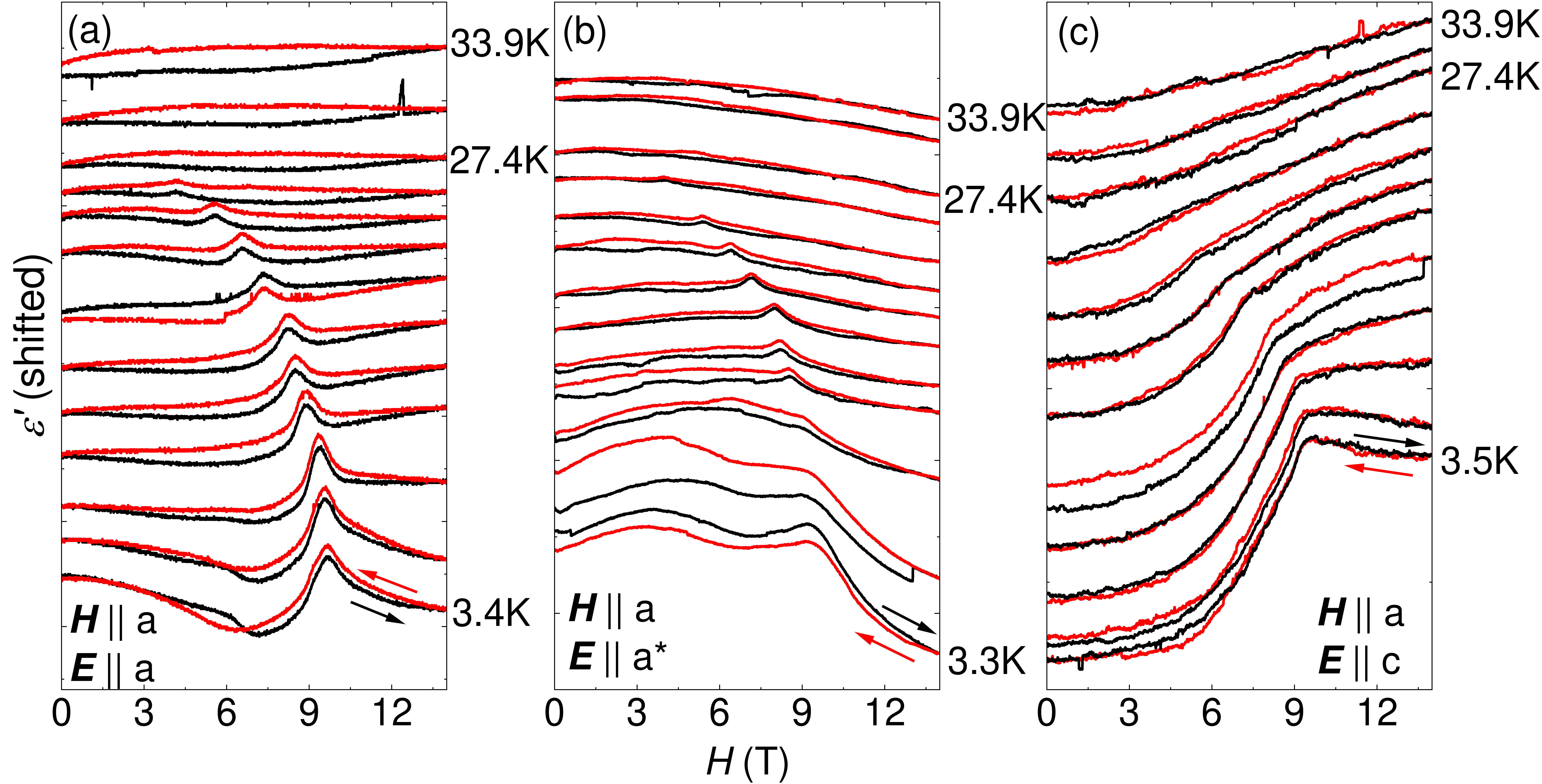}%
\caption {Dielectric constant as a function of magnetic field (${\bf H}$) when ${\bf H} \parallel$ $a$-axis taken at various temperatures. The electric field is applied along (a) $a$-, (b) $a^*$-, and (c) $c$-axes, respectively. \label{SUPP-DC-Ha}}
\end{figure*}

\begin{figure*}
\includegraphics[width=\textwidth]{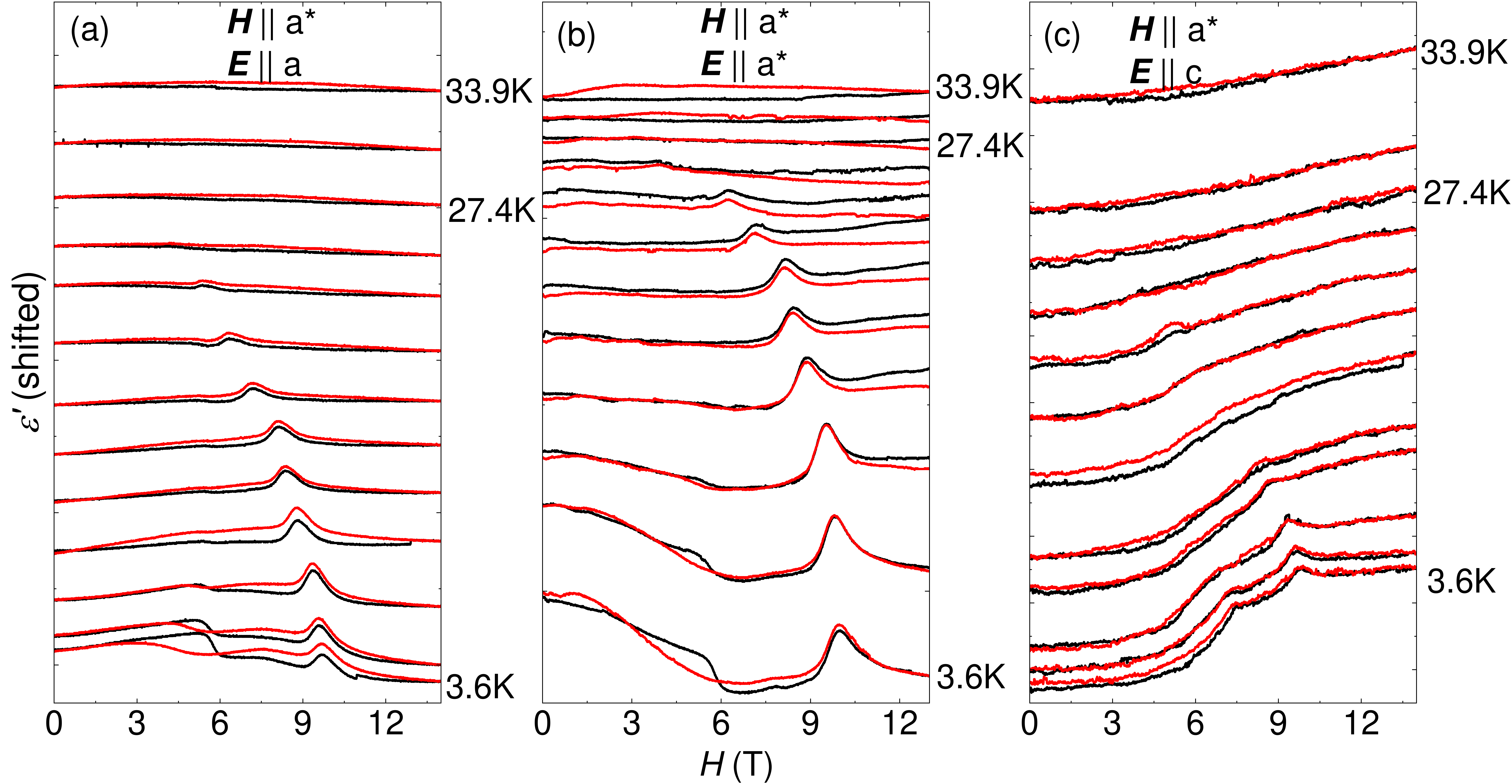}%
\caption {Dielectric constant as a function of magnetic field (${\bf H}$) when ${\bf H} \parallel$ $a^*$-axis taken at various temperatures. The electric field is applied along (a) $a$-, (b) $a^*$-, and (c) $c$-axes, respectively. \label{SUPP-DC-Hastar}}
\end{figure*}

\begin{figure}
\includegraphics[width=\linewidth]{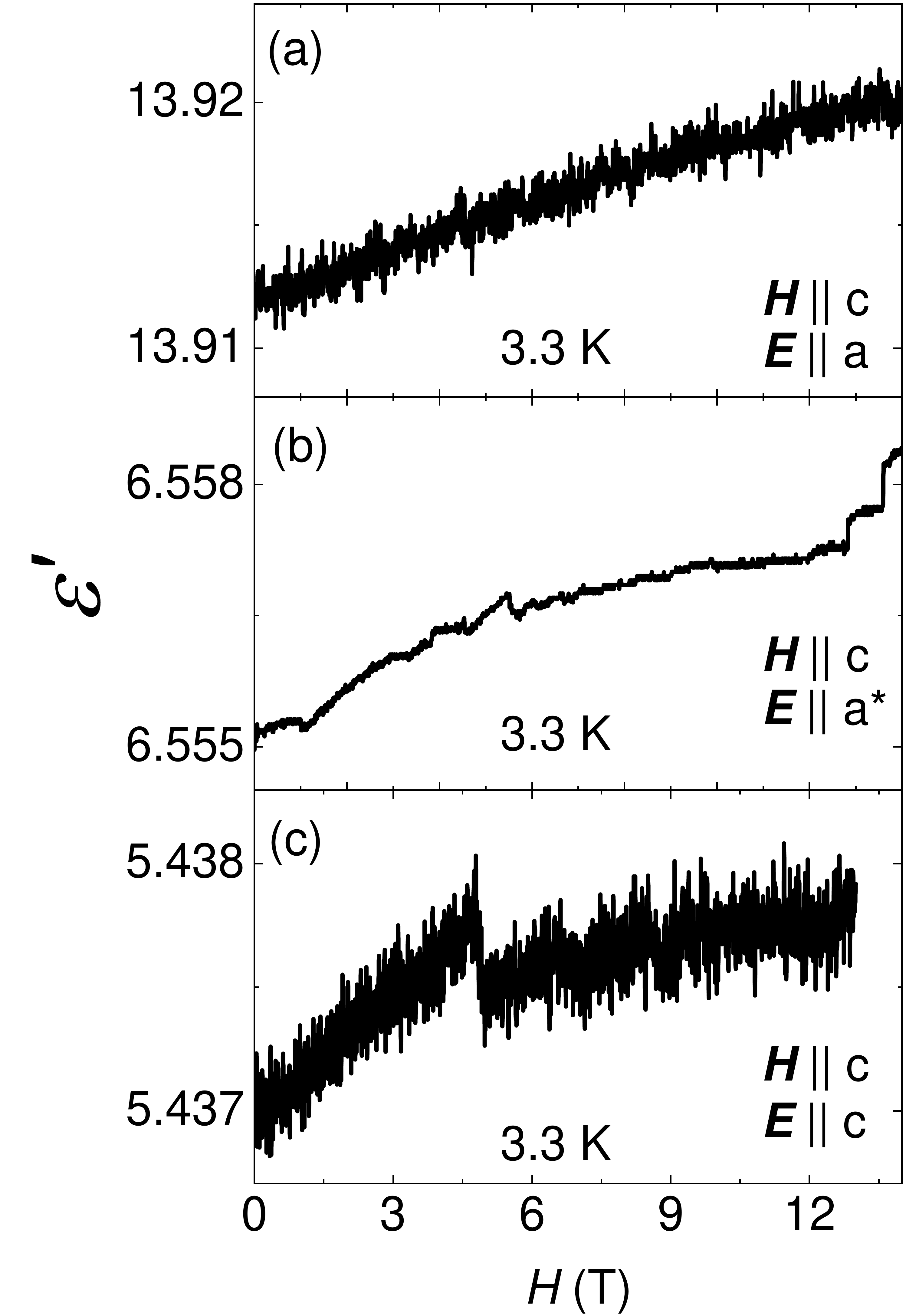}%
\caption {Dielectric constant as a function of magnetic field (${\bf H}$) when ${\bf H} \parallel c$-axis taken at 3.3 K. The electric field is applied along (a) $a$-, (b) $a^*$-, and (c) $c$-axes, respectively. \label{SUPP-DC-Hc}}
\end{figure}

The complete data set of dielectric constant as a function of magnetic field are plotted in Figs.~\ref{SUPP-DC-Ha}, \ref{SUPP-DC-Hastar}, and \ref{SUPP-DC-Hc} for ${\bf H} \parallel a$, ${\bf H} \parallel a^*$ and ${\bf H} \parallel c$, respectively. The hysteretic behavior of $H_1$ is more clearly shown in Figs.~\ref{SUPP-DC-Ha} and \ref{SUPP-DC-Hastar}. Note that in Fig.~\ref{SUPP-DC-Hc}, no pronounced feature is observed up to 14 T.

\subsection{Thermal expansion}
\begin{figure}
\includegraphics[width=\linewidth]{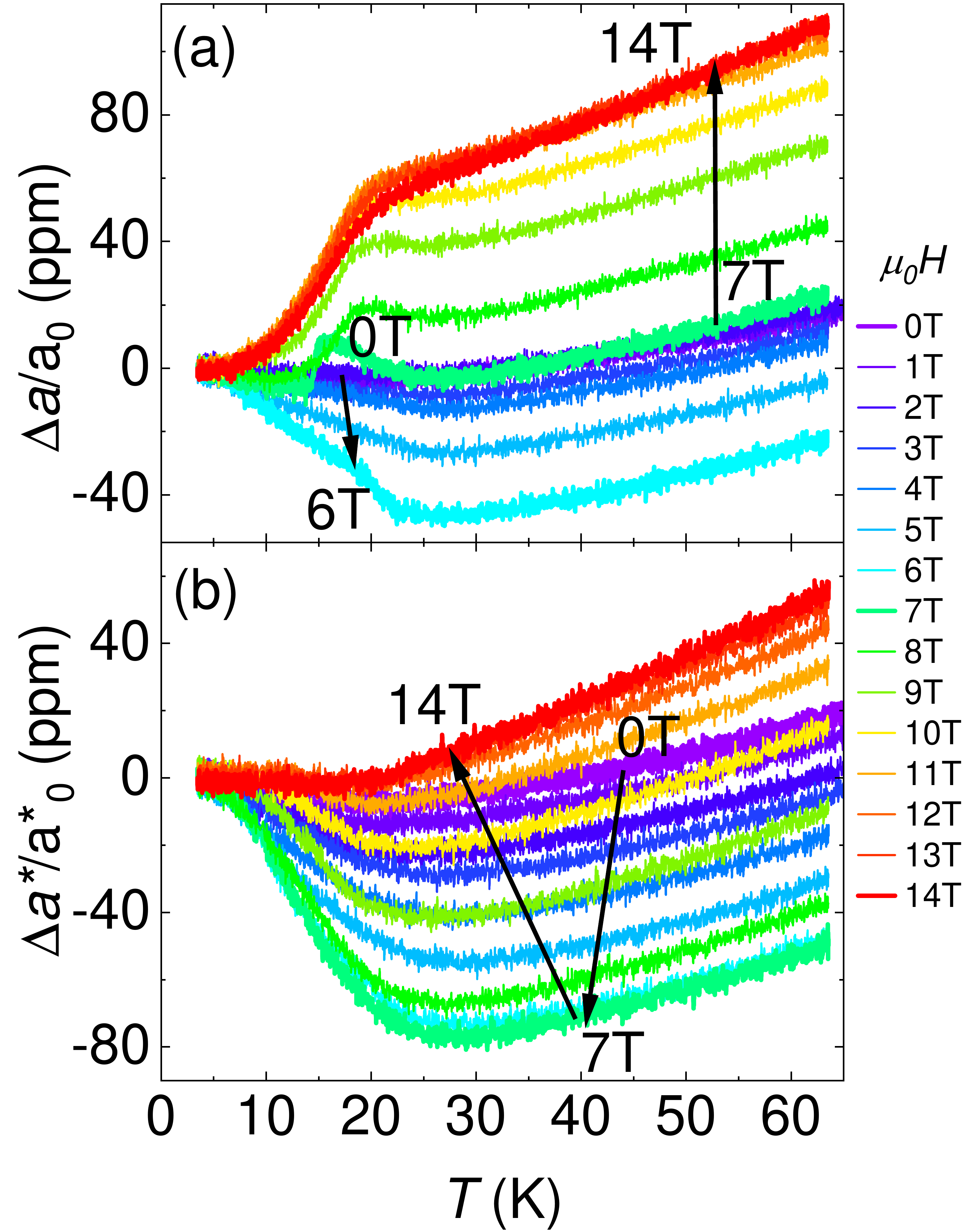}%
\caption {Complete data set of thermal expansion as a function of temperature measured at various magnetic fields. The arrows illustrate the evolution of the curves with increasing magnetic field. \label{SUPP-MSvsT}}
\end{figure}

\begin{figure*}
\includegraphics[width=\linewidth]{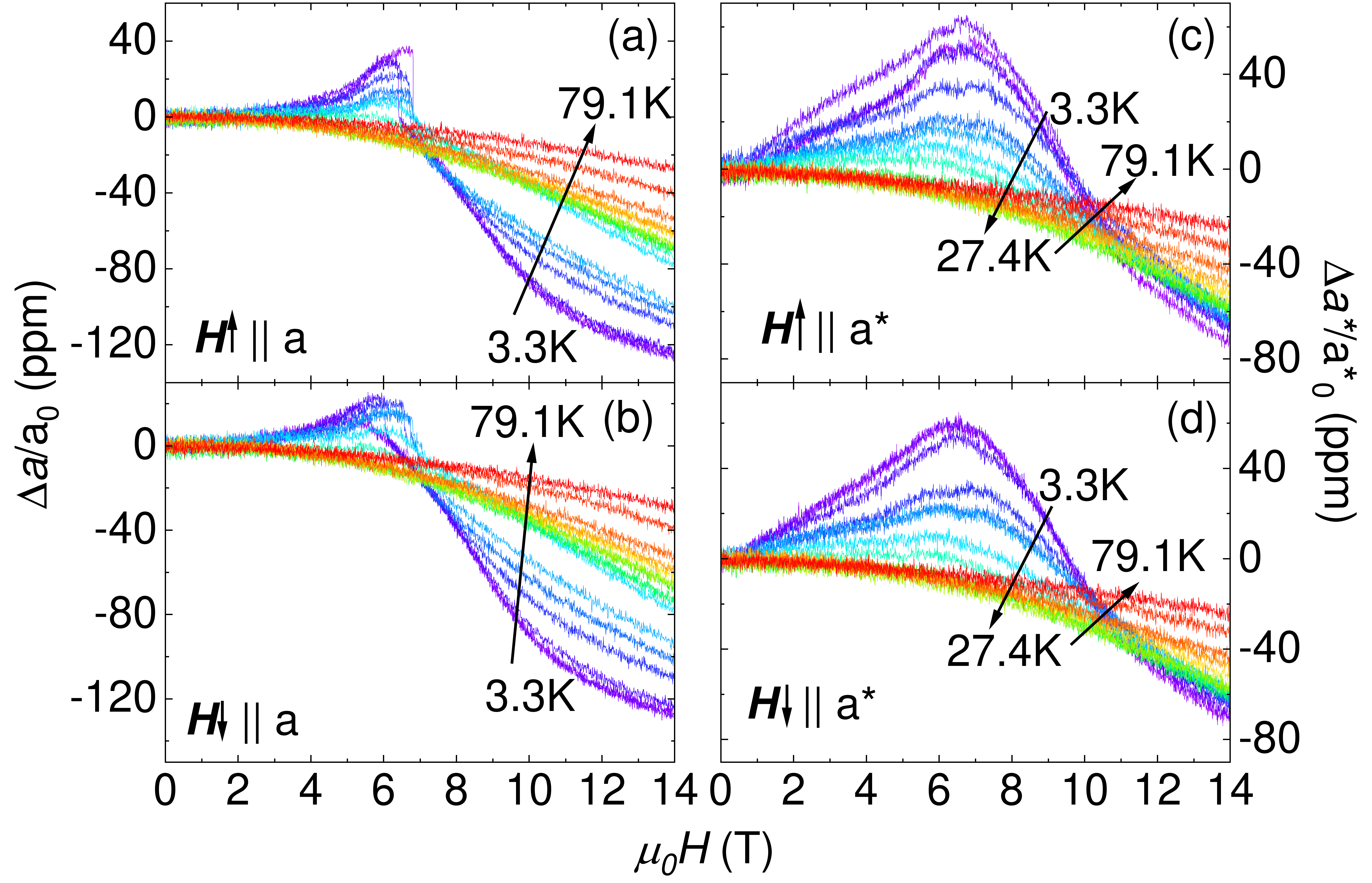}%
\caption {Complete data set of magnetostriction measured at various temperatures. The arrows illustrate the evolution of the curves with increasing temperature. \label{SUPP-MSvsH}}
\end{figure*}

The complete data set of thermal expansion as a function of temperature and magnetostriction are shown in Figs.~\ref{SUPP-MSvsT} and \ref{SUPP-MSvsH}, respectively. Temperature dependent thermal expansion exhibits a non-monotonic magnetic field dependence with 6 - 7 T as the turning point, roughly consistent with the $H_1$ phase transition. The magnetostriction along $a$-axis shows a monotonic temperature dependence. The sudden jump between the obviously two sets of curves occurs at $T_{\text{N}}$. On the other hand, the $a^*$-axis magnetostriction shows a non-monotonic temperature dependence with $T_{\text{N}}$ as the turning point.

\subsection{Magnetocaloric effect}
\begin{figure*}
\includegraphics[width=\textwidth]{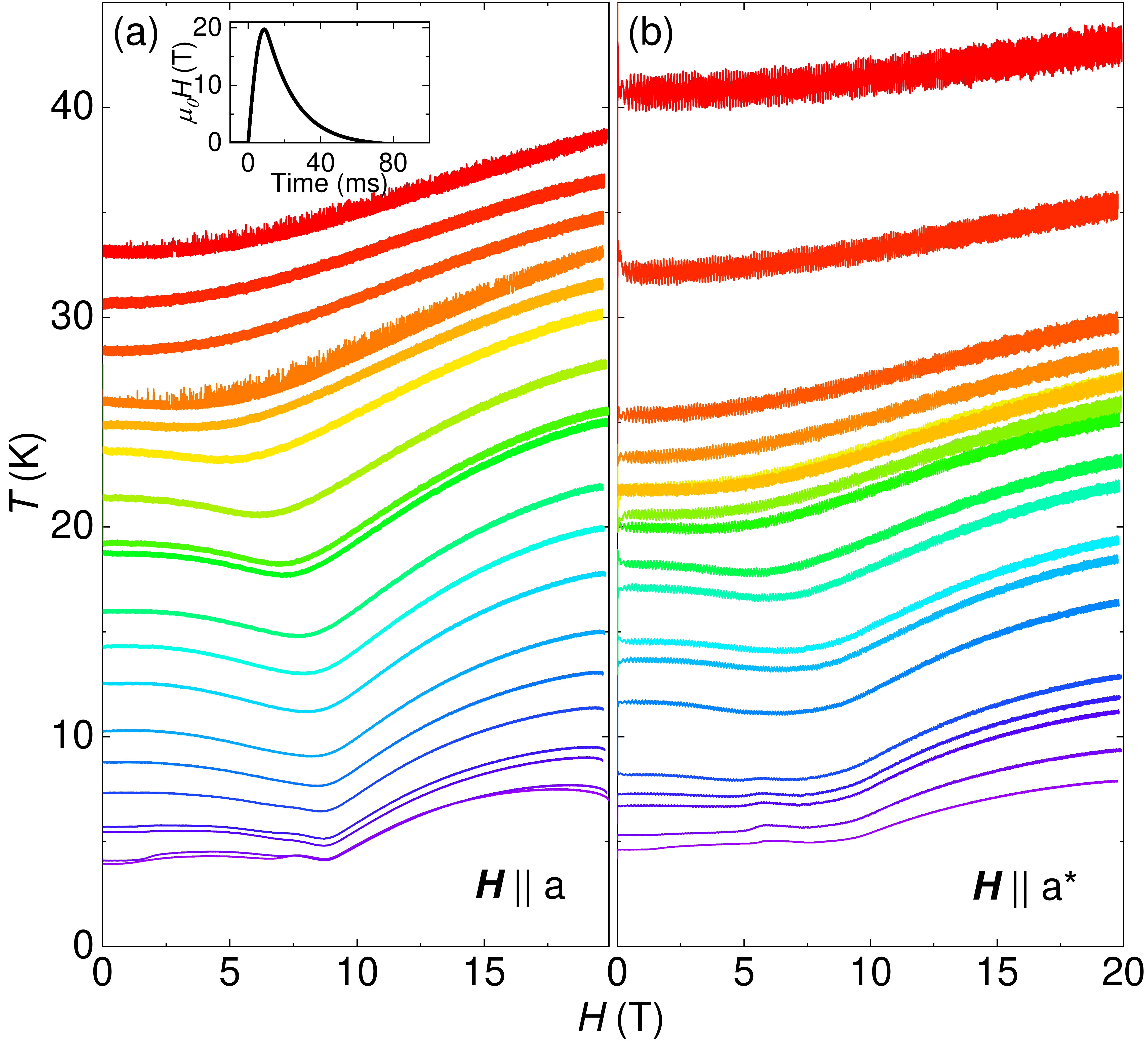}%
\caption {Complete data set of sample temperature as a function of magnetic field from magnetocaloric effect measurements when (a) ${\bf H} \parallel$ $a$-axis and (b) ${\bf H} \parallel$ $a^*$-axis. The inset of (a) illustrates the magnetic field as a function of time during one pulse using the 65 T pulsed magnet. \label{SUPP-MC}}
\end{figure*}

The complete data set of magnetocaloric effect measurements is shown in Fig.~\ref{SUPP-MC}. For clarity, only the up-sweep field is plotted.

\bibliography{Na2Co2TeO6}